\documentclass[useAMS,usenatbib]{mn2e}

\usepackage{amsmath}
\usepackage{natbib, txfonts, latexsym}
\usepackage{graphicx}

\renewcommand{\d}{{\rm d}} 

\newcommand{\owls}{\textit{OWLS}}
\newcommand{\msun}{${\rm M}_{\odot}$}

\newcommand{\per}{$^{-1}$}

\title[Properties of simulated galaxy populations at $z=2$ -- I]{Physical properties of simulated galaxy populations at $z=2$ -- \\I. Effect of metal-line cooling and feedback from star formation and AGN}
\author[Marcel R. Haas et al.]{Marcel~R.~Haas$^{1, 2, 3}$\thanks{E-mail: mhaas@physics.rutgers.edu (MRH)},
Joop~Schaye$^2$, C.~M.~Booth$^{4, 5, 2}$, Claudio~Dalla~Vecchia$^6$,
\newauthor Volker~Springel$^{7, 8}$,
Tom~Theuns$^{9, 10}$ and 
Robert~P.~C.~Wiersma$^2$\thanks{Current address: Atomic Energy of Canada Limited, Chalk River Laboratories, Chalk River, Ontario, K0J1J0, Canada} \\
\\
$^1$Department of Physics and Astronomy, Rutgers University, 136 Frelinghuysen Rd., Piscataway, NJ 08854, USA \\
$^{2}$Leiden Observatory, Leiden University, P.O. Box 9513, NL-2300 RA, Leiden, The Netherlands \\
$^{3}$Space Telescope Science Institute, 3700 San Martin Drive, Baltimore, MD 21218, USA \\
$^4$Department of Astronomy and Astrophysics, The University of Chicago, Chicago, IL 60637, USA \\
$^5$Kavli Institute for Cosmological Physics and Enrico Fermi Institute, The University of Chicago, Chicago, IL 60637, USA \\
$^{6}$Max Planck Institute for Extraterrestrial Physics, Gissenbachstra\ss{}e, 85748 Garching, Germany \\
$^7$Heidelberger Institut f\"{u}r Theoretische Studien, Schloss-Wolfsbrunnenweg 35, 69118 Heidelberg, Germany\\
$^8$Zentrum f\"ur Astronomie der Universit\"at Heidelberg, Astronomisches Recheninstitut, M\"{o}nchhofstr. 12-14, 69120 Heidelberg, Germany \\
$^9$Institute for Computational Cosmology, Department of Physics, University of Durham, Science Laboratories, South Road, Durham DH1 3LE, UK \\
$^{10}$Department of Physics, University of Antwerp, Campus Groenenborger, Groenenborgerlaan 171, B-2020 Antwerp, Belgium \\
}

\begin{document}

\date{Submitted to MNRAS}

\pagerange{\pageref{firstpage}--\pageref{lastpage}} \pubyear{2013}

\maketitle

\label{firstpage}


\begin{abstract}
We use hydrodynamical simulations from the \owls\ project to investigate the dependence of the physical properties of galaxy populations at redshift 2 on metal-line cooling and feedback from star formation and active galactic nuclei (AGN). We find that if the sub-grid feedback from star formation is implemented kinetically, the feedback is only efficient if the initial wind velocity exceeds a critical value. This critical velocity increases with galaxy mass and also if metal-line cooling is included. This suggests that radiative losses quench the winds if their initial velocity is too low. If the feedback is efficient, then the star formation rate is inversely proportional to the amount of energy injected per unit stellar mass formed (which is proportional to the initial mass loading for a fixed wind velocity). This can be understood if the star formation is self-regulating, i.e.\ if the star formation rate (and thus the gas fraction) increase until the outflow rate balances the inflow rate. Feedback from AGN is efficient at high masses, while increasing the initial wind velocity with gas pressure or halo mass allows one to generate galaxy-wide outflows at all masses. Matching the observed galaxy mass function requires efficient feedback. In particular, the predicted faint-end slope is too steep unless we resort to highly mass loaded winds for low-mass objects. Such efficient feedback from low-mass galaxies ($M_\ast \ll 10^{10}\,{\rm M}_\odot$) also reduces the discrepancy with the observed specific star formation rates, which are higher than predicted unless the feedback transitions from highly efficient to inefficient just below $M_\ast \sim 5 \times 10^{9}$ M$_\odot$.
\end{abstract}




\section{Introduction}

Simulating the growth of dark matter (DM) haloes from initially small density perturbations through to the present day has become well established. Even the complex, non-linear stage of structure formation can be predicted by means of high-resolution gravitational $N$-body simulations. The distribution of DM haloes derived from these simulations agrees very well with observations. The formation and evolution of galaxies is, however, much less well understood. Modeling the baryonic component is much more difficult than simulating the DM due to the collisional nature of the gas and the wealth of phenomena that need to be taken into account (cooling, star formation, feedback, etc.).

There are two popular approaches to tackle this challenging task. In semi-analytic models, simple descriptions of the behaviour of the baryonic component, as a function of the DM halo mass, merging history and environment, describe the evolution of gas and stars \citep[e.g.][]{kauffmann99, somervilleprimack99, cole00}. The freedom to choose functional forms and parameter values combined with the ability to run large numbers of models, ensure that reproducing observations is usually within reach. While this approach has great advantages, such as the ability to make mock galaxy surveys that are sufficiently realistic to reveal observational biases, there are also significant drawbacks. The large number of parameters can make it difficult to identify the key physical processes. More importantly, the ability to reproduce observations with a model that uses unphysical functional forms or unrealistic parameter values to describe physical processes can easily result in erroneous conclusions and misplaced confidence.

Complimentary to the semi-analytic models are models in which no explicit functional forms are assumed for physical processes. The so-called `abundance matching' techniques link theoretically constructed dark matter halo populations to observed galaxy populations. These techniques rely on few assumptions, such as a correspondence between galaxy luminosity and dark matter halo mass \citep[e.g.][]{kravtsov04, valeostriker04}. Although such models are relatively easy to construct and give a potentially robust match between dark matter properties and galaxy observables, all baryonic processes that govern galaxy formation are hidden in the assumed correspondence, so even though the consequences of galaxy formation physics can be characterised, it can not be probed directly.

The other approach is to follow both the dark matter (DM) and the baryonic components by hydrodynamic simulation. While the DM is nearly always simulated using particles, baryons can either be modeled with Eulerian methods \citep[discretizing the volume in an (adaptive) grid, e.g.][]{ryu90, cen90} or using the Lagrangian approach also used for the DM \citep[discretizing the mass using particles, e.g.][]{evrard88, hernquistkatz89, thomascouchman92}. Here, the freedom is limited to the parametrization of unresolved sub-grid processes. The high computational expense associated with full numerical simulations and the reduced level of freedom in the sub-grid modelling mean that, thus far, numerical simulations have been much less successful in reproducing observations of galaxy populations than semi-analytic models. Compared with the semi-analytic method, the advantages of the simulation approach include the much reduced (though still present) risk of getting the right answers for the wrong reasons, the ability to ask more detailed questions due to the tremendous increase in resolution, and the fact that not only galaxies, but also the intergalactic medium (IGM) is modeled.

Following the baryonic component in direct simulation is numerically expensive. There is therefore a trade-off between simulated volume and numerical resolution. Whereas in this work we will simulate a representative cosmological volume, in order to get a large sample of galaxies, other works use a zoom-in technique to obtain high resolution in a couple of regions of interest (typically a handful of galaxies), e.g. \citet{agertz09, ceverino10, ceverino12}. Such simulations, through their higher resolution, justify the use of more detailed physical models, at the price of a much smaller sample of simulated galaxies.

As many processes related to the baryons are not (well) resolved by even the highest resolution simulations, they are dealt with in the so-called sub-grid models. Among these are radiative cooling, the temperature and pressure of the multiphase gas at high densities (in the rest of the paper loosely called ``the interstellar medium (ISM)'') and the formation of stars, the energy and momentum fed back by these stars into the ISM/intra-cluster medium (ICM)/IGM, stellar mass loss and the growth of supermassive black holes and associated feedback processes. 

In this work, we employ cosmological, hydrodynamical simulations to investigate a number of basic baryonic properties of haloes, including: the (specific) star formation rate (SFR); the stellar, gas and baryon fractions; the gas-consumption timescale and the galaxy stellar mass function. Reproducing observations in detail is not the main goal of this paper and so we have not attempted to tune our models or to optimize the sub-grid implementations to match any particular data-set. Instead, we focus on understanding \emph{how} different physical mechanisms shape the galaxy population.

We make use of a subset of the simulations from the \textit{OverWhelmingly Large Simulations} project \owls\ \citep{owls}, a large ($\sim 50$ simulations) set of cosmological, smoothed particle hydrodynamics (SPH) simulations, each of the same volume of the universe, run with a wide variety of different prescriptions for sub-grid physical processes. The large variety of input physics in the \owls\ runs allows us to investigate properties of haloes and their relation to the physical and numerical parameters. In this paper, we investigate the effect of feedback from star formation and from accreting supermassive black holes, as well as metal-line cooling. In a companion paper \citep[][hereafter Paper II]{PaperII} we investigate the effects of the assumed cosmology, the reionization history, the treatment of the unresolved, multiphase ISM, the assumed star-formation law and the stellar initial mass function. These ingredients are all present in the simulations presented here, but the parameters are not varied. A subset of these simulations have been used by \citet{sales09, sales10} to investigate the dependence of the angular momentum and sizes of galaxies on the various feedback ingredients.

This work (together with Paper II) complements that of \citet{owls}, where the cosmic star formation histories predicted by the \owls\ models were analyzed. The global SFR density can be decomposed into a DM halo mass function, which is determined by the cosmology, and the statistical distribution of the SFR as a function of halo mass. Here we will study the latter, which is astrophysically more relevant than the global SFR density as it removes the main effect of cosmology (the mass function) and allows us to investigate how the effects of the various baryonic processes vary with mass. Whilst we will add a dimension to the work of \citet{owls} by investigating the dependence on mass, we will remove another one in order to keep the scope of the study manageable. Thus, we limit ourselves to $z=2$ and to the high-resolution series presented in \citet{owls} (these runs were halted at this redshift). 

This paper is structured as follows:  In Sec.~\ref{sec:owls_sims} we describe the reference simulation used in this study (Sec.~\ref{sec:ref}), how we define and extract haloes (Sec.~\ref{sec:identification}) and how we compare the simulations to observations (Sec.~\ref{sec:comparing}).  In Sec.~\ref{sec:refprops} we describe how the properties of the galaxies in our reference simulation depend on both halo mass (Sec.~\ref{sec:halomass}) and galaxy stellar mass (Sec.~\ref{sec:stellarmass}).  In Sec~\ref{sec:physicsvars} we describe how galaxy properties depend upon the physics included in the simulation.  In this paper we consider the effects of: metal-line cooling (Sec.~\ref{sec:cooling}), constant-energy supernova (SN) winds (Sec.~\ref{sec:constwindE}), the effect of a top-heavy IMF in high-density gas (Sec.~\ref{sec:dblimf}), ``momentum-driven'' winds (Sec.~\ref{sec:wmom}), winds decoupled from the hydrodynamics (Sec.~\ref{sec:whydrodec}), thermally coupled SN-driven winds (Sec.~\ref{sec:wthermal}), and the effect of feedback from active galactic nuclei (AGN) (Sec.~\ref{sec:agn}).  Finally, in Sec.~\ref{sec:conclusions} we summarize our findings and conclude. In Appendix~\ref{sec:physprop_convergence} we study the numerical convergence of our results using simulations of different volumes and with different numerical resolution while in Appendix~\ref{sec:physprop_halo} we show that our results are insensitive to our particular choice of halo finder.


\section{Numerical techniques} \label{sec:owls_sims}
For a detailed discussion of the full set of \owls\ models we refer the reader to \citet{owls}. Here we briefly summarize the reference model. Throughout this paper we refer to this reference simulation as `\textit{REF}'. 

\subsection{The reference simulation} \label{sec:ref}
We ran our simulations in periodic boxes of 25 co-moving $h^{-1}$Mpc with $512^3$ dark matter and baryonic particles (which originally are collisional `gas' particles, but can be converted into collisionless `star' particles in the course of the simulation). We evolved the particles using an extended version of the N-Body Tree/SPH code \textsc{Gadget3} \citep[last described in][]{gadget2}. The simulation particles have masses of $8.68 \times 10^{6} h^{-1}$ \msun\ for dark matter and $1.85 \times 10^{6} h^{-1}$ \msun\ for baryons (initially; the baryonic particle masses change in the course of the simulation due to mass transfer from stars back to the gaseous phase). In Appendix~\ref{sec:physprop_convergence} we show that our results are reasonably well converged with respect to the resolution and box size of the simulation. The gravitational softening length is initially fixed at 1/25 the inter-particle spacing in co-moving coordinates (1.95 co-moving $h^{-1}$kpc), but below $z=2.91$ it is fixed at 0.5 $h^{-1}$kpc in proper units. 

\textsc{cmbfast} \citep{seljakzaldariagga96} was used to generate initial conditions, that were evolved forward in time using the \citet{zeldovich70} approximation from an initial glass-like state. The simulation is started at $z=127$. The value of the cosmological parameters are $\Omega_{\rm m} = 0.238$, $\Omega_\Lambda=0.762$, $\Omega_{\rm b} = 0.0418$, $h$ = 0.73, $\sigma_8 = 0.74$ and $n_{\rm s} = 0.951$. These values are derived from the WMAP 3-yr data and largely consistent\footnote{The only significant discrepancies are in $\sigma_8$, which is 8 per cent, or $2.3\sigma$, lower than the value favoured by the WMAP 7-year data and the Hubble parameter, which is 1$\sigma$ below the 7-yr value} with the 7-yr WMAP data \citep{Komatsu11}. Because we will study the relative differences between different subgrid models, the actual values of the cosmological parameters are not of major importance. The variation in sub-grid physics implementations is the main power of the \owls\ set of simulations. The rest of this paper deals with variations of the sub-grid models for SN and AGN feedback, and their influence on the galaxy population. Therefore, we will first describe the parameters and subgrid models used in the reference simulation.

The simulation explicitly follows the 11 elements H, He, C, N, O, Ne, Mg, Si, S, Ca and Fe. Radiative cooling and heating are calculated element-by-element in the presence of the Cosmic Microwave Background and the \citet{haardtmadau} model for the UV/X-ray background radiation from quasars and galaxies, as described in \citet{wiersma09cooling}. In these calculations, the gas is assumed to be in photo-ionization equilibrium and optically thin.

In the centers of haloes the pressure is so high that the gas is expected to be in multiple phases, with cold and dense molecular clouds embedded in a warmer, more tenuous gas. This multi-phase structure is not resolved by our simulations (and the simulations lack the physics to describe these phases), so we impose a polytropic effective equation of state for particles with densities $n_{\textrm{\scriptsize H}} > 10^{-1}$ cm$^{-3}$. These particles are also assumed to be star forming, as this is the density required to form a cold interstellar gas phase in the disk plane \citep[under the assumption of hydrostatic equilibrium in the disk, this central density corresponds to a column density in Hydrogen of $\sim 10^{21}$ cm$^{-2}$][]{schaye04}. We set the pressure of these particles to $P \propto \rho^{\gamma_{\textrm{\scriptsize eff}}}$, where $\gamma_{\textrm{\scriptsize eff}}$ is the polytropic index and $\rho$ is the physical proper mass density of the gas. In order to prevent spurious fragmentation due to a lack of numerical resolution we set $\gamma_{\textrm{\scriptsize eff}} = 4/3$, as then the ratio of the Jeans length to the SPH kernel and the Jeans mass are independent of density \citep{schayedallavecchia08}. The normalization of the polytropic equation of state is such that for atomic gas with primordial composition, the energy per unit mass corresponds to $10^4$K, namely ($P/k = 1.08 \times 10^{3}$ K cm$^{-3}$ for $n_{\textrm{\scriptsize H}} = 10^{-1}$ cm$^{-3}$). The implementation of star formation is stochastic, as described in \citet{schayedallavecchia08}, with a pressure dependent SFR, obtained from local hydrostatic equilibrium and the observed Kennicutt-Schmidt law \citep[KS-law][]{kennicutt98}. 

For the same 11 elements that we use for the cooling, we follow the production by AGB stars and by Type Ia and Type II (including Type Ib,c) SNe, as described in \citet{wiersma09chemo}. The star particles are assumed to be simple stellar populations (SSPs) with a \citet{chabrier03} initial mass function (IMF).  SN feedback is implemented kinetically.  After a short delay of 30 Myr, corresponding to the maximum lifetime of stars that end their lives as core-collapse SNe, newly formed star particles inject kinetic energy into their surroundings by kicking a fraction of their SPH neighbours in random directions. Each SPH neighbour $i$ of a newly formed star particle $j$ has a probability of $\eta m_j/\sum^{N_{\rm ngb}}_{i=1}m_i$ of receiving a kick with a velocity $v_\textrm{w}$. Our reference model uses $\eta=2$ and $v_\textrm{w}=600$\,km/s, which for our assumed IMF corresponds to $\sim40$\% of the available energy from SNe being injected as winds (where all stars more massive than 6\msun\ are assumed to explode in a SN with $10^{51}$ ergs of energy). See Table~\ref{tab:simulations} for the parameters that are varied in this paper.

\begin{table*}
\caption{Overview of the simulations and the input physics that is varied in this paper. Bold face indicates departures from the reference model. The first column gives the name of the simulation; the second column denotes the type of SN feedback, either kinetic, thermal or none. The third column shows the wind velocity for kinetic feedback models (the circular velocity is defined as $v_c = \sqrt{G M_\textrm{vir} / R_\textrm{vir}}$ and the velocity dispersion $\sigma$ is related to the gravitational potential $\Phi$: $\sigma = \sqrt{-\Phi/2}$); the fourth column indicates, for kinetic feedback models, the wind mass loading. The fifth column indicates whether the winds are decoupled from the hydrodynamics, the sixth column indicates whether metal-dependent cooling is followed in the simulation and the seventh column indicates the stellar IMF(s). The eighth column shows which simulations include AGN feedback and the last column specifies the section in which each simulation is discussed.}
\label{tab:simulations}      
\centering                          
\begin{tabular}{l l l l l l l l l}        
\hline
\hline
Name 				& Kinetic/thermal	        & $v_\textrm{wind}$	             & $\eta = \frac{\dot{M}_\textrm{wind}}{\dot{M}_*}$				& Winds 		& $Z$        & IMF		          & AGN 		& Sect. \\
				& SN feedback		        & (km/s)			     &					& decoupled?	        & Cooling    &			          & feedback?           & \\
\hline                                                                                                                                                               
\emph{REF}			& Kinetic			& 600 				     & 2 				& no			& yes        & Chabrier 	          & no 		& All \\
\emph{NOSN\_NOZCOOL}            & \textbf{None}                 & \textbf{n.a.}             & \textbf{n.a.}                          & \textbf{n.a.}                   & \textbf{no}& Chabrier                   & no          & \ref{sec:cooling} \\
\emph{NOZCOOL}                  & Kinetic                       & 600                                & 2                                & no                    & \textbf{no}& Chabrier                   & no          & \ref{sec:cooling} \\
\emph{WML4} 		        & Kinetic			& 600      			     & \textbf{4}			& no			& yes        & Chabrier 	          & no 		& \ref{sec:constwindE} \\
\emph{WML8V300} 		& Kinetic			& \textbf{300}			     & \textbf{8}			& no			& yes        & Chabrier 	          & no 		& \ref{sec:constwindE} \\
\emph{WML4V424} 		& Kinetic			& \textbf{424}			     & \textbf{4}			& no			& yes        & Chabrier 	          & no 		& \ref{sec:constwindE} \\
\emph{WML1V848} 		& Kinetic			& \textbf{848}		             & \textbf{1}			& no			& yes        & Chabrier 	          & no 		& \ref{sec:constwindE} \\
\emph{WDENS} 			& Kinetic			& $\mathbf{\sim c_s \propto \rho^{1/6}}$ & $\mathbf{\sim \rho^{-1/3}}$	& no                    & yes        & Chabrier 	          & no 		& \ref{sec:constwindE} \\
\emph{DBLIMFML14}		& Kinetic			& 600 			             & \textbf{2, 14} 		        & no			& yes        & \textbf{Chabrier, top-heavy}& no      & \ref{sec:dblimf} \\
\emph{DBLIMFV1618}		& Kinetic			& \textbf{600, 1618}	             & 2 				& no			& yes        & \textbf{Chabrier, top-heavy}& no      & \ref{sec:dblimf} \\
\emph{DBLIMFCONTSFV1618}        & Kinetic			& \textbf{600, 1618}	             & 2 				& no			& yes        & \textbf{Chabrier, top-heavy}& no      & \ref{sec:dblimf} \\
\emph{WVCIRC} 			& Kinetic		& $\mathbf{5 v_c /\sqrt2}$	             & $\mathbf{150/(\sqrt2 v_c)}$	& no		        & yes        & Chabrier 	          & no 		& \ref{sec:wmom} \\
\emph{WPOTNOKICK} 		& Kinetic			& $\mathbf{3\sigma}$	             & $\mathbf{150/(\sqrt2 v_c)}$	& no		        & yes        & Chabrier 	          & no 		& \ref{sec:wmom} \\
\emph{WPOT} 			& Kinetic			& $\mathbf{5\sigma}$	             & $\mathbf{150/(\sqrt2 v_c)}$	& no		        & yes        & Chabrier 	          & no 		& \ref{sec:wmom} \\
\emph{WHYDRODEC} 		& Kinetic			& 600			             & 2				& \textbf{yes}	        & yes        & Chabrier 	          & no 		& \ref{sec:whydrodec} \\
\emph{WTHERMAL} 		& \textbf{Thermal}	& \textbf{n.a.}			             & \textbf{n.a.}			& \textbf{n.a.}	        & yes        & Chabrier 	          & no 		& \ref{sec:wthermal} \\
\emph{AGN} 			& Kinetic			& 600				     & 2				& no			& yes        & Chabrier 	          & \textbf{yes}& \ref{sec:agn} \\
\hline
\hline
\end{tabular}
\end{table*}

\subsection{Halo identification}  \label{sec:identification}
Haloes are identified using a Friends-of-Friends (FoF) algorithm, which links together all dark matter particles which are closer to each other than the linking parameter ($b=0.2$ times the mean inter-particle distance). FoF identifies iso-overdensity contours of $\delta\equiv(\rho-\bar{\rho})/\bar{\rho} \simeq 3/(2\pi b^3)\simeq 60$ \citep{davis85, laceycole94}. Baryonic particles are linked to their nearest dark matter particle and belong to the same group, if any. 

Following the convergence tests presented in Appendix~\ref{sec:physprop_convergence}, we only include haloes that contain at least 100 star particles when considering halo properties as a function of stellar mass and we use a minimum of 2000 dark matter particles when we plot properties against halo mass. These two cuts produce nearly identical halo samples in the reference simulation and ensure that only well resolved haloes are considered. In Appendix~\ref{sec:physprop_halo} we confirm that our results are robust to changes in the definition of halo mass used. Physical properties of haloes (e.g. stellar mass and SFR) are just the sum of the properties of all constituent particles (in case of fractions like baryon fraction and specific SFR, it is the ratio of the sums of numerator and denominator).

Whenever we show the correlation between two halo properties, the plot consists of lines that connect the medians of bins evenly spaced in the quantity plotted along the horizontal axis. Each bin contains at least 30 data points. If there are fewer than 30 points in a given bin, it is extended until it includes 30 objects. The last bin may contain between 0 and 30 objects. We bin the data starting from the high-mass end. There, the difference in mass for two consecutive haloes is much larger than at the low-mass end, and in this way we are sure that the value of the mass at the high-mass end of the plots is always the mean of the mass of the 15th and 16th most massive systems. 

\subsection{Comparing simulations to observations} \label{sec:comparing} 
In this section we describe how we compare our simulated galaxy population to observed stellar masses and SFRs. To convert observationally inferred stellar masses and SFRs from the cosmology assumed in the literature to our cosmology, we multiply them by the square of the ratio of luminosity distances, $[d_{L, \textrm{\scriptsize our cosm}} (z) / d_{L, \textrm{\scriptsize obs cosm}} (z)]^2$.  The subscripts `our cosm' and `obs cosm' denote our cosmology and the cosmology under which the observations are transformed into masses/SFRs, respectively. Using the observational data sets compared to in this paper, this ratio varies from very close to 1 to $\sim 2.6$.

Note that we are using FoF halos, so all satellites are added to the central galaxy. We show in Appendix~\ref{sec:physprop_halo} that it makes very little difference to treat the satellites separately.

We compare our simulated SFRs to those observed by \citet{daddi07}, who measured obscured and unobscured star formation by taking SFRs from both the UV and IR for $K$-selected $sBzK$ galaxies \citep[star forming, see ][]{daddi04} in the GOODS fields at $z \sim 2$. The median of the observed SFR as a function of stellar mass is well fit by $\textrm{SFR} = 250 \times (M_*/10^{11} \,$\msun$)^{0.9}$.  The IMF assumed in the observations is the \citet{salpeter55} IMF, whereas our stellar masses and SFRs are based on the \citet{chabrier03} IMF. We therefore divide the observationally inferred SFRs by a factor 1.65, which is the asymptotic (reached after only $10^8$ yr) ratio of the number of ionizing photons per unit stellar mass predicted by \citet{bruzualcharlot03} for a constant SFR.
 
For stellar masses, the IMF conversion factor is more sensitive to the age of the population and the observed rest-frame wavelength. As the light in most wavelength bands is dominated by massive stars and the high-mass ends of both the Salpeter and Chabrier IMFs are power laws with very similar power law indices, we use the same factor of 1.65 as we used for the SFRs. For very old populations observed in red wavelength bands (tracing stellar continua rather than dust emission) the conversion factor should be different. We verified that the K-band mass-to-light ratio is about a factor 1.65 smaller for a Chabrier than for a Salpeter IMF for SSPs and constantly star forming populations, for the full range of ages and metallicities available in the \citet{bruzualcharlot03} population synthesis package. We therefore also divide by a factor of 1.65 to convert stellar masses from the Salpeter to the Chabrier IMF.

We compare to the galaxy stellar mass function of \citet{marchesini09}, a combined sample, using the deep near-infrared Multi-wavelength Survey by Yale-Chile, the Faint Infrared Extragalactic Survey and the Great Observatories Origins Deep Survey-Chandra Deep Field South surveys at $z\sim2$. Specifically, we compare to the $1/V_{{\rm max}}$ results of \citet{marchesini09}, including all of their uncertainties, with the exception of bottom-light IMFs. The reason for this choice is that models with bottom-light IMFs dominate the systematic errors and represent a more extreme assumption than the variations in the other quantities. Additionally, theoretical justifications for using bottom-light IMFs exist, thus far, only at high redshift \citep{dave08, vandokkum08, wilkins08a}. We obtain the error bars on the observed data points by considering all of the sources of random errors included in the observational data points, which include: Poisson errors on the number counts, cosmic variance and the random errors from the use of photometric redshifts. We add the sources of random error in quadrature and, in addition, linearly add the maximum of the systematic errors in the same mass bins, just as \citet{marchesini09}. The systematic errors include the systematic component in the errors from photometric redshifts, errors arising from different population synthesis packages \citep[\citet{marchesini09} tested][Charlot \& Bruzual, in prep.]{bruzualcharlot03, maraston05}, varying the metallicities of the stellar populations, and the use of different extinction curves (Milky Way from \citealt{allen76}, SMC from \citealt{prevot84}, \citealt{bouchet85} and \citealt{calzetti00}). 

Because our redshift of interest ($z=2$) corresponds to the boundary between two of the redshift bins of \citet{marchesini09}, which are $1.3<z<2$ and $2<z<3$, we weigh the averaging to the sizes of the redshift intervals (weights 1.2\,Gyr and 0.8\,Gyr respectively), which gives results consistent with the $z=2$ results of the Newfirm Medium-Band Survey \citep{marchesini10}. Additionally, the observed mass bins are not exactly the same size in both redshift intervals, although the differences are very small. Observed mass bins are of size 0.3 dex (at $1.3<z<2$) and 0.29 dex (at $2<z<3$).  We interpolate the observed mass bins to a constant size of 0.3 dex. The resulting $1/V_{{\rm max}}$ estimate of the analysis of \citet{marchesini09} is shown as the yellow shaded regions in the bottom-right panels of Fig.~\ref{fig:All_sims}, and Figs.~\ref{fig:sims_cooling} -- \ref{fig:agn}, labelled (I). 

\citet{marchesini09} use a diet Kroupa IMF, for which the correction factor to our Chabrier IMF is very small ($\sim 1.03$, diet Kroupa being slightly more massive for the same flux). As this number is also derived from population synthesis packages, which come along with their own uncertainties, we chose not to convert masses for the small difference in IMFs. We do correct the masses for the difference in luminosity distances as described earlier. Number densities also need to be converted, as the volume at a given redshift is different for different angular diameter and co-moving distances. Therefore, the number density ($\phi_*$) is corrected for the ratio of volume elements (which is a function of the assumed cosmology). 

When we compare our simulated galaxies to observations of the molecular gas mass in galaxies, we use a sub-set of the compilation used in \citet{genzel10}. As their total range of redshifts is very large, we chose to use only the two sub-sets directly above and below $z=2$. The near-IR long-slit H$\alpha$ sample from \citet{erb06} contains 11 galaxies with a mean redshift of 2.3 and is originally drawn from the BX selected sample of \citet{steidel04} and \citet{reddy05}. The 4 galaxies with a mean redshift of 1.5 that we include are star-forming BzK galaxies from \citet{daddi08, daddi10}. For both samples the molecular gas masses come from CO measurements with the Plateau de Bure interferometer and the stellar masses come from optical/UV SED fitting, under the assumption of a Chabrier IMF. We therefore only have to correct for the difference in luminosity distance, as explained above. Our simulations do not explicitly calculate the molecular fraction in the high-density gas. Nevertheless, the galaxies we compare to are at the very highest stellar mass end of our simulated galaxies. For those galaxies the molecular gas fraction (by mass) is thought to be very high due to high pressures in the ISM, so we assume that all gas on the imposed equation of state is molecular. A recent study by \citet{narayanan12} suggests that with an improved conversion factor from CO to H$_2$, gas fractions of these galaxies are likely somewhat lower than quoted by \citet{genzel10}. 


\section{Physical properties of simulated galaxies as a function of dark matter halo mass}\label{sec:refprops}

\begin{figure*}
\centering
\includegraphics[width=0.33\linewidth]{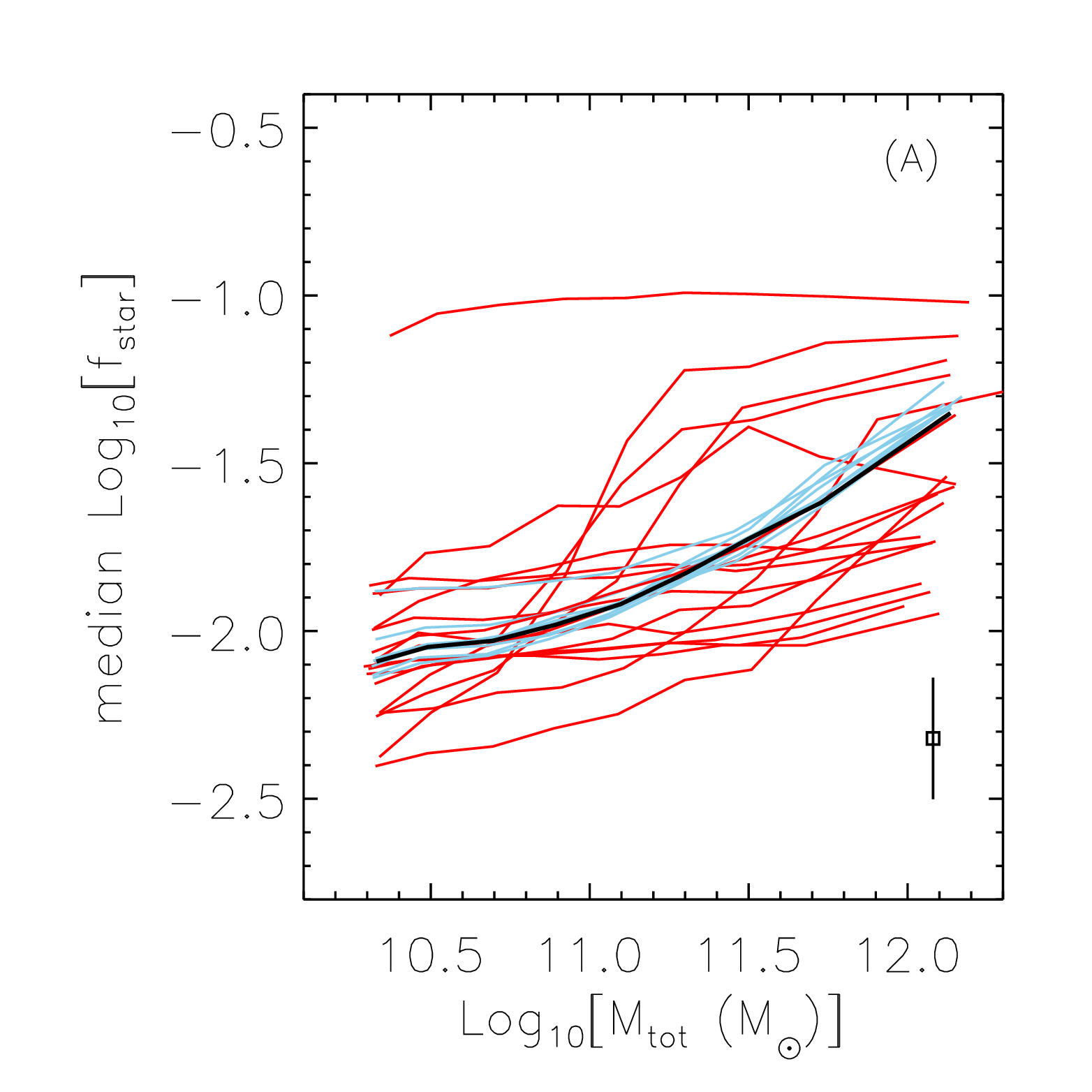}
\includegraphics[width=0.33\linewidth]{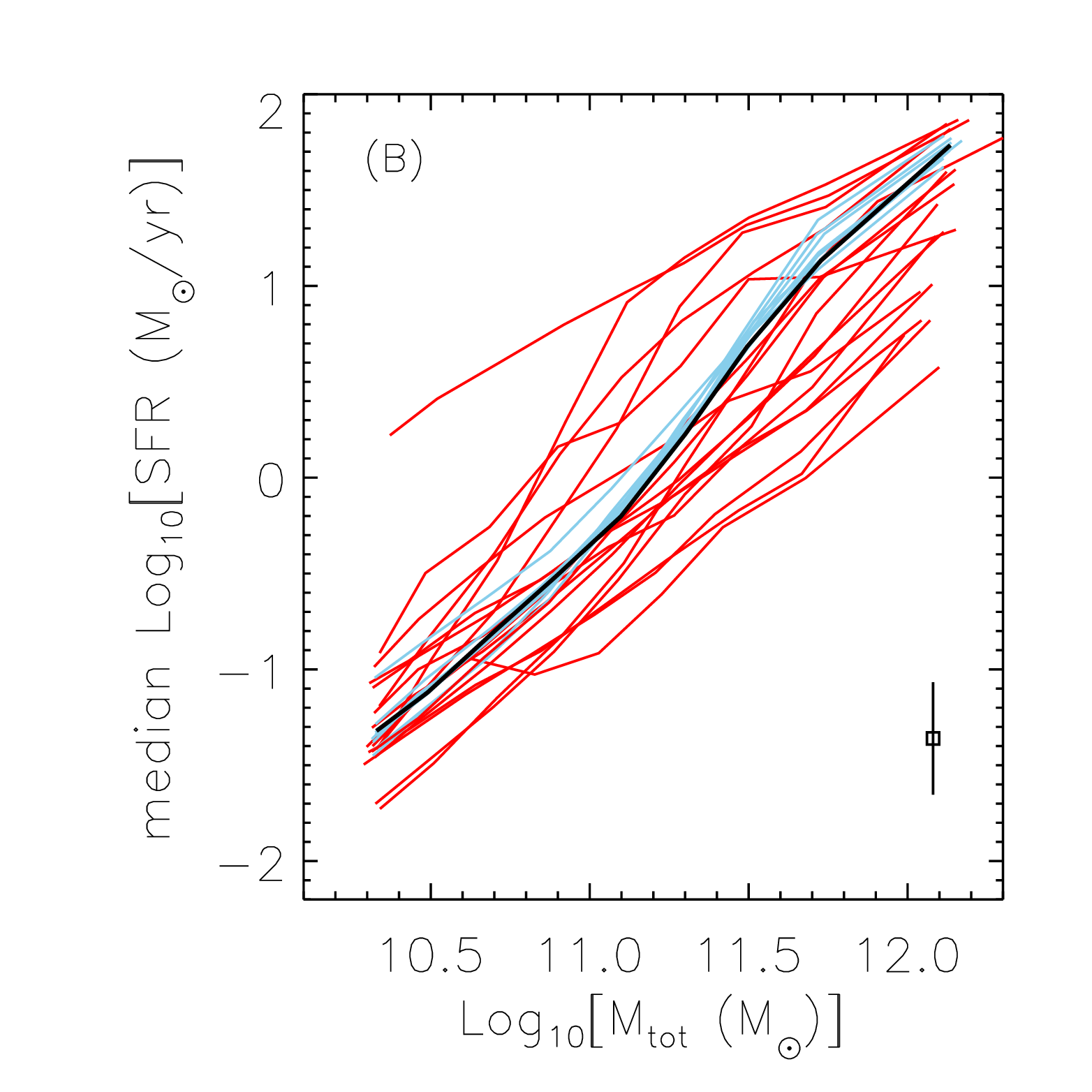}
\includegraphics[width=0.33\linewidth]{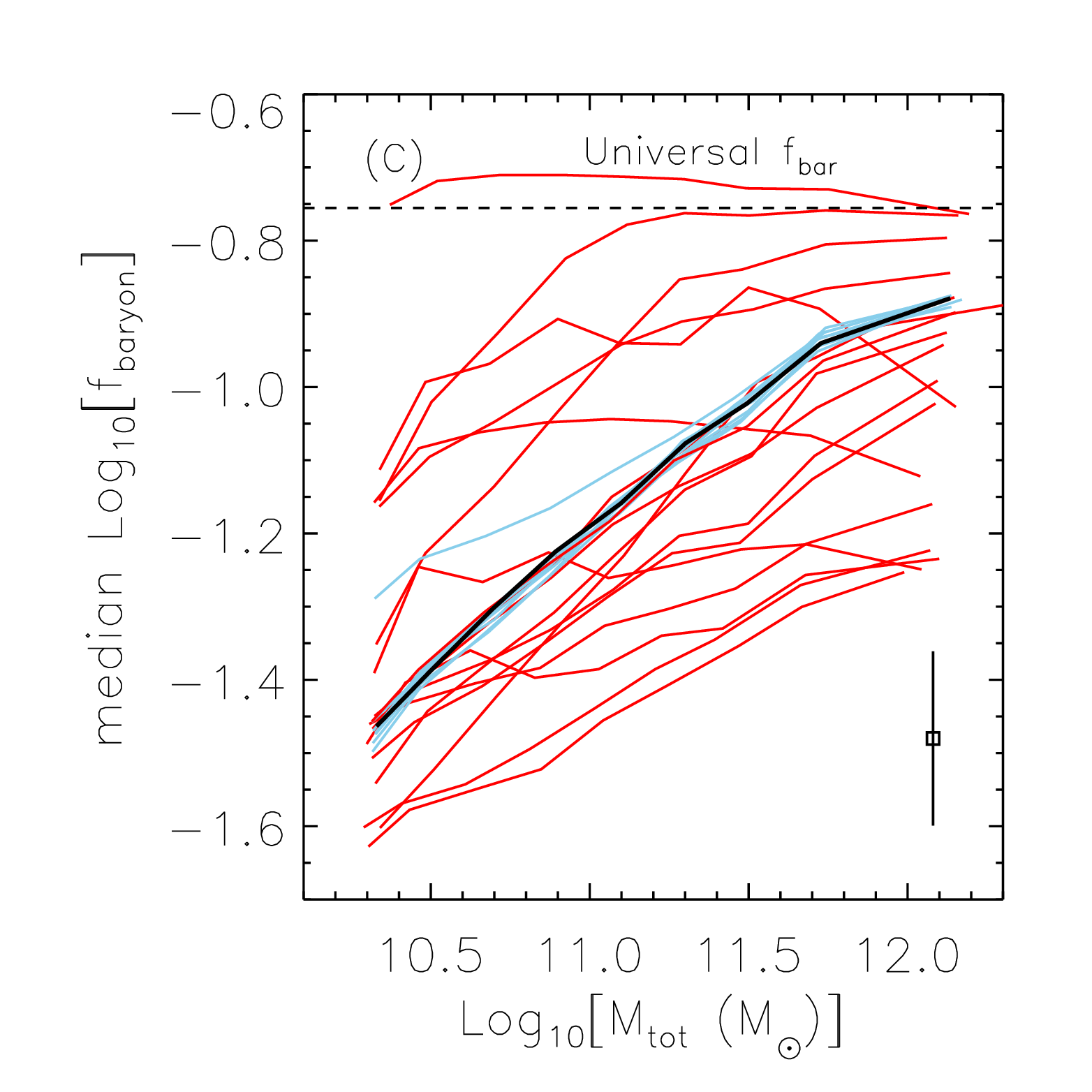} \\
\includegraphics[width=0.33\linewidth]{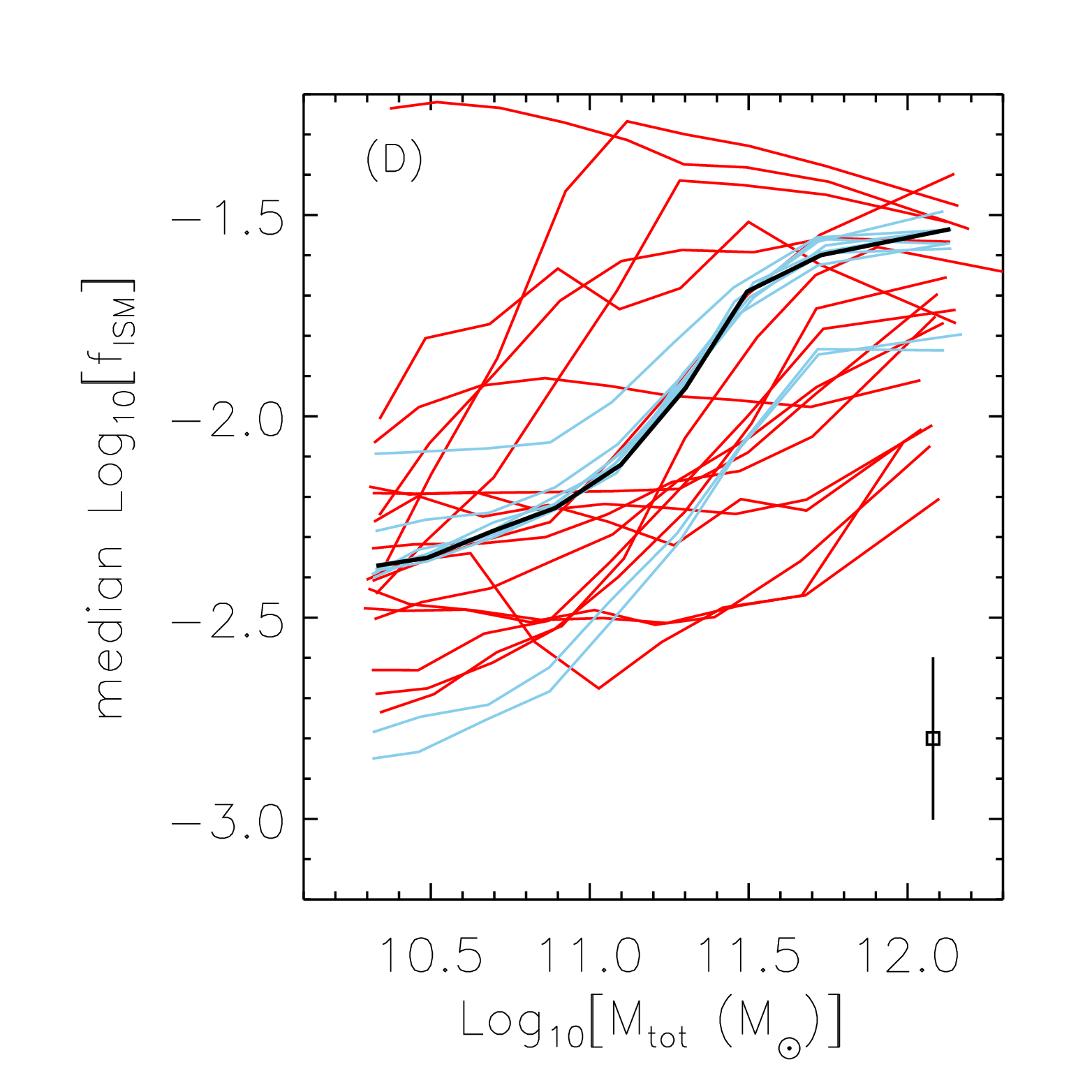}
\includegraphics[width=0.33\linewidth]{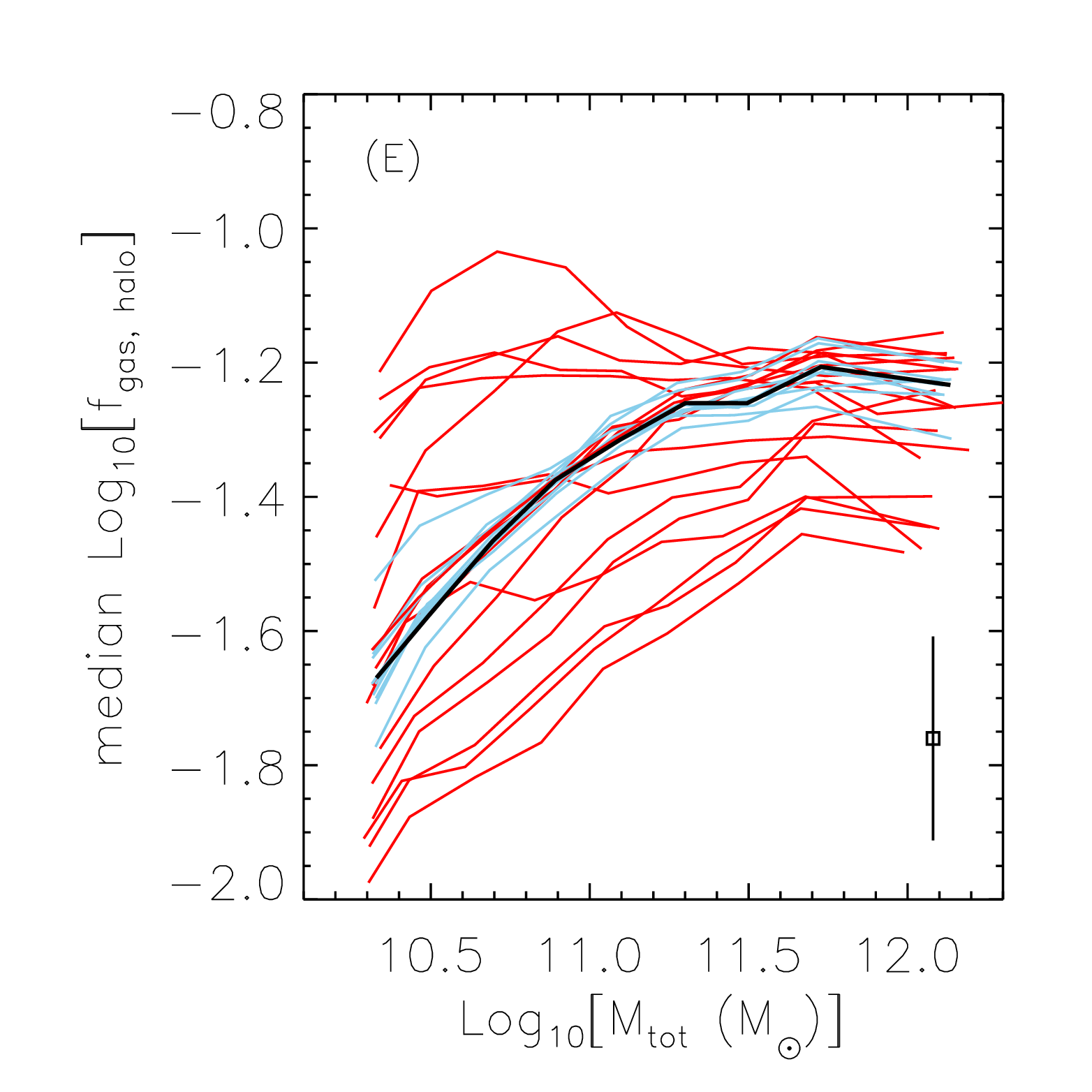}
\includegraphics[width=0.33\linewidth]{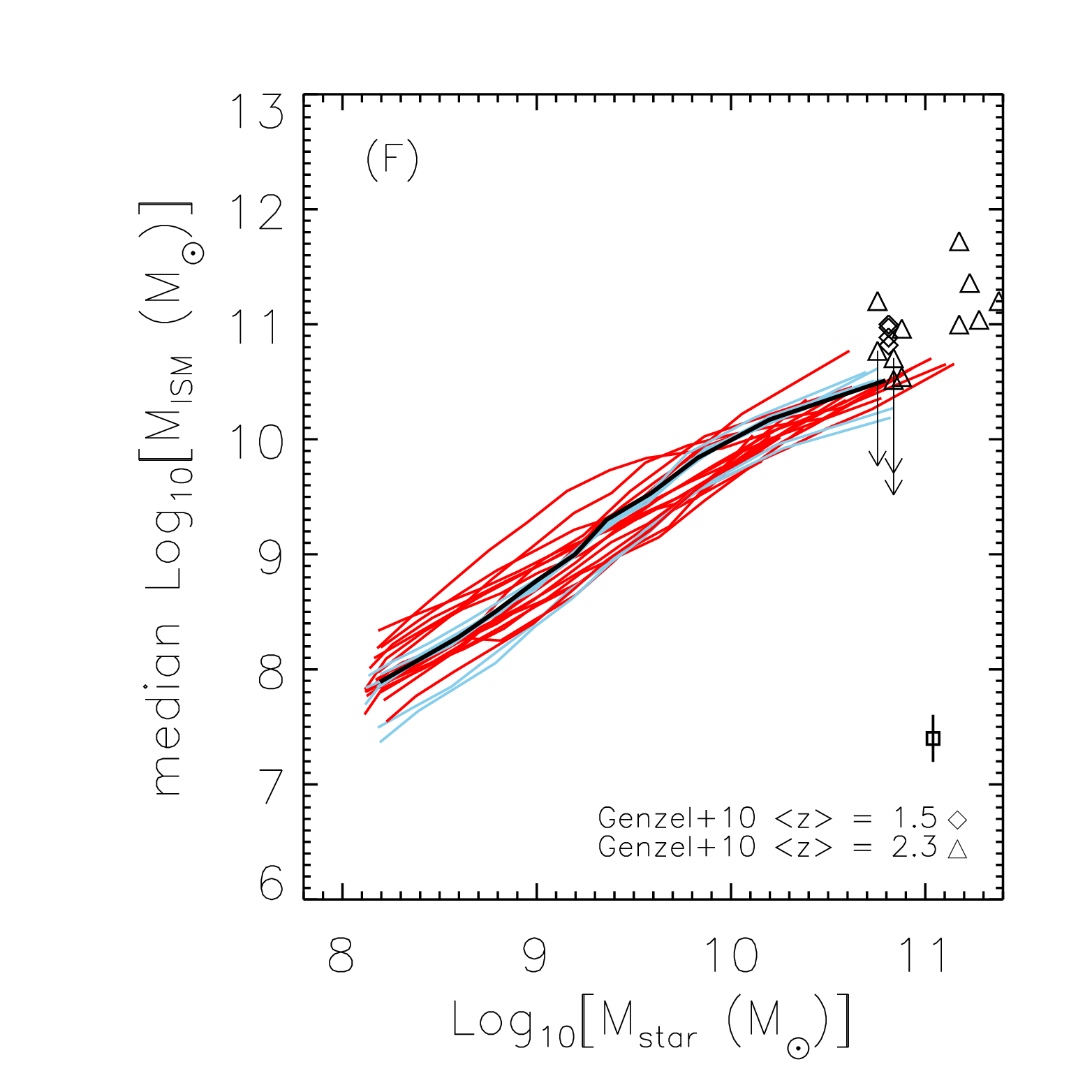} \\
\includegraphics[width=0.33\linewidth]{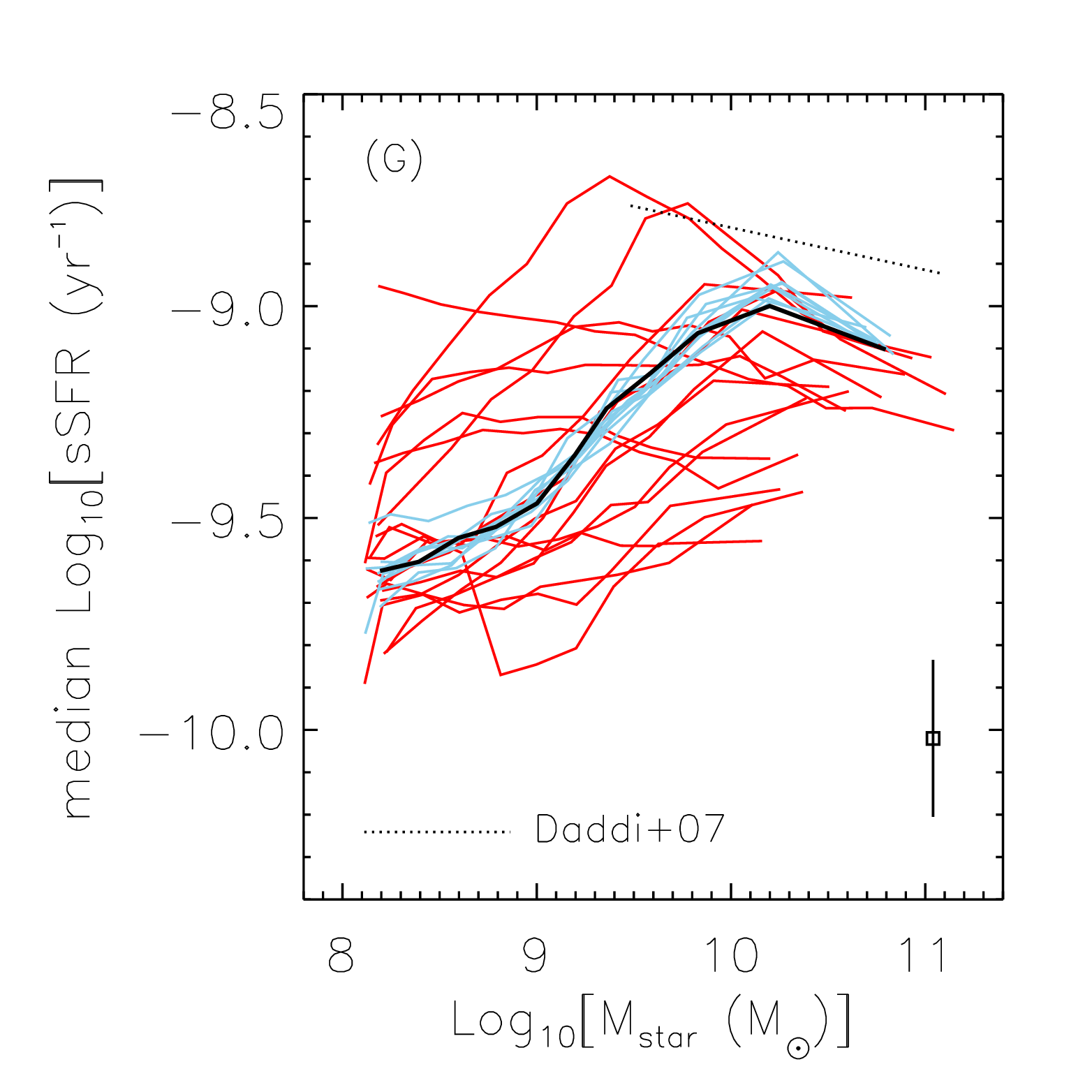}
\includegraphics[width=0.33\linewidth]{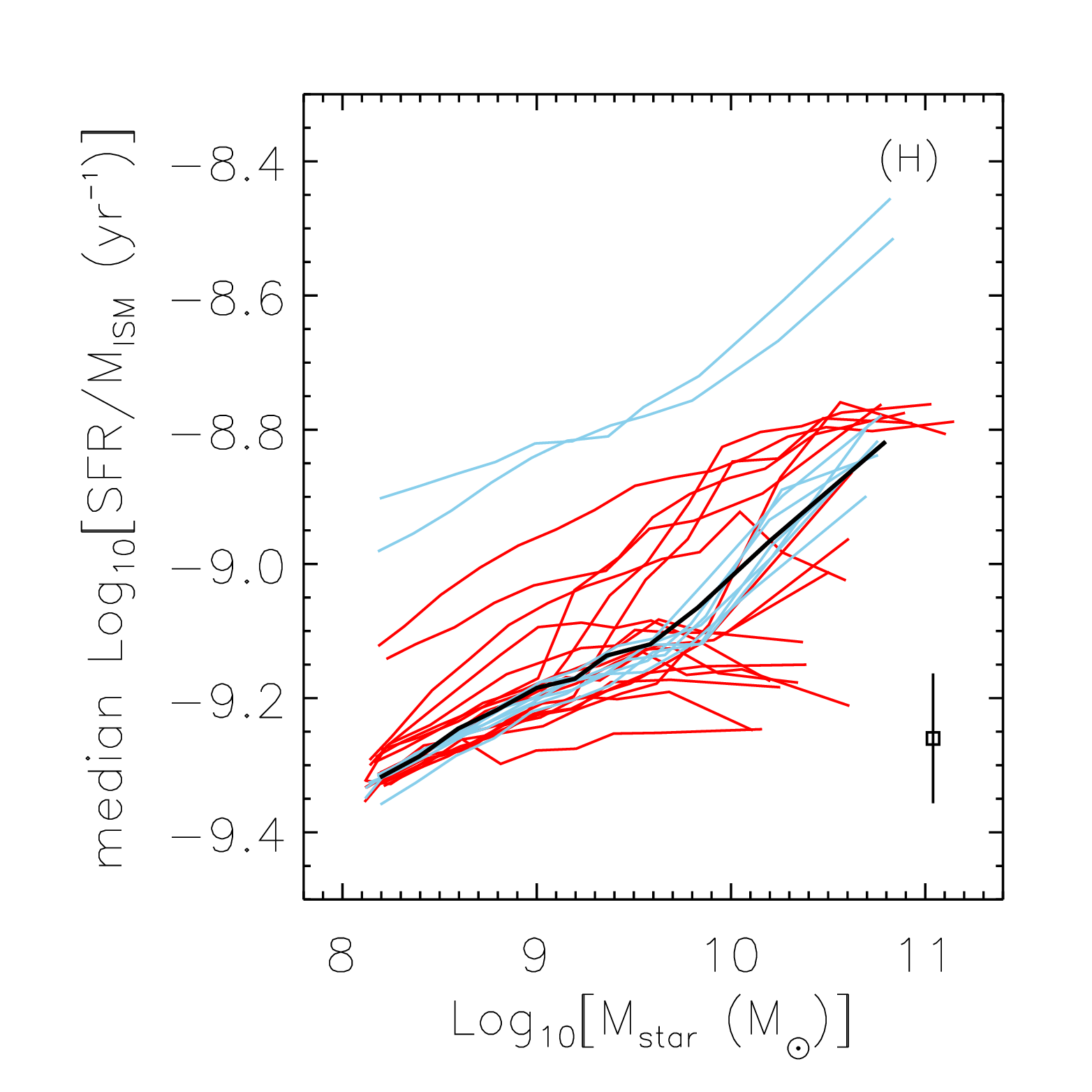}
\includegraphics[width=0.33\linewidth]{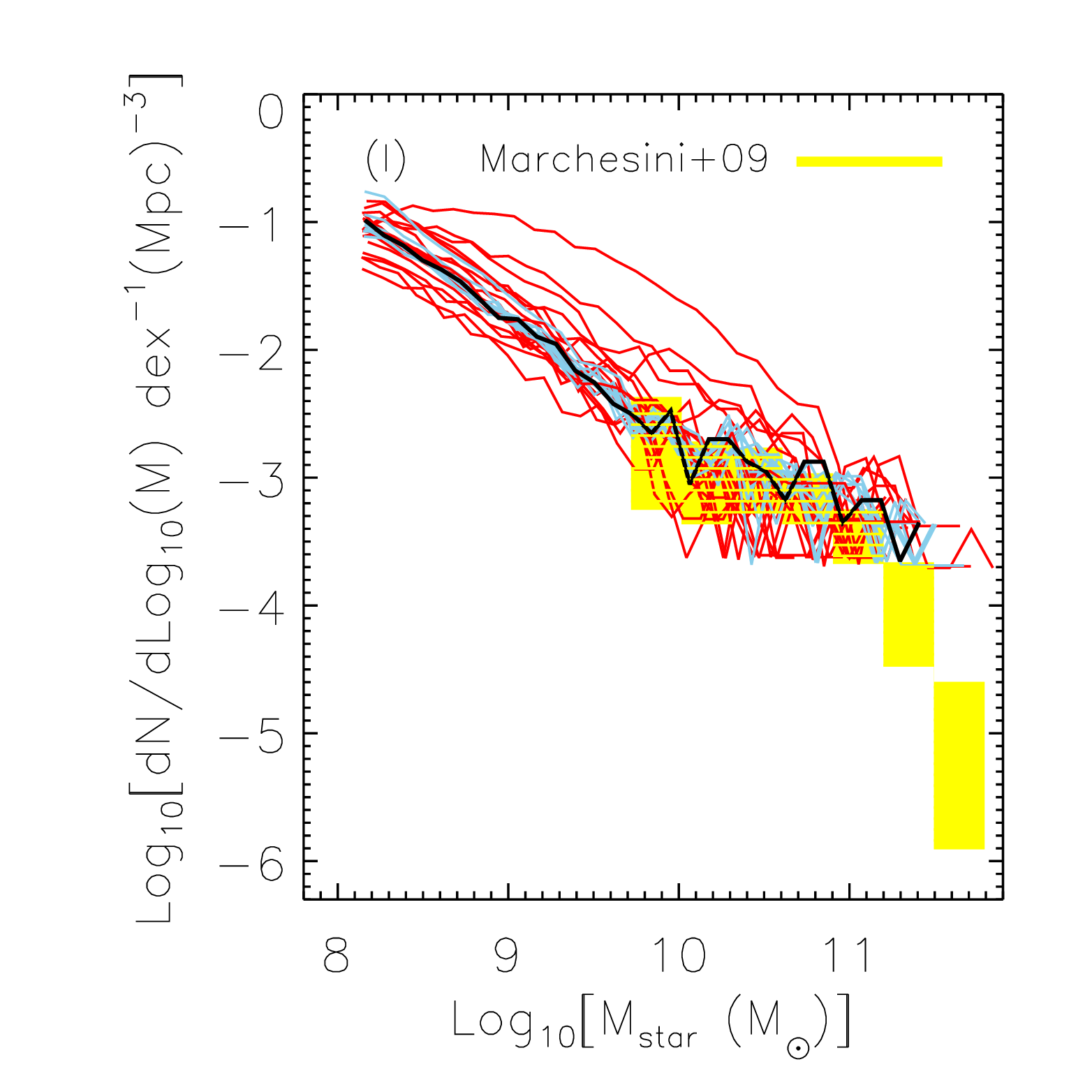}
\caption{Median relations between halo properties in all the simulations used in this work (red lines) and in Paper II (blue lines). The reference model is shown as a black curve in each panel.  In Sec.~4 we consider subsets of simulations in more detail. In the first five panels we show, as function of total halo mass, medians of stellar mass fraction ($f_\textrm{star} = M_\textrm{star} / M_\textrm{tot}$, panel A), SFR (panel B), baryon fraction ($f_\textrm{baryon} = M_\textrm{baryon} / M_\textrm{tot}$, panel C), fraction of star-forming gas ($f_\textrm{ISM} = M_\textrm{ISM} / M_\textrm{tot}$, panel D) and gas  fraction ($f_\textrm{gas, halo} = (M_\textrm{gas, total} - M_\textrm{ISM}) / M_\textrm{tot}$, panel E). The next four panels show, as a function of stellar mass, medians of the molecular gas mass in the ISM (panel F), specific SFR (sSFR = SFR/$M_{*}$, panel G), the inverse of the gas consumption timescale (SFR/$M_\textrm{ISM}$, panel H) and galaxy number density (the galaxy stellar mass function, panel I). As described in the text, we show medians in bins along the horizontal axes for all haloes that satisfy the convergence criteria that apply to that specific panel.  The horizontal, dashed line in panel (C) shows the universal baryon fraction for our chosen cosmology, the data points in panel (F) show a sub-set of the compilation studied in Genzel et al. (2010), the dotted black line in panel (G) shows the stellar mass - sSFR relation from the GOODS field (Daddi et al. 2007) and the shaded yellow region in panel (I) shows the galaxy stellar mass function of Marchesini et al. (2009). The error bars in the lower right corners show the mass-weighted mean scatter of galaxies in the reference simulation, as explained in the text.} 

\label{fig:All_sims} 
\end{figure*}

Fig.~\ref{fig:All_sims} shows, as a function of total halo mass the nine different galaxy properties we consider in this paper: medians of stellar mass fraction ($f_\textrm{star} = M_\textrm{star} / M_\textrm{tot}$, panel A), SFR (panel B), baryon fraction ($f_\textrm{baryon} = M_\textrm{baryon} / M_\textrm{tot}$, panel C), fraction of star-forming gas ($f_\textrm{ISM} = M_\textrm{ISM} / M_\textrm{tot}$, panel D) and gas mass fraction ($f_\textrm{gas, halo} = (M_\textrm{gas, total} - M_\textrm{ISM}) / M_\textrm{tot}$, panel E). Then, as a function of stellar mass:  medians of the molecular gas mass in the ISM (panel F), specific SFR (sSFR = SFR/$M_{*}$, panel G), the inverse of the gas consumption timescale (SFR/$M_\textrm{ISM}$, panel H) and galaxy number density (the galaxy stellar mass function, panel I).

In each panel of Fig.~\ref{fig:All_sims}  we show the properties of the galaxies in the reference simulation with a black curve.  The blue curves show results from the simulations analyzed in Paper II, which include changes in the cosmology, ISM physics, star-formation law and IMF.  The red curves show the simulations discussed in this paper (see section~\ref{sec:physicsvars}), which comprise changes in the feedback processes that are modeled. The red and blue lines are included in Fig.~\ref{fig:All_sims} to give a visual impression of the magnitude of the effects that we are considering. Here, we note that the magnitude of the differences between simulated galaxy properties due to uncertainties in the sub-grid physics are larger than the differences that arise from changing the hydrodynamics method from SPH to a (moving-) mesh code \citep{springel10arepo}, as e.g. shown by \citet{vogelsberger12, keres12} and \citet{scannapieco12}. The relative accuracy of different numerical techniques can nevertheless be important, but an investigation of this issue falls outside the scope of this paper.

The error bar in the lower right corner of each panel (except the stellar mass function in panel H) indicates the typical scatter of galaxies around the median in the reference simulation. The error bar is a mass-weighted (by halo mass in panels A-E and by stellar mass in panels F-H) mean $\sigma_\textrm{tot}$ of the standard deviation of all galaxies in all bins $\sigma_\textrm{bin}$: $\sigma_\textrm{tot} = \sum_\textrm{bins} \sigma_\textrm{bin} M_\textrm{bin} / \sum_\textrm{bins} M_\textrm{bin}$. We note that in all panels this scatter is dominated by the lower mass bins, where the scatter is typically much larger (and the bins are smaller and thus more numerous), and possibly partly due to resolution effects. Also, at the high mass end the `scatter' in the data is partially caused by the (almost exclusively positively sloped) relation between the dependent variable and mass. The scatter of galaxies around the median is typically smaller than the difference between the median relations in the most extreme simulations. A difference between two simulations that is smaller than the scatter is nevertheless still meaningful.

In the remainder of this section we describe and explain the trends seen in the properties of the galaxies in the reference simulation, and comment briefly on the effects of the physics variations.  Note that the first five panels (galaxy properties as a function of DM halo mass) and the last four panels (galaxy properties as a function of stellar mass) are resolved down to slightly different resolution limits, as discussed in Appendix~\ref{sec:physprop_convergence}.

\subsection{Properties as a function of halo mass}\label{sec:halomass}
In panel (A) of Fig.~\ref{fig:All_sims} we plot the stellar mass fraction as a function of total halo mass.  Stellar mass and total mass are tightly (and almost linearly) correlated so differences between the models are emphasized by plotting the ratio of the masses.  In the reference model, the fraction of the total mass locked into stars increases smoothly as a function of halo mass, from $\sim$1\% in the smallest resolved haloes ($\log_{10}(M_{\rm tot}/$\msun$)\sim 10.5$), to 4\% in the highest mass haloes.  It is difficult to make a quantitative comparison with galaxy formation efficiencies quoted in much of the literature \citep[e.g.][]{guo10}, as the definition of halo mass used in this study is different from the spherical overdensity halo definitions used in these works, but the qualitative trend is the same. As we move towards higher halo masses, a fractionally larger proportion of the mass is locked into stars.  It is clear from inspection of panel (A) that the changes in physics considered in this paper (red curves) have  a large effect on the $M_{\rm halo}-M_*$ relation (much bigger than the simulations discussed in Paper II; blue curves).  We discuss the reasons for this sensitivity in Sec.~\ref{sec:physicsvars}.  The one simulation that is a strong outlier, with much higher stellar mass at fixed halo mass, is the simulation that neglects SN feedback and, as such, contains no mechanism to prevent the catastrophic cooling of gas into stars. Most observational studies do not go out as far as $z=2$ and rely on extrapolations beyond the observed galaxy population, especially at the low mass end. \citet{yang12} study the stellar mass -- halo mass relation around $z=2$, but their lowest observed point is at $M_\textrm{halo} \sim 10^{12}$\msun, which is our very highest mass bin. At that mass, the reference simulation slightly overpredicts the mass in stars (a factor of $\sim 2$), as expected from the inefficient SN feedback that will be discussed at length below. Some simulations with very efficient feedback underproduce stars at $M_\textrm{halo} \sim 10^{12}$\msun, as compared to these subhalo abundance matching techniques. 

In panel (B) we show the integrated SFR inside haloes as a function of their total mass.  In all simulations the SFR is an increasing function of halo mass in the mass range probed here ($10.5<\log_{10}(M_{\rm tot}/$\msun$)<12.0$). As in panel (A), the vast majority of the physics variations considered in Paper II have very little effect, but the variations discussed here are very important. This implies that the sub-grid physics employed to model the ISM, other than that related to feedback, does not affect the large-scale, stellar properties of the galaxies.  The outlier among the red curves in this panel is again the simulation that neglects SN feedback, which shows that feedback is an important component of the sub-grid models.

Panel (C) shows the baryon fractions of the haloes as a function of halo mass. The universal baryon fraction appropriate for our default cosmology ($\Omega_{\textrm b}/\Omega_{\textrm m}=0.18$) is over-plotted as a horizontal, dashed line.  In the default simulation, feedback processes are able to eject baryons from the halo very easily in low-mass haloes, leading to low baryon fractions (0.6 dex below universal) for $M_\textrm{tot} \sim 10^{10.5}$\msun, whereas they become gradually less efficient as the halo mass increases. Again, the physics changes presented in this paper have a much more significant effect on this relation than those discussed in Paper II.  Panels (D) and (E) show the fraction of the mass that is in gas in the ISM (which \emph{is} sensitive to the physics variations in Paper II) and the rest of the gas in the haloes, respectively. In general, both are increasing functions of the total mass. 

\subsection{Properties as a function of stellar mass}\label{sec:stellarmass}
The mass in the ISM as a function of the galaxy stellar mass is slightly sub-linear, as depicted in panel (F). This reflects the fact that at higher masses, gas is more efficiently transformed into stars. Our simulations slightly under-predict the molecular gas mass as a function of stellar mass when compared to the sub-sample of the galaxies studied by \citet{genzel10}.

Both the stellar mass (panel A) and SFR (panel B) of galaxies are correlated almost linearly with halo mass, suggesting, that there is a linear relation between the stellar mass and the SFR.  This linear trend masks important differences between the simulations, so to examine galaxy properties in more detail, we use the galaxy specific SFR (sSFR$=$SFR$/M_*$).

Panel (G) shows the median sSFR as a function of galaxy stellar mass.  In this panel, the black, dotted line shows the observations of \citet{daddi07}.  The scatter in the data is not shown, but is constant at approximately 0.2 dex.  The scatter in the simulation data points is similar, although somewhat smaller ($\sim 0.1$ to $0.15$ dex).  We plot the observed relation over the mass range that is observed, and do not extrapolate beyond there.

The trend seen in the simulations is such that at low masses the sSFR is an increasing function of mass, but at $\sim 10^{10}$ \msun\, the trend reverses and the sSFR decreases with increasing stellar mass. At the highest masses the relation is declining more steeply than observed.  As discussed in Sec.~\ref{sec:halomass}, most of the physics variations discussed in this paper have little effect on the integrated stellar properties of the haloes. The underestimate of the SFR is likely connected to the underestimate of molecular mass shown in panel (F).

Because SFRs are expected to be more strongly influenced by the mass of available star-forming gas than by the amount of stars already formed, we define a second normalized SFR, the ratio of the galaxy SFR to its star-forming gas mass. This is the inverse of the time needed to convert the present reservoir of star forming gas into stars, assuming that the present SFR is maintained. We plot the inverse of the gas consumption time scale as a function of stellar mass in panel (H). The gas consumption time is a monotonically decreasing function of stellar mass in the simulations. In the cores of the more massive haloes, gas reaches higher densities and pressure, so gas is converted more efficiently into stars. 

In panel (I) we examine the galaxy stellar mass function: the number density of galaxies as a function of their stellar mass.  The yellow, shaded regions show the data from \citet{marchesini09}.  When the observational uncertainties are taken into account, most of the simulated mass functions fall well within the observed range. We note that, at low masses, our simulated stellar mass function is steeper than most faint-end slopes of derived Schechter function parametrizations at similar redshifts \citep[e.g.][]{conselice05, marchesini09, marchesini10}, but that this discrepancy exists largely outside the range of stellar masses where the mass function has been observed. Newer results are complete down to only slightly lower masses \citep[e.g.][who are complete down to $10^{9.5}$\msun]{mortlock11}. Our simulation box size is too small to form the rare objects that populate the exponential cutoff of the stellar mass function.

Comparing the behaviour of the reference simulation between panels (G) and (I) reveals an interesting behaviour: although the galaxy sSFR is a factor of a few too low at $z=2$ over the observed range, the simulations form enough galaxies of all masses, and possibly too many low-mass systems \citep[a similar behaviour was also noted by][]{choinagamine12}. This may indicate that the observations are not internally consistent, such that the integral of observed SFRs do not add up to the observed current stellar mass after correcting for stellar mass loss. A perhaps more likely explanation is that galaxy formation in the simulation is too efficient in low mass systems at high redshift and that other physics variations are required to flatten the stellar mass function and to get the SFRs at $z=2$ to agree with observed values.


\section{Isolating the effects of the input physics}\label{sec:physicsvars}

\begin{figure*}
  \centering
   \resizebox{0.9\hsize}{!}{\includegraphics{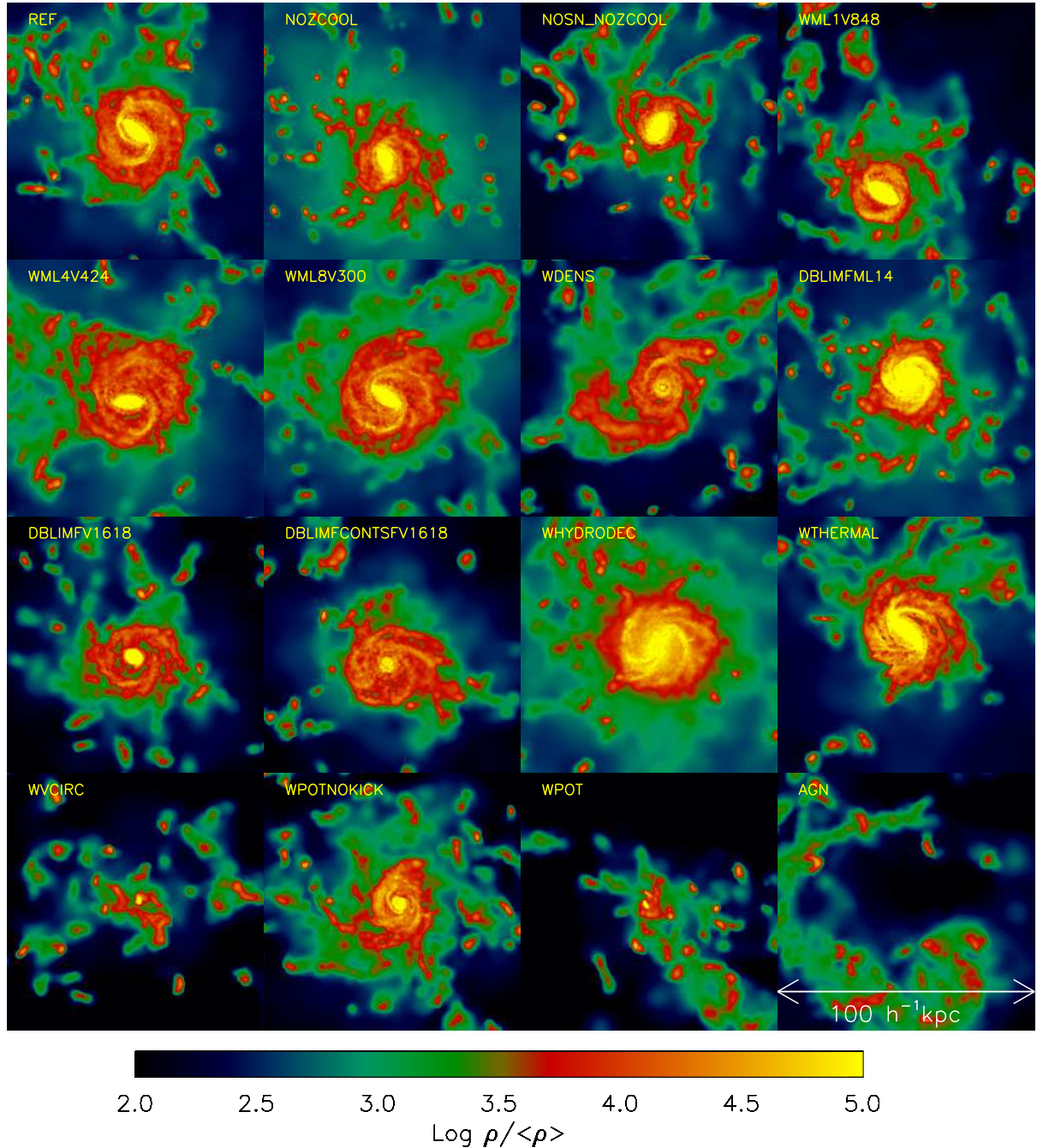}}
    \caption{A graphical representation of a galaxy in a halo of $10^{12.5}$ \msun\ in 15 of our simulations at redshift 2. The colour coding denotes the gas density in a slice of 100 $h^{-1}$ kpc thickness, divided by the mean density of the universe. All frames are 100 co-moving $h^{-1}$kpc on a side and are centered on the position of the galaxy in the `\textit{REF}' simulation. All frames have a thickness of 100 co-moving $h^{-1}$kpc. The orientation of the line of sight is along the z-axis, which is almost perfectly aligned with the angular momentum vector of all material inside 10\% of the virial radius of this galaxy in the `\textit{REF}' simulation.  It is clear that different feedback prescriptions can have a large effect on the size of the galactic disk. Note that `\textit{WML4}' is not included here. That simulation is also discussed in Paper II, and it is included there.}
   \label{fig:prettypics}
\end{figure*}

In this section we discuss each of the variations of the input physics. Table~\ref{tab:simulations} summarizes the simulations discussed in this paper and indicates in which subsection each one is discussed. Bold face values indicate differences between each simulation and `\textit{REF}'. In this paper we discuss how metal-line cooling and the inclusion of kinetic SN feedback affect galaxy properties (Sec.~\ref{sec:cooling}); how the parameters of the kinetic SN feedback model affect the results (Sec.~\ref{sec:constwindE}); the effect of a top-heavy IMF at high pressures (Sec.~\ref{sec:dblimf}); winds with \lq momentum-driven\rq\, scalings that depends on galaxy properties (Sec.~\ref{sec:wmom}); the effects of decoupling the winds from the hydrodynamics (Sec.~\ref{sec:whydrodec}); the effect of injecting SN energy thermally (Sec.~\ref{sec:wthermal}); and the effects of strong, AGN feedback (Sec.~\ref{sec:agn}).

A graphical representation of the gas density of a galaxy formed in a representative set of models is shown in Fig.~\ref{fig:prettypics}. The galaxy resides in a halo of total mass $\sim 10^{12.5}$ \msun. It was first identified in the `\textit{REF}' simulation, where its position (defined as the centre of mass of all particles within 10\% of the virial radius) is determined. The line of sight is along the z-axis, which is almost perfectly aligned with the angular momentum vector of the gas within 10\% of the virial radius ($\cos(\phi) = 0.994$). For the other simulations the image is centered on the same position, showing the remarkable similarity in the positions and orientations of the galaxies.  It is immediately clear that changing the feedback model can have a large effect on the galaxy morphology.  Going so far as almost completely destroying the galaxy disk in the strongest cases (`\textit{AGN}' and `\textit{DBLIMFV1618}').

\subsection{Metal-line cooling} \label{sec:cooling}

\begin{figure*}
\centering
\includegraphics[width=0.33\linewidth]{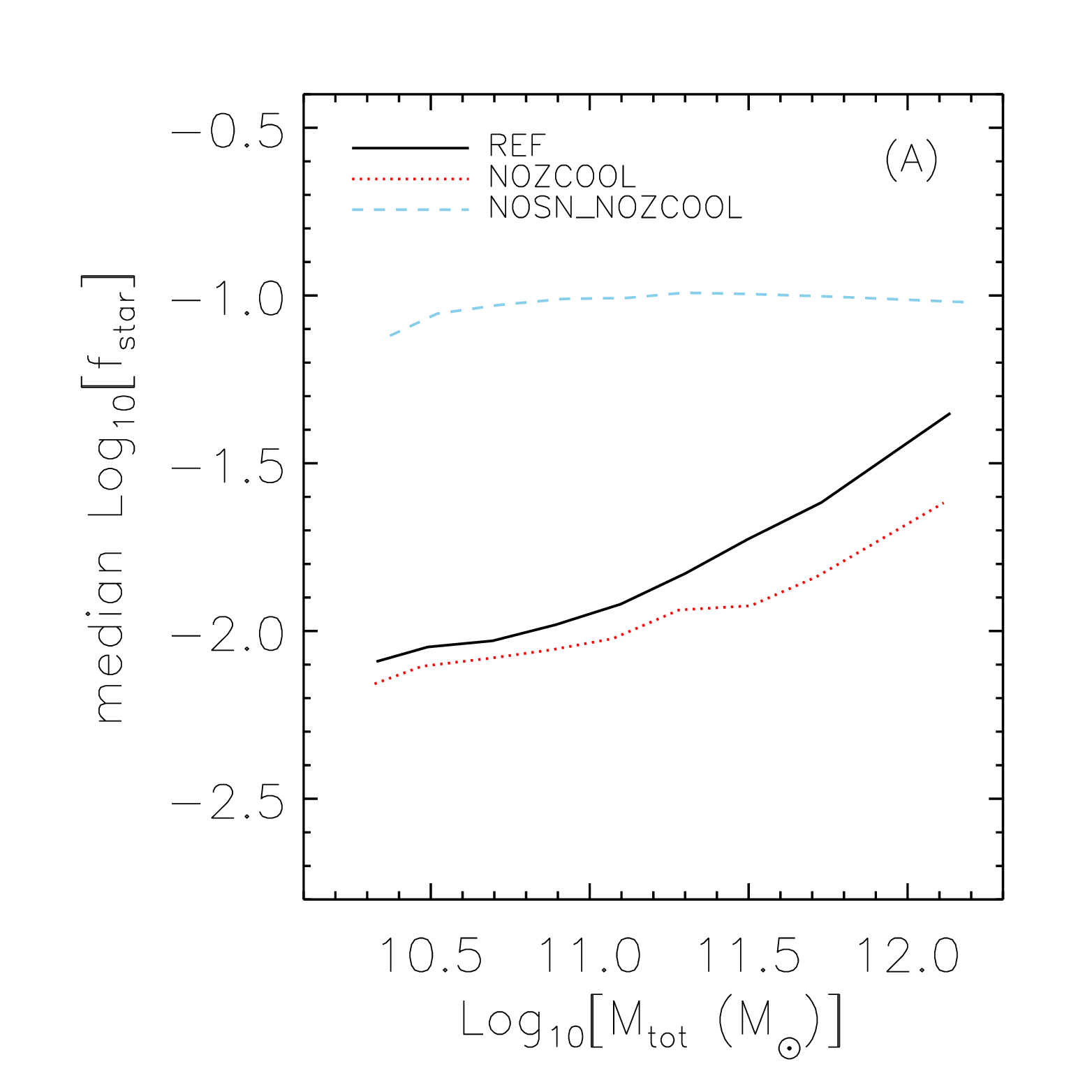}
\includegraphics[width=0.33\linewidth]{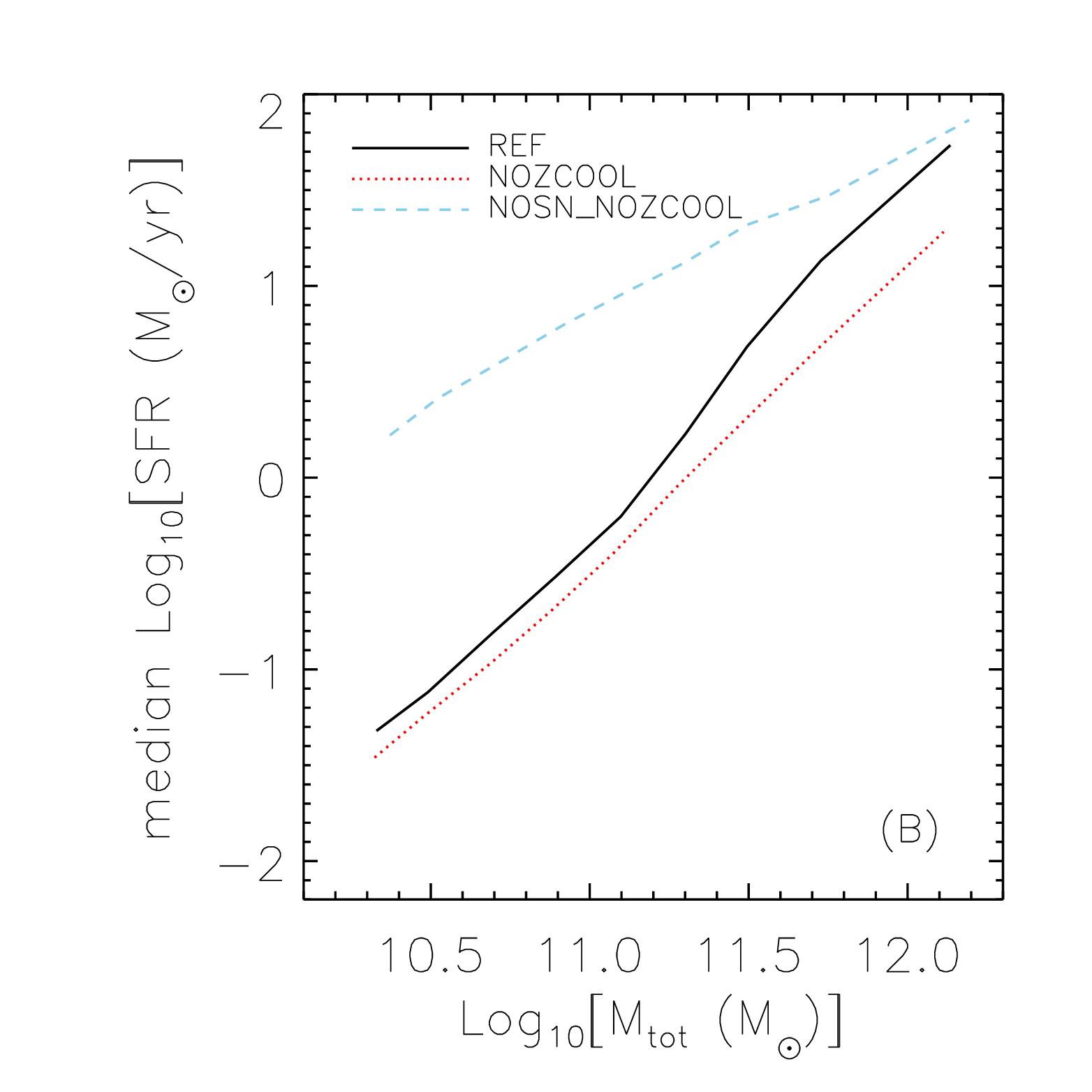}
\includegraphics[width=0.33\linewidth]{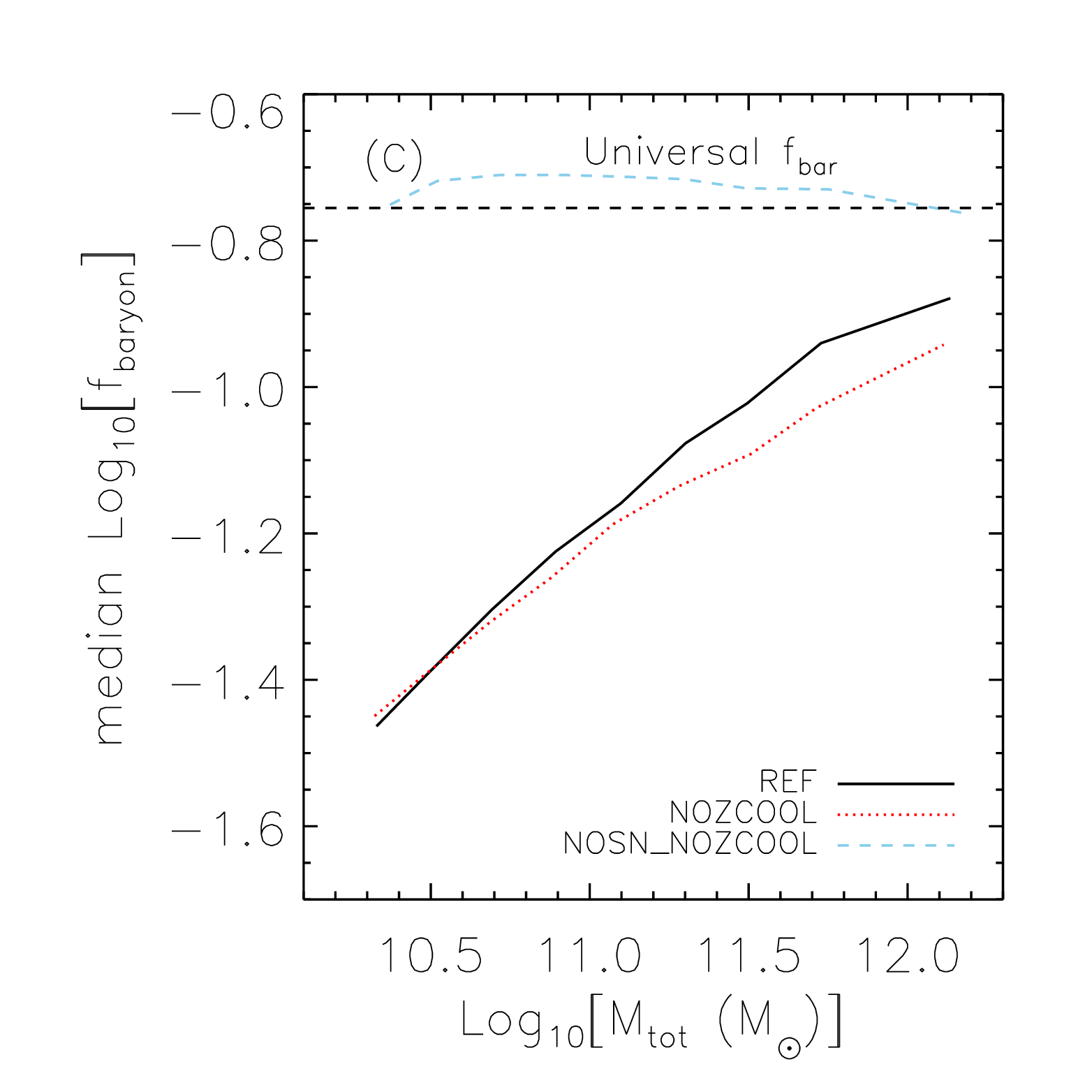} \\
\includegraphics[width=0.33\linewidth]{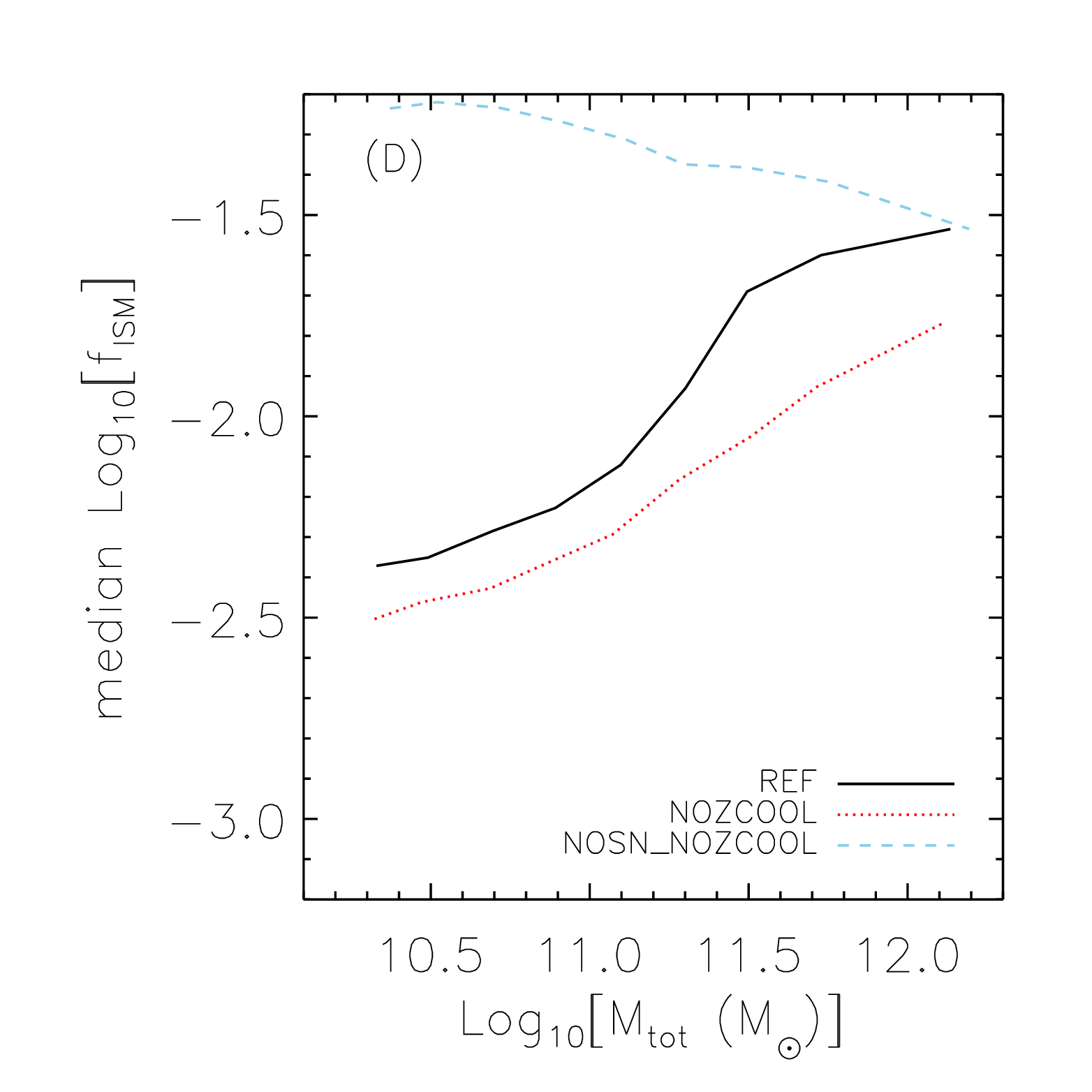}
\includegraphics[width=0.33\linewidth]{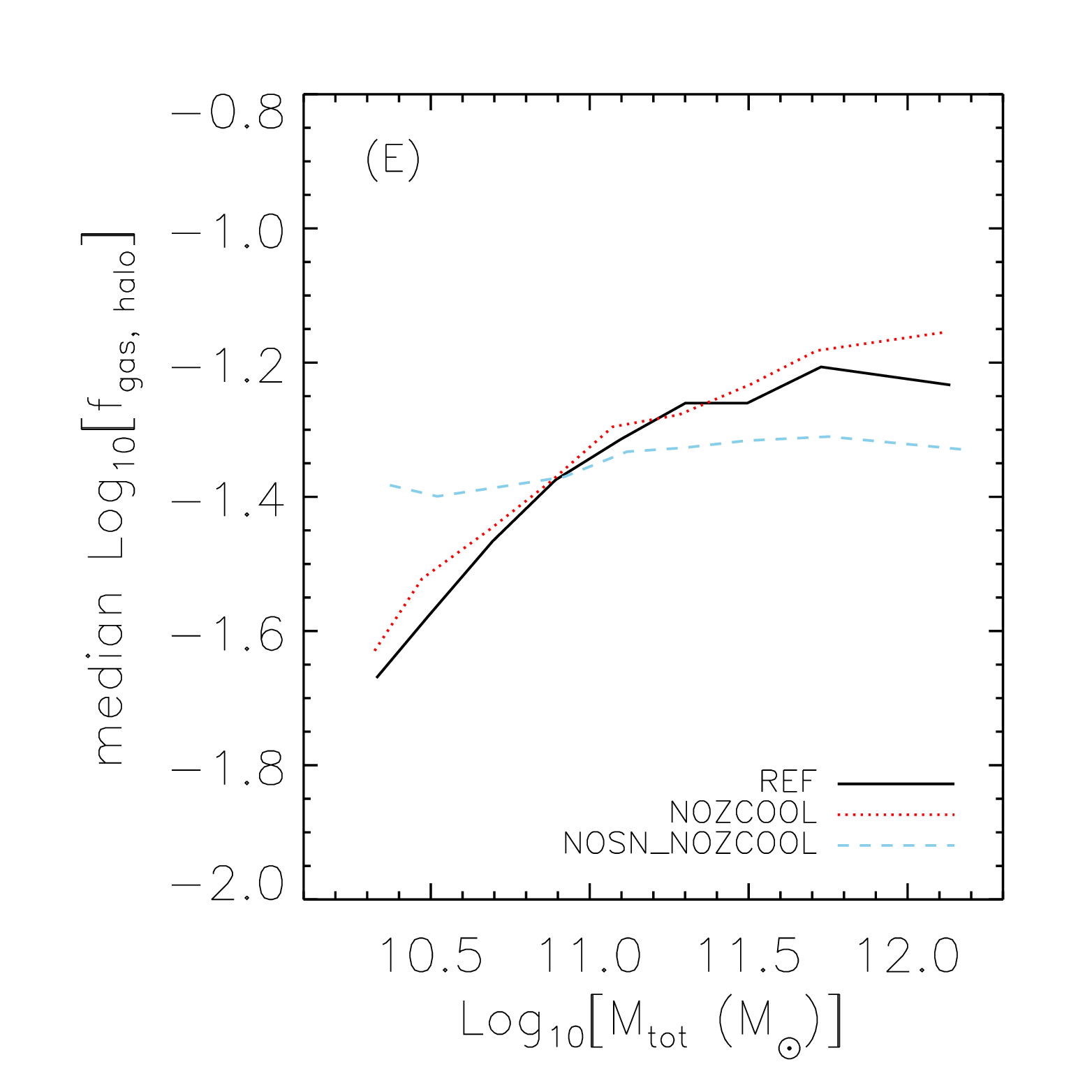}
\includegraphics[width=0.33\linewidth]{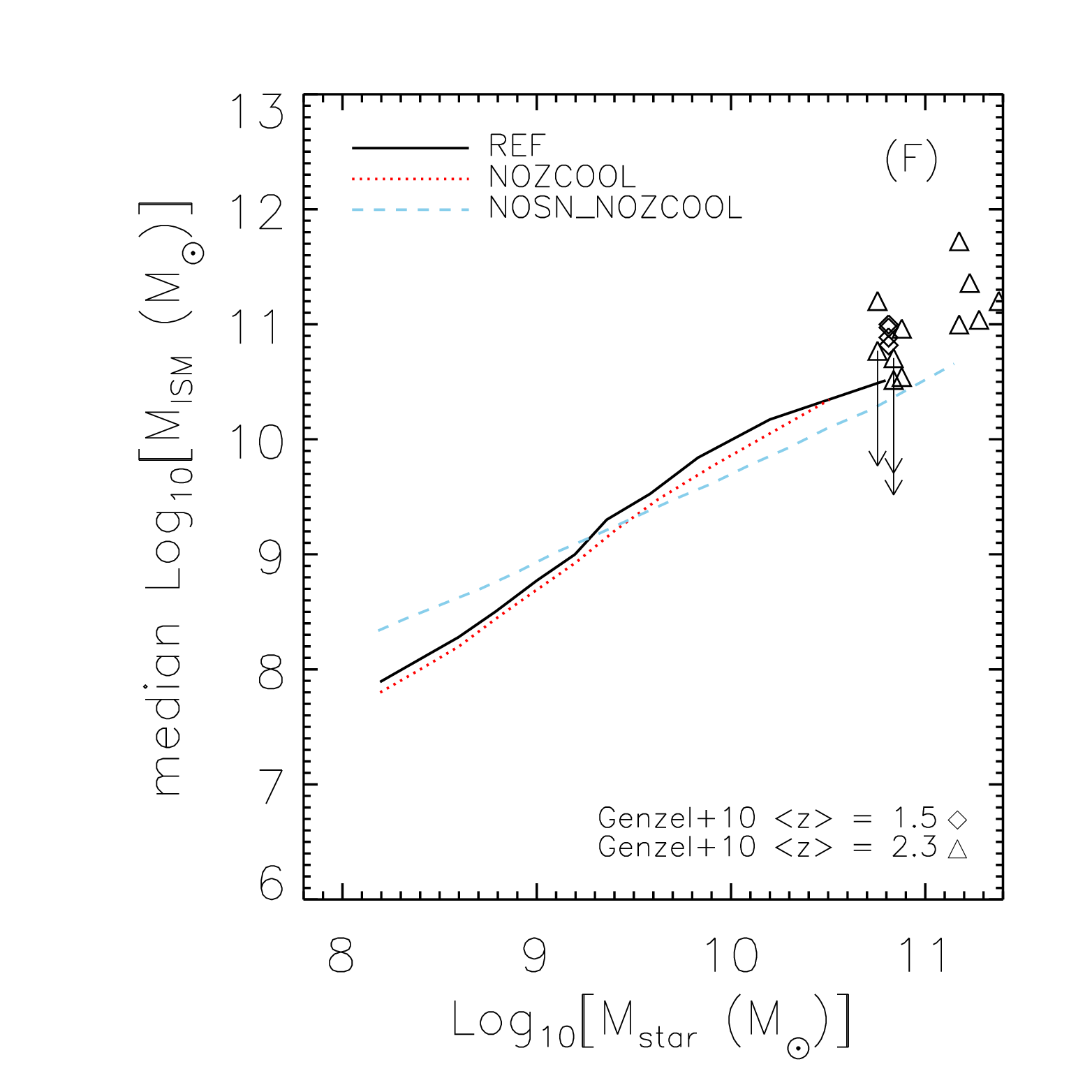} \\
\includegraphics[width=0.33\linewidth]{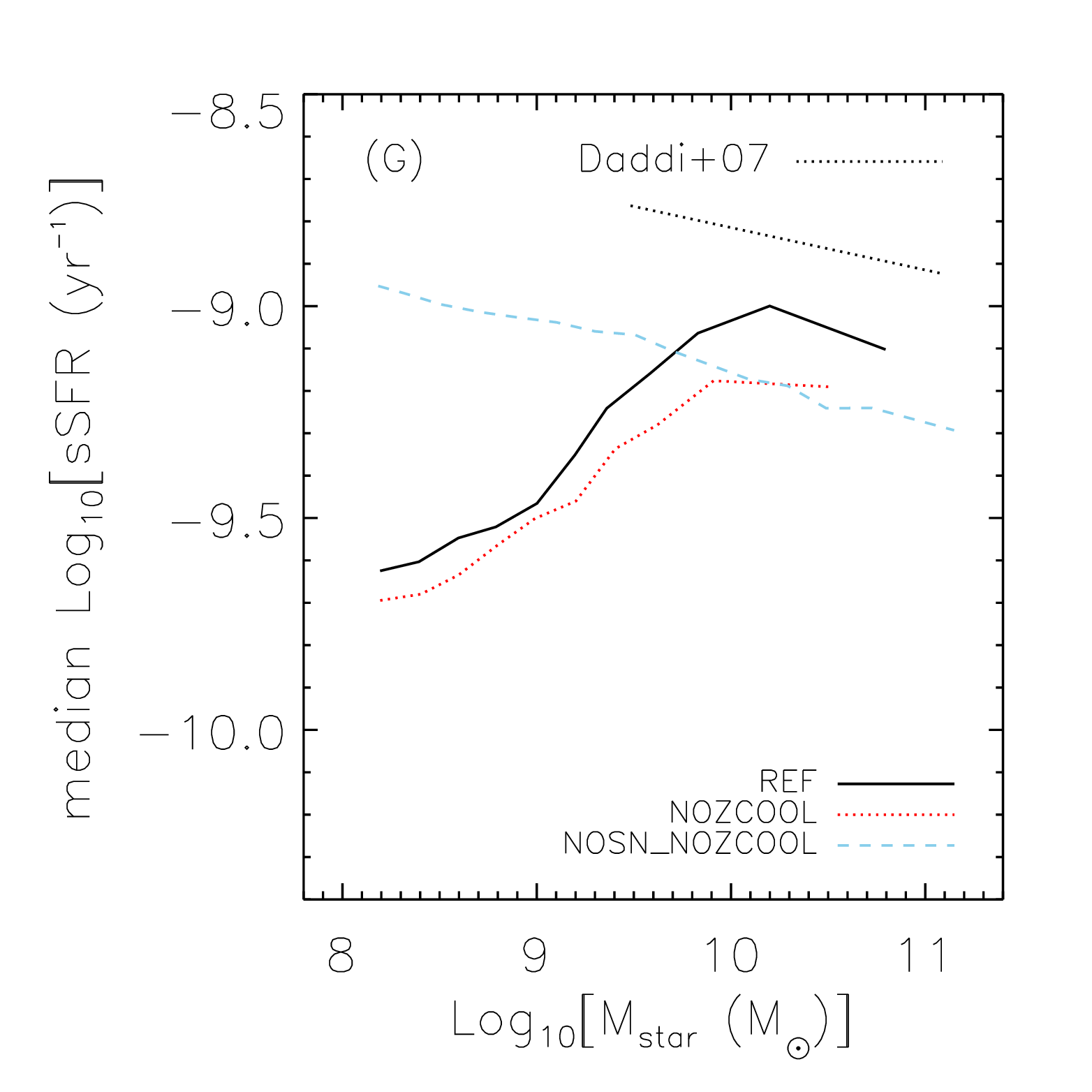}
\includegraphics[width=0.33\linewidth]{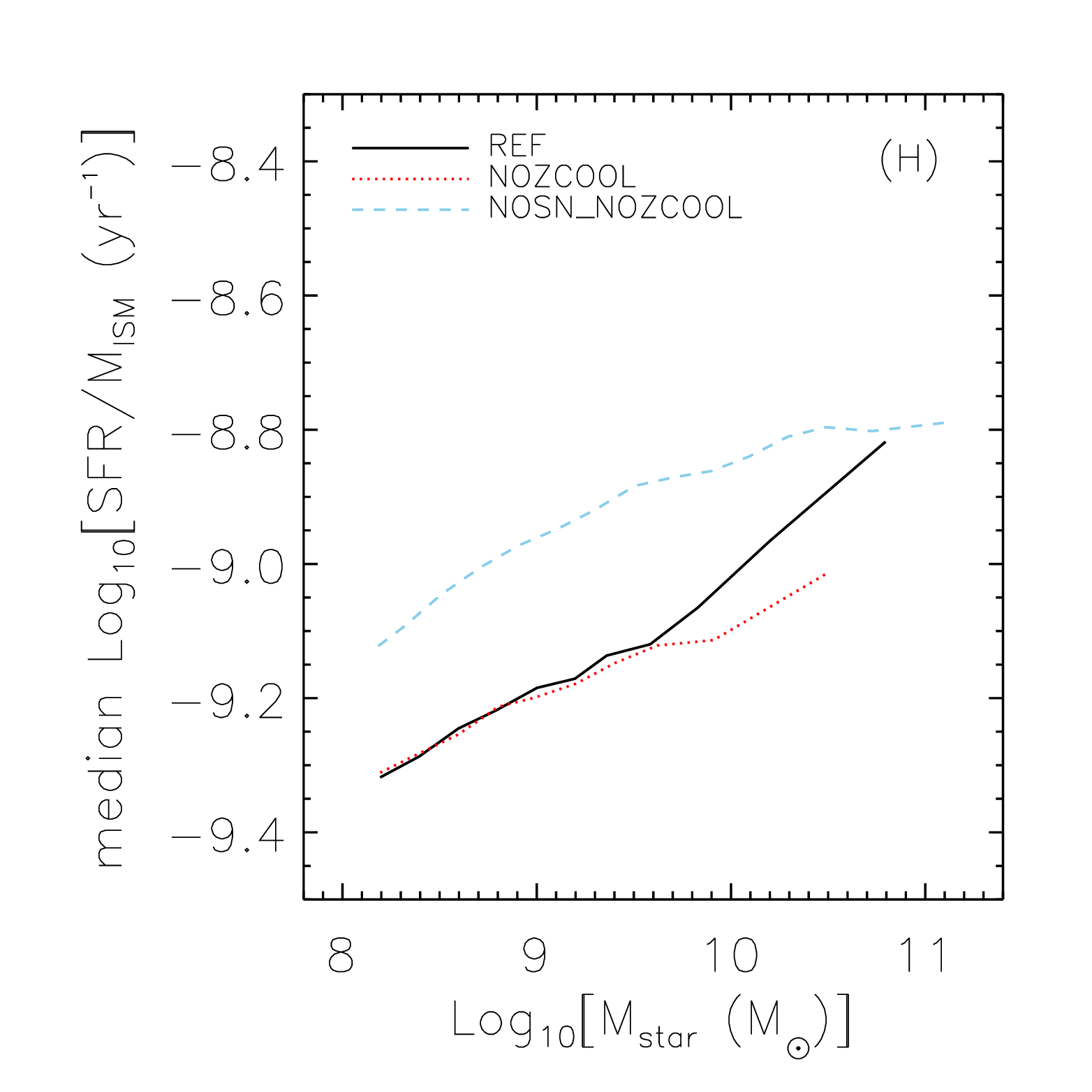}
\includegraphics[width=0.33\linewidth]{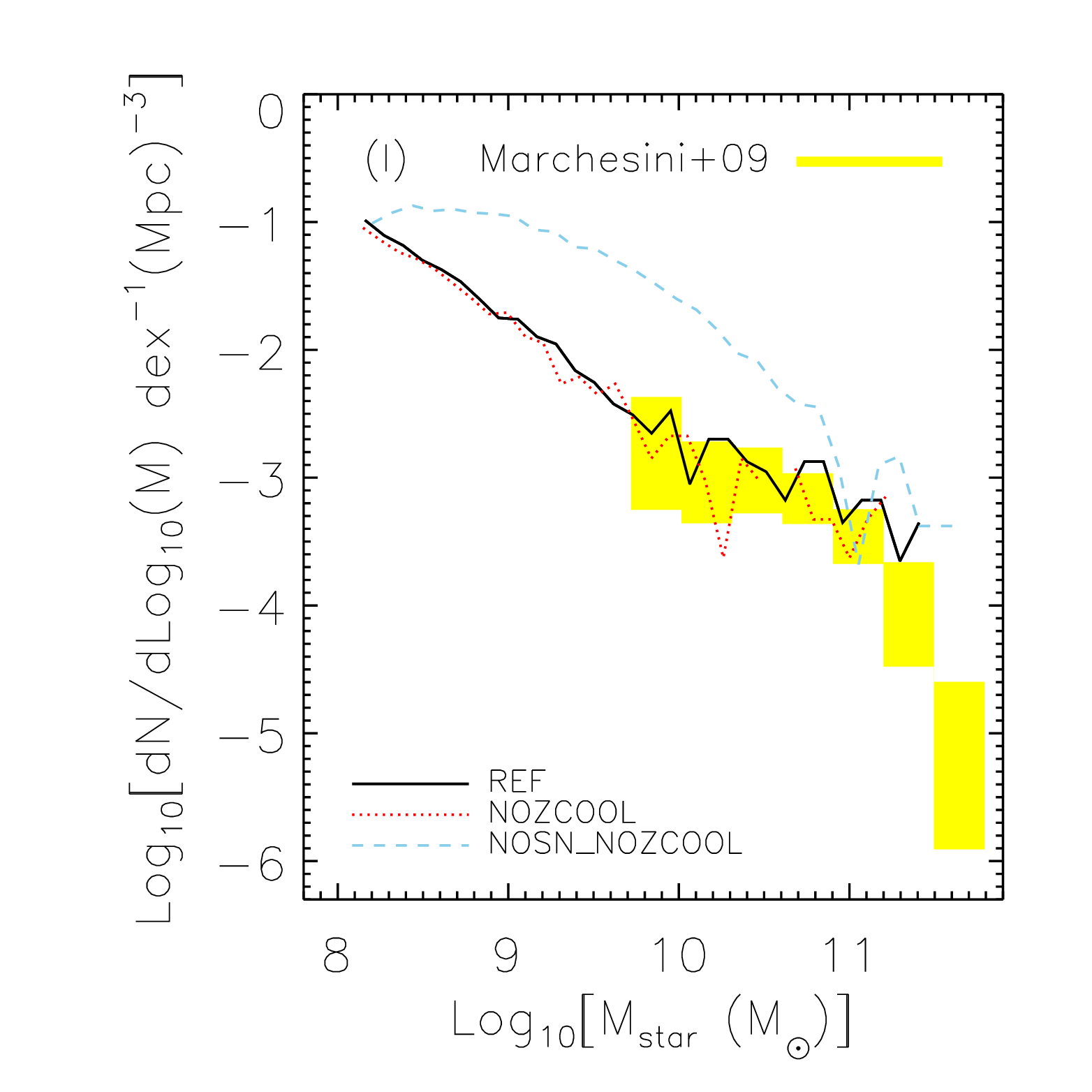} \\
\caption{As Fig.~\ref{fig:All_sims}, but showing only the subset of simulations in which the metal-line cooling and/or kinetic SN feedback are turned off. The solid, black curve shows the `\textit{REF}' simulation.  The red, dotted line shows the effect of turning off only metal-line cooling (`\textit{NOZCOOL}'). The blue, dashed curve shows the effect of switching off both metal-line cooling and SN feedback (`\textit{NOZCOOL\_NOSN}'). The effect of the SN feedback can thus be isolated by comparing the blue, dashed and red, dotted curves.} 
\label{fig:sims_cooling} 
\end{figure*}

\noindent In this set of simulations we investigate how gas cooling through metal lines affects the galaxy population. Metal-line cooling is the dominant mechanism by which enriched gas can cool in the temperature range $10^5\,{\rm K}<T<10^6\,{\rm K}$ \citep[e.g.][]{wiersma09cooling}. Metals are also the main coolants for gas at temperatures well below $10^4$ K, but we do not include those phases of the ISM in these simulations. If gas shock heats to high temperatures while accreting onto galaxies, neglecting metal-line cooling will greatly decrease the supply of gas that can cool out of haloes to fuel the galaxy \citep[see][for a comprehensive discussion of galaxy fueling in these simulations]{vandevoort11}.  In order to isolate these effects we compare the `\textit{REF}' simulation to one in which cooling through metal-lines was switched off (`\textit{NOZCOOL}'), and to a further simulation in which both cooling through metal-lines and SN feedback were switched off (`\textit{NOZCOOL\_NOSN}').  The morphology of a typical massive galaxy in these simulations can be seen in Fig.~\ref{fig:prettypics}. Turning off metal-line cooling reduces the extent of the gaseous disk in this massive system as less gas cools out of the halo into the galaxy.  Neglecting SN feedback leads to a hugely centrally concentrated galaxy as there is no mechanism to eject low-angular momentum material or to stop the fragmentation, and associated angular momentum losses.

To isolate the effect of metal-line cooling, we can compare the `\textit{REF}' simulation (solid, black curve) to the `\textit{NOZCOOL}' simulation (dotted, red curve) in Fig.~\ref{fig:sims_cooling}.  Panel (B) of Fig.~\ref{fig:sims_cooling} shows the effect of metal-line cooling on the SFRs of galaxies. In general, metal-line cooling increases SFRs, because cooling rates increase with increasing metallicities. The magnitude of the differences between the simulations increases with halo mass because both the fraction of gas that is shock heated when it enters the galaxy \citep[e.g.][]{birnboimdekel03, keres05, ocvirk08, vandevoort11} and the halo virial temperature increase with mass, making the effect of metal-cooling more pronounced.  The same trends are evident in other gas properties as a function of halo mass and stellar mass. Without metal-line cooling, less gas is able to cool onto the galaxy (panel D), and hence stellar masses (panel A) are lower.

As a function of stellar mass, we see that sSFRs are higher (panel G) and gas consumption timescales are shorter (panel H) in the `\textit{REF}' than in the `\textit{NOZCOOL}' simulation, as gas can cool more quickly into the galaxy when metal-line cooling is allowed, leading to higher gas densities and more efficient star formation. At high masses, winds are inefficient and more metal-rich gas piles up, increasing the difference between the simulations. The stellar mass functions of the two simulations (panel I) are almost identical up to $M_{\rm star}=10^{9.5}$\,\msun, but above this mass the `\textit{NOZCOOL}' simulation lies systematically 0.1-0.2 dex below `\textit{REF}'.  This occurs because the normalization of the $M_{\rm star}-M_{\rm tot}$ relation is decreased in `\textit{NOZCOOL}', and so a given stellar mass corresponds to a larger, rarer halo.

Turning our attention now to the effect of SN feedback by comparing the `\textit{NOZCOOL}' simulation (dotted, red curve) to the `\textit{NOZCOOL\_NOSN}' simulation (dashed, blue curve), we see that neglecting SN feedback leads to dramatic changes in the galaxy population as with no energy injection there is no process that can prevent gas from cooling directly into the galaxy.

Notably, panel (C) demonstrates that, unlike all of the other simulations we consider, `\textit{NOZCOOL\_NOSN}' has baryon fractions \emph{above} the universal value. This is due to the fact that star formation and cooling allow more gas to be pulled into the potential well of the galaxy. Non-radiative simulations predict baryon fractions of $\sim0.8$ times the universal value (\citealt{crain07}, but see \citealt{kravtsov05} who find a value much closer to one in a non-radiative mesh-based simulation), as pressure support forces more gas outside of haloes.

Interestingly, we note that (unlike all of the other simulations considered here), the slope of the $M_{\rm star}-{\rm sSFR}$ relation (panel G) in the `\textit{NOZCOOL\_NOSN}' has an almost identical slope to that observed, although with a normalization that is lower by a factor of approximately two. That the slope of the observed $M_{\rm star}-{\rm sSFR}$ relation is so similar to that of the simulation without feedback may indicate that perhaps feedback is ineffective at high galaxy masses, as is the case for the reference simulation above $10^{10}$\msun\ (see also Paper II). The similarity to the observed relation is striking.  The galaxies in the observations are selected to be star forming, so one would naively expect the star formation in those galaxies to be regulated by feedback.


\subsection{Winds with constant energy per unit stellar mass formed} \label{sec:constwindE}

\begin{figure*}
\centering
\includegraphics[width=0.33\linewidth]{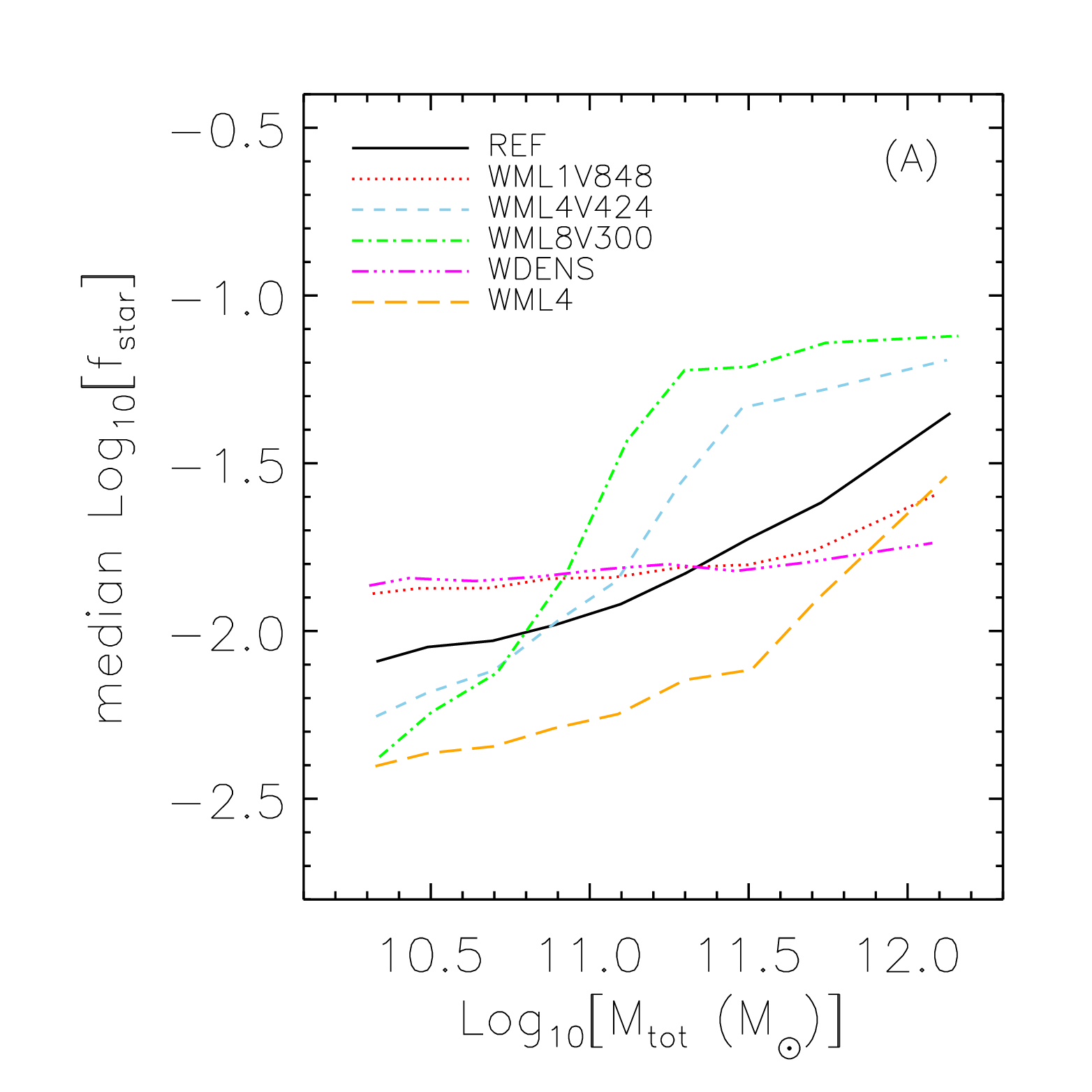}
\includegraphics[width=0.33\linewidth]{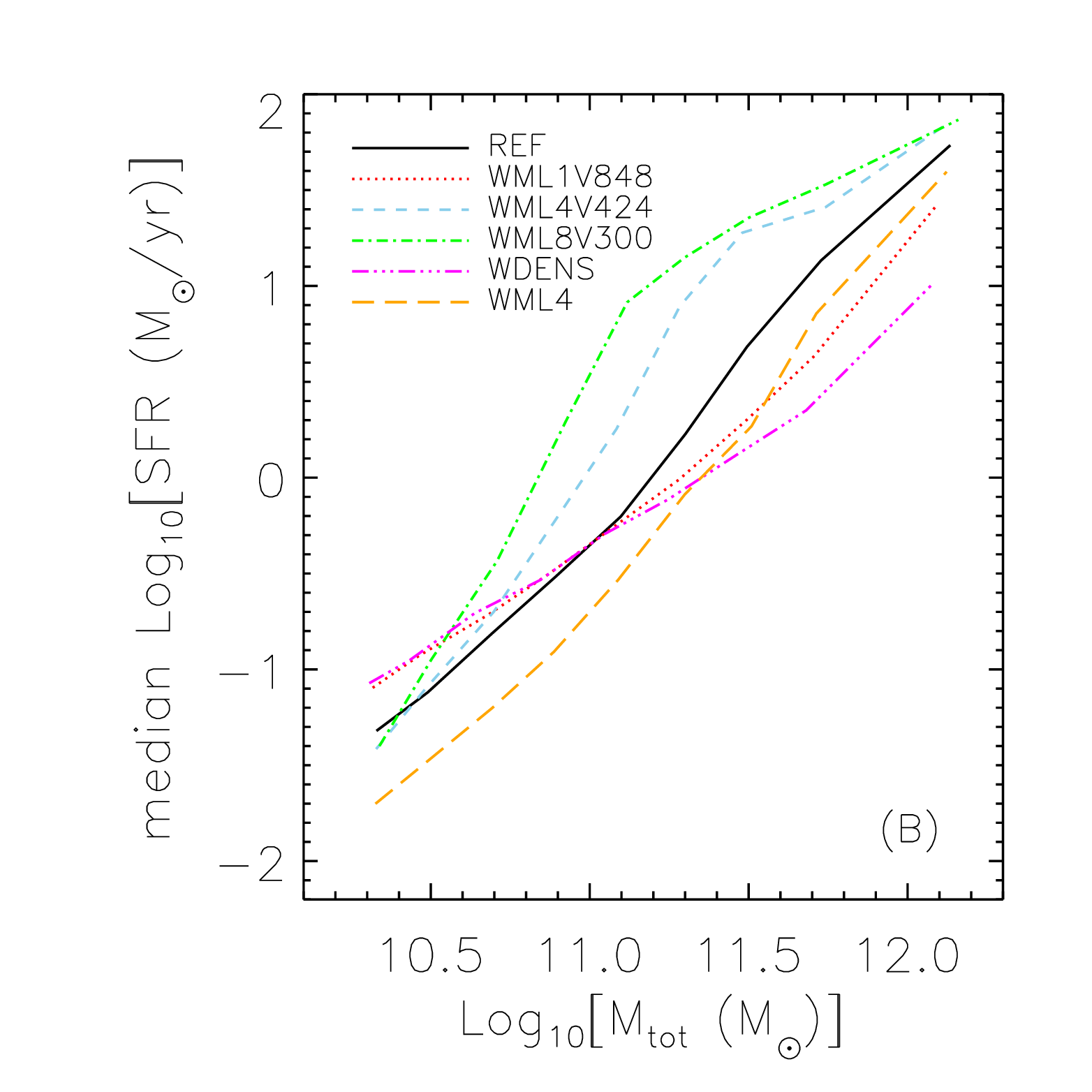}
\includegraphics[width=0.33\linewidth]{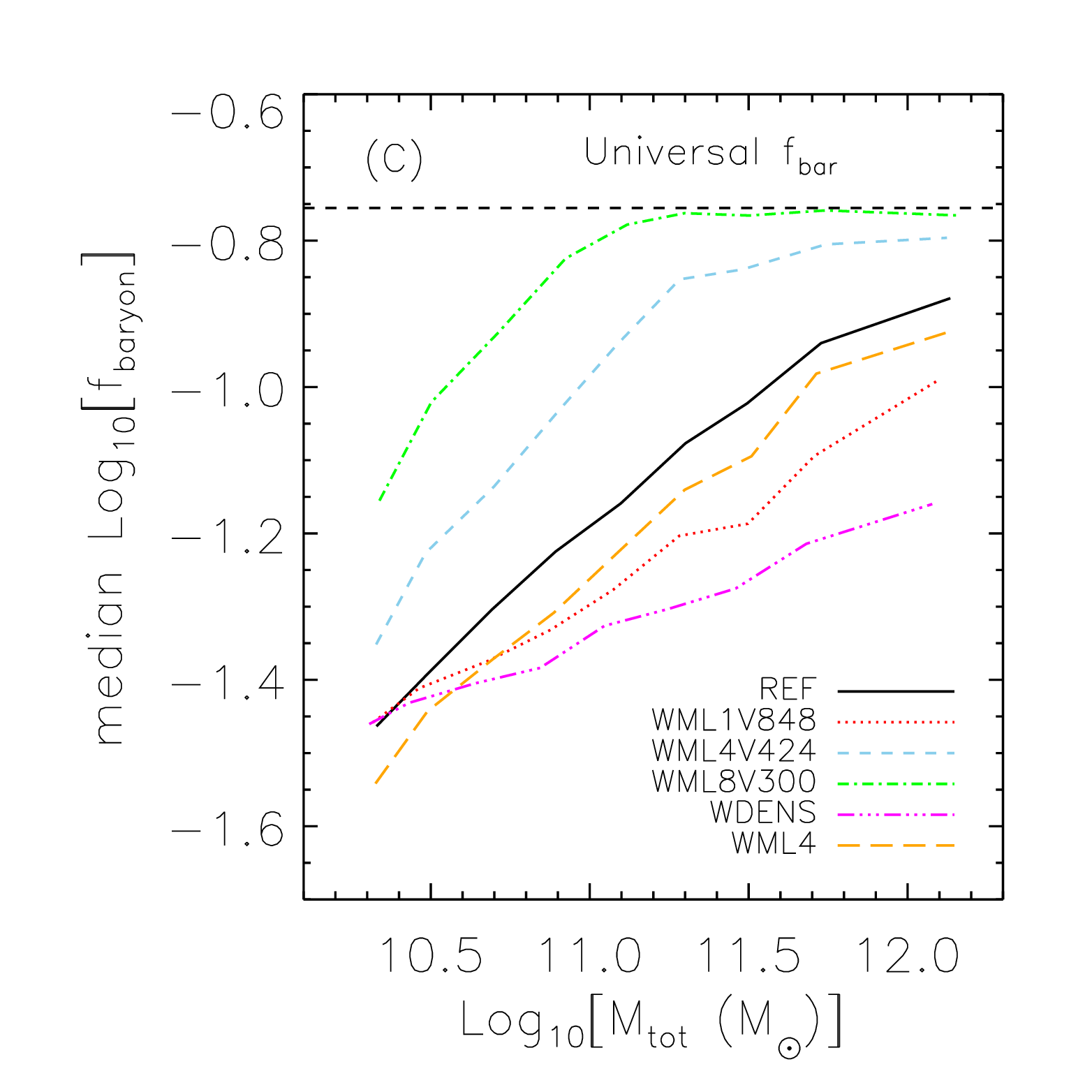} \\
\includegraphics[width=0.33\linewidth]{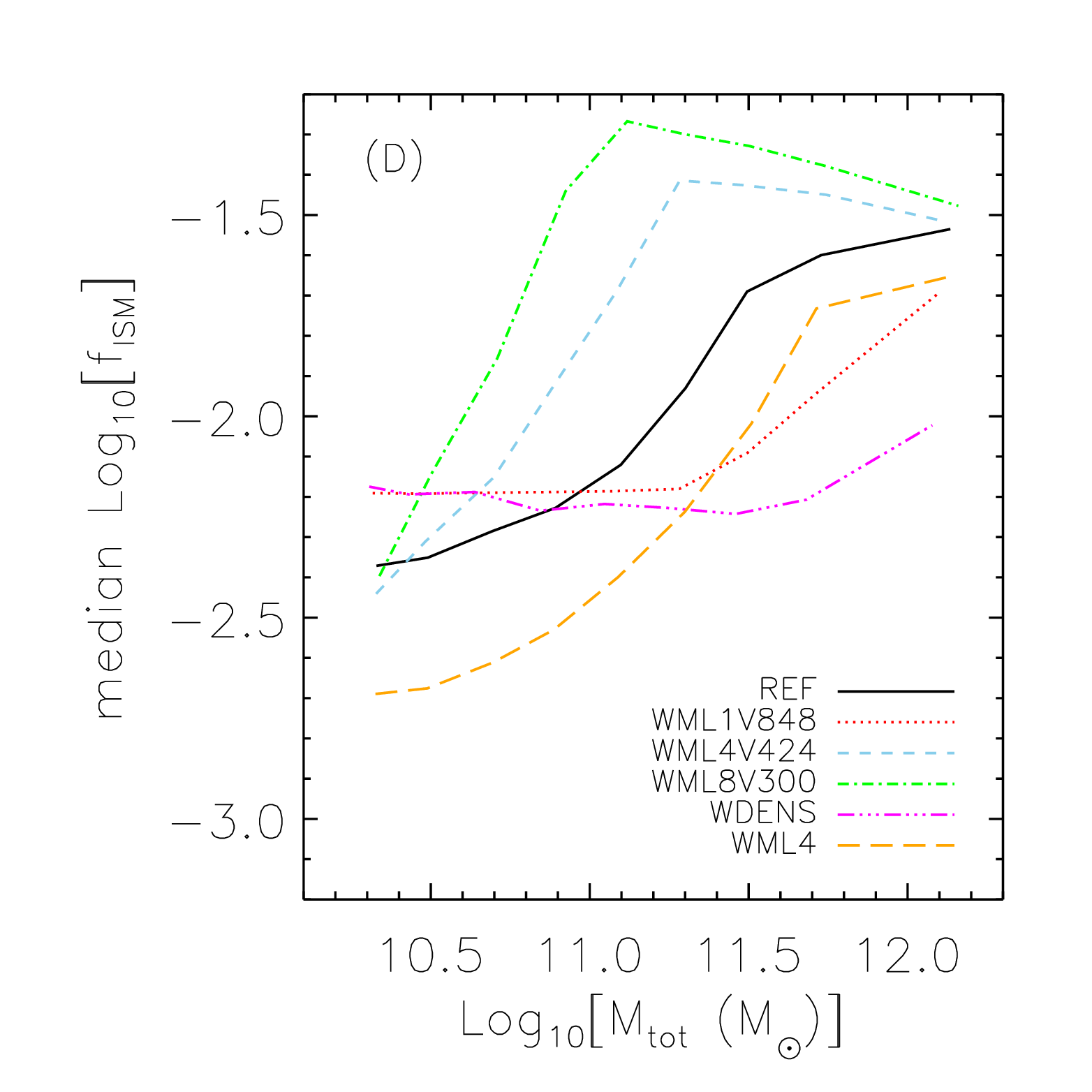}
\includegraphics[width=0.33\linewidth]{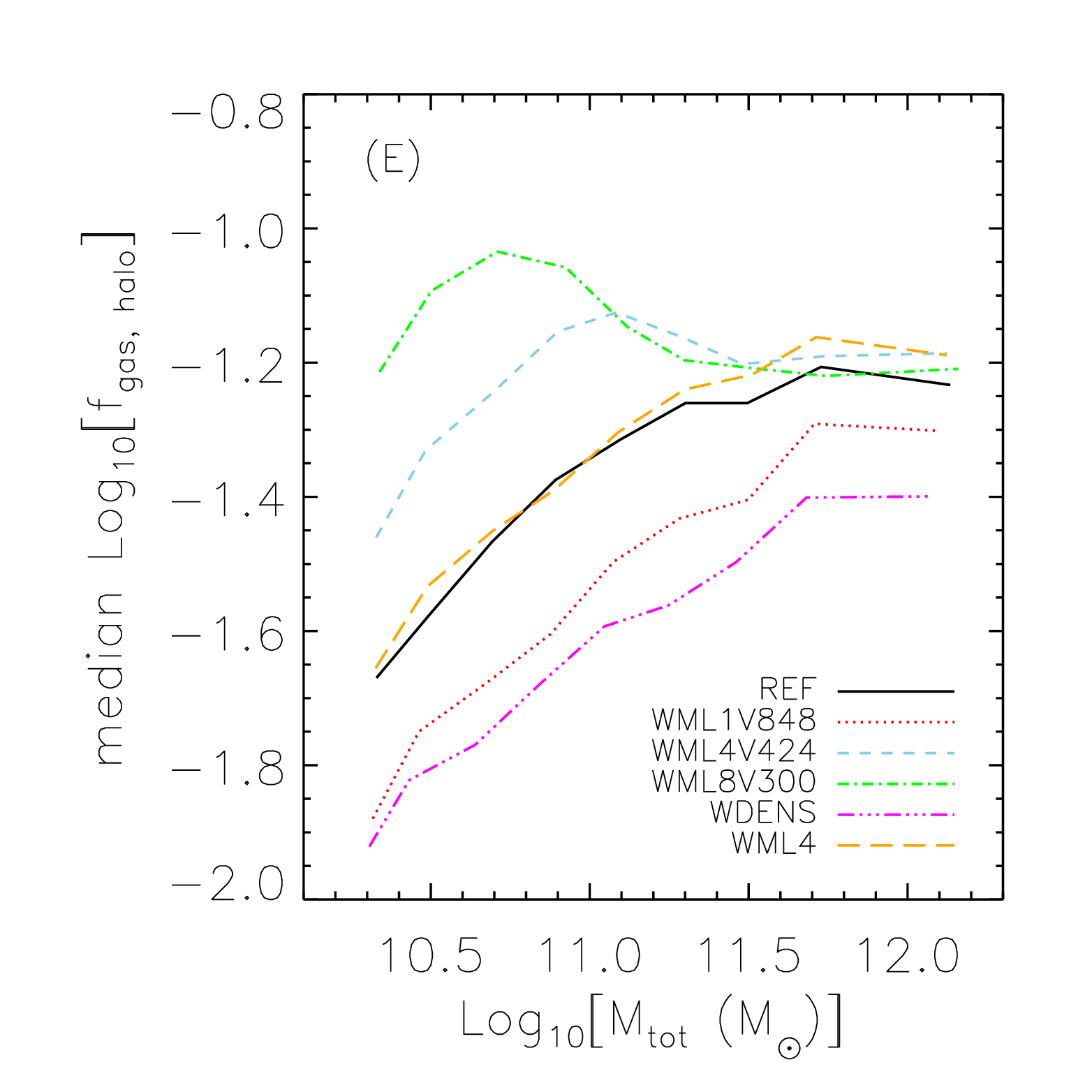}
\includegraphics[width=0.33\linewidth]{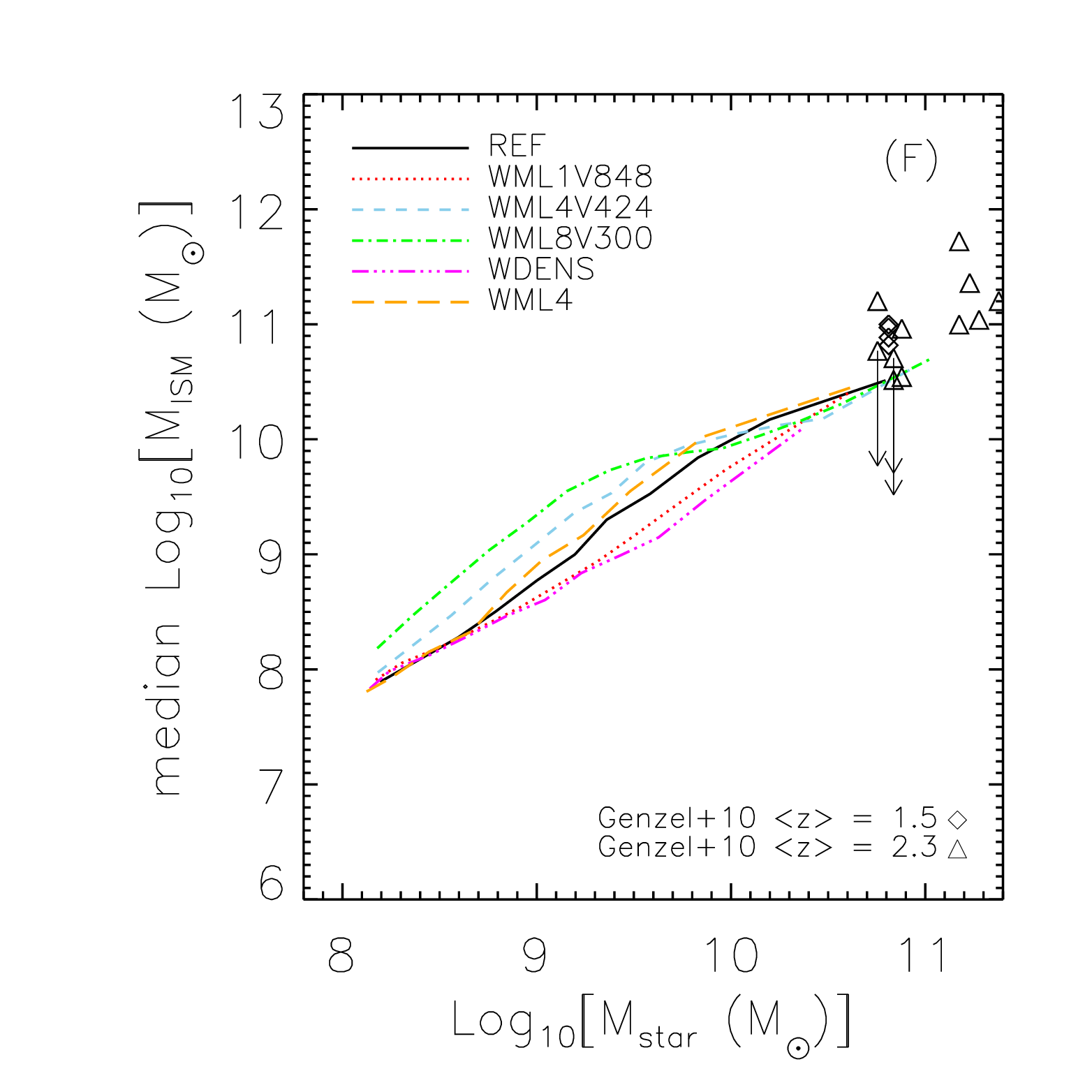} \\
\includegraphics[width=0.33\linewidth]{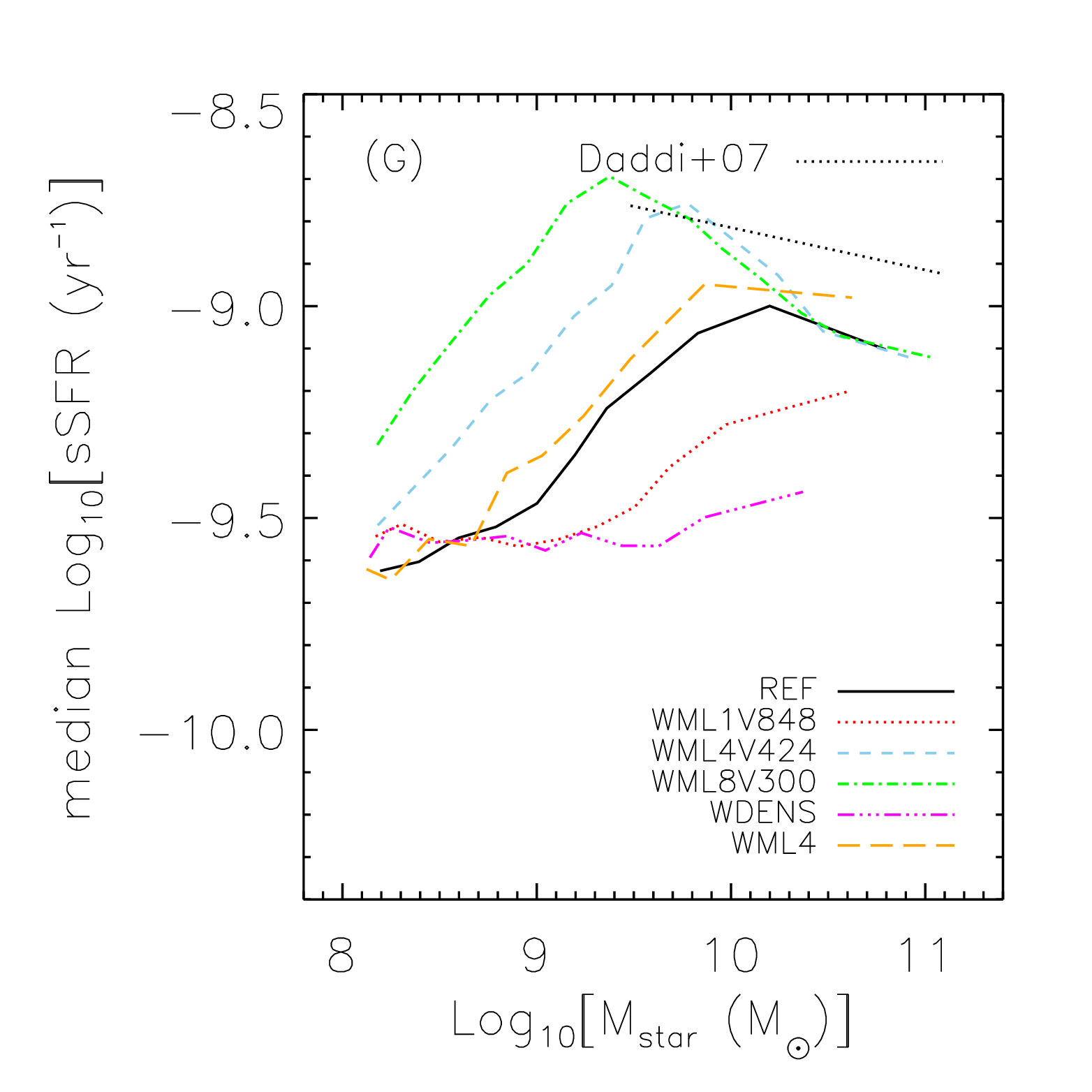}
\includegraphics[width=0.33\linewidth]{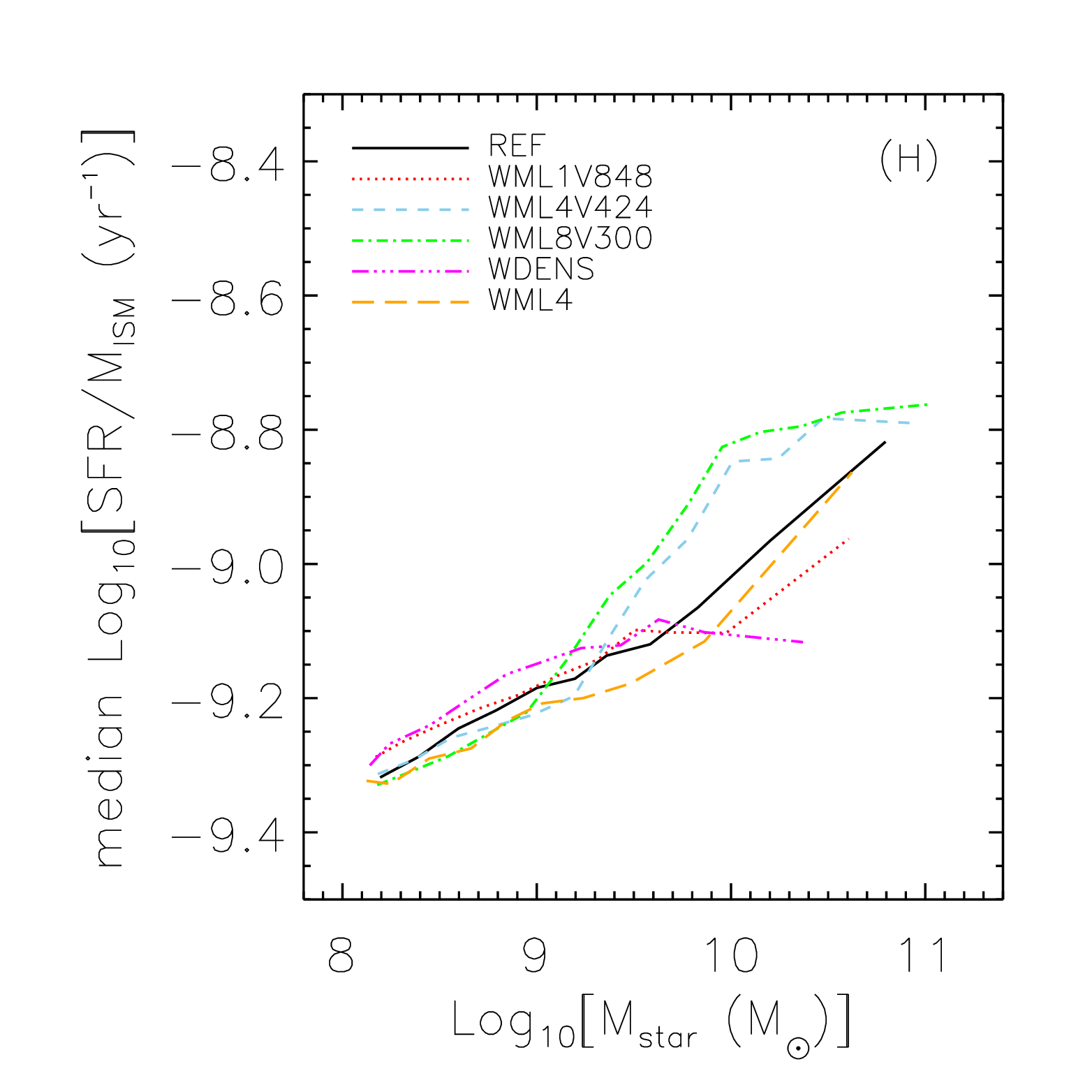}
\includegraphics[width=0.33\linewidth]{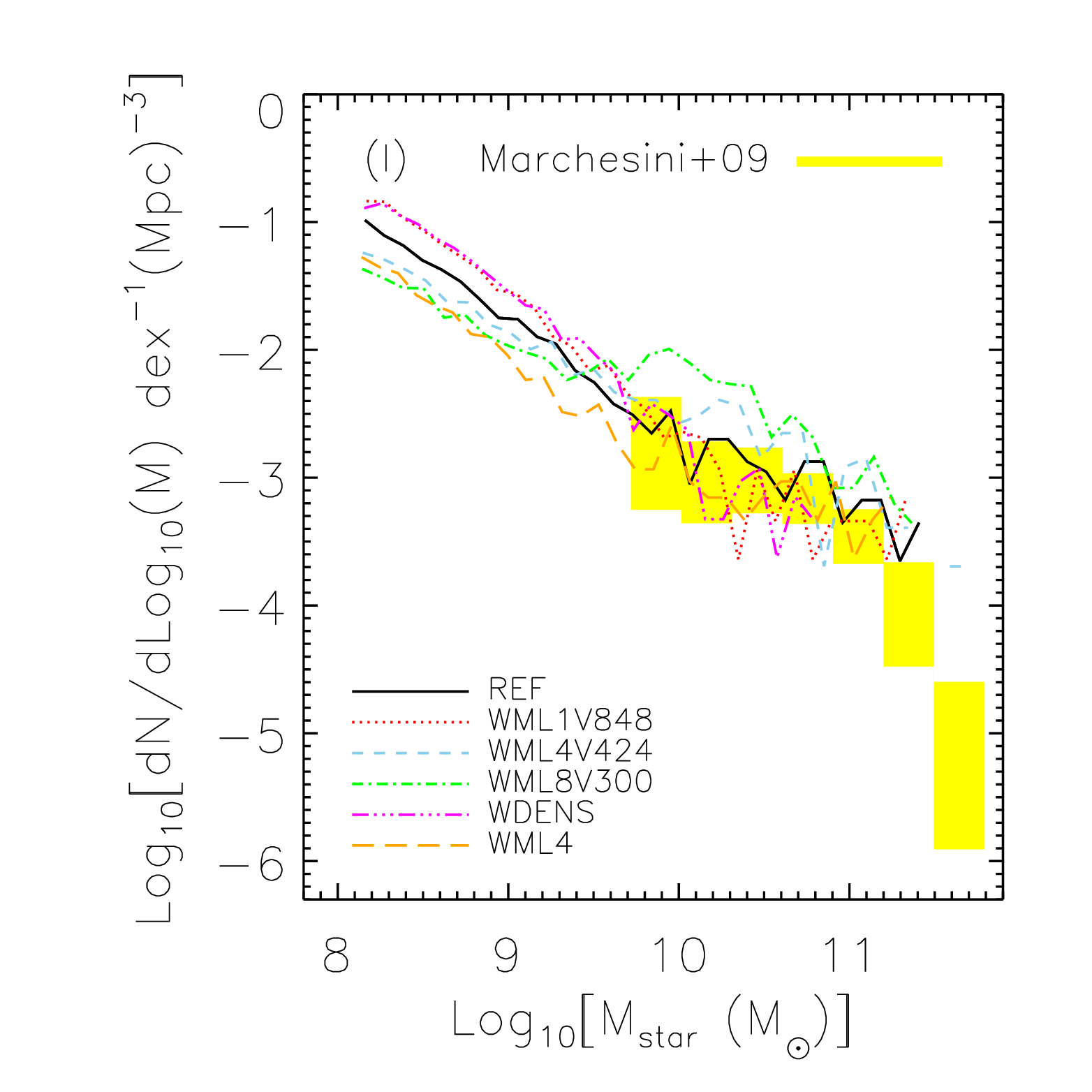} \\
\caption{As Fig.~\ref{fig:All_sims}, but showing only a set of simulations in which the initial wind velocity and the initial wind mass loading are varied. The reference simulation (black, solid curve) has a mass loading of $\eta = 2$ and a wind velocity of 600 km s\per. The simulations shown by the red-dotted, blue-dashed and green-dot-dashed lines show simulations that use the same total energy per unit stellar mass but different combinations of the mass loading and wind velocity.  The simulations shown have mass-loadings of 1 (`\textit{WML1V848}'; red, dotted curve), 4 (`\textit{WML4V424}'; blue, dashed curve) and 8 (`\textit{WML8V300}'; green, dot-dashed curve).  The magenta dot-dot-dot-dashed line represents a simulation (`\textit{WDENS}') which has a mass loading and velocity dependent on the local density, such that the energy in the wind is still the same and the initial velocity is proportional to the local sound speed.  Changing the parameters of the kinetic feedback model can change the galaxy stellar properties by up to an order of magnitude, even when the energy that is injected is kept constant.  The halo mass above which SN feedback becomes inefficient at removing gas from the galaxy (as seen by a steeper rise in the galaxy SFR with halo mass) depends primarily on the SN-driven wind velocity.} 
\label{fig:constwindE} 
\end{figure*}

\noindent Winds driven by feedback from star formation have become a fundamental ingredient of the galaxy formation picture by preventing too much gas from being locked into stars. In the `\textit{REF}' simulation, such winds are assumed to be driven by SNe and are implemented kinetically. That is, the effect of SNe is to \lq kick\rq\, some amount of gas.  This model contains two free parameters: the initial mass-loading $\eta = \dot{M}_\textrm{wind} / \dot{M}_*$ and the initial wind velocity $v_\textrm{w}$, which remain essentially unconstrained.  We therefore compare a series of four simulations that use the same SN energy per unit stellar mass formed (i.e. $\eta v_\textrm{w}^2$ is the same as for `\textit{REF}'), but distribute it differently between the velocity and mass loading. The mass loadings in the four simulations are 1, 2 (i.e. `\textit{REF}'), 4 and 8, with corresponding velocities of 848, 600, 424 and 300 km$/$s, respectively.  The parameters of the wind model are contained in the simulation name.  For example, the simulation `\textit{WML1V848}' uses a mass-loading of 1 and a wind velocity of $848$ km s$^{-1}$. A fifth simulation with kinetic feedback with the same feedback energy, `\textit{WDENS}', uses a wind velocity that scales in proportion to the local gas sound-speed ($\propto\rho^{1/6}$) and a mass loading that scales such that the total injected energy is constant ($\propto\rho^{-1/3}$).  The normalization of the parameters in the `\textit{WDENS}' simulation is such that the wind velocity and mass loading are the same as in the reference model if the gas density equals the star formation threshold, i.e.\ $n_{\textrm{\scriptsize H}} = 0.1$ cm$^{-3}$. For illustrative purposes, we also consider a simulation in which the available SN energy relative to `\textit{REF}' is doubled (by doubling the mass-loading to 4 and keeping the wind velocity the same).  This simulation is termed `\textit{WML4}'.

We first compare the set of simulations in which we hold the amount of SN energy that is injected per unit stellar mass constant, but vary how the energy is distributed between the initial mass loading and wind velocity.  We show in Fig.~\ref{fig:constwindE} that the simulation with the highest wind velocity (`\textit{WML1V848}; red, dotted curve) suppresses SF much more effectively in high-mass objects than the simulations with lower wind velocities (e.g.\ `\textit{WML8V300}').  However, at low galaxy masses, this trend reverses and it is the simulations with the lowest wind velocities that suppress SF most efficiently (panels A, B and D).  At the low-mass end, the amount of suppression is set by the amount of gas that is kicked, where models with large mass-loadings are able to inject large amounts of gas into the wind, much of which escapes into the halo or beyond, leading to these models being most effective.  However, in massive haloes a large wind velocity is required to drive gas out of the galaxy, so the models with large mass-loadings become incapable of driving winds and form more stars in high-mass objects.  There exists a clear transition mass above which, for a given wind velocity, SN feedback is no longer able to drive gas from the galaxy \citep[see also][]{springelhernquist03new}.  This is visible both in the properties of the galaxies (panels F--H), and the properties of the galaxy haloes (panels A--E), where above a mass (that depends on the initial wind velocity) the results tend towards those in a simulation that includes no feedback (Fig.~\ref{fig:sims_cooling}; blue, dashed curve). 

\citet{dallavecchiaschaye08} showed that the reason why effective feedback in more massive galaxies requires higher velocity winds is not directly related to the requirement that the winds overcome the gravitational potential. Instead, they found that the outflows are already quenched in the ISM and only if gas pressure forces are taken into account. Because the pressure in the ISM increases with the depth of the potential well, more massive galaxies require higher velocity winds. A deeper understanding was provided by \citet{dallavecchiaschaye12}, who demonstrated analytically, and confirmed numerically, that the outflows are quenched due to radiative losses in the shock-heated ISM. Higher velocities imply higher post-shock temperatures and hence longer cooling times. As the ISM of more massive galaxies is denser, higher velocities are required for effective feedback in more massive galaxies. They pointed out that cosmological simulations overestimate the radiative losses due to their finite resolution and showed that models with different wind parameters, but the same amount of energy per unit stellar mass, make converging predictions if the resolution is sufficiently high. The resolution of our simulations is low compared to the analytic estimates of the required resolution provided by \citet{dallavecchiaschaye12}, which explains our finding that galaxy properties are sensitive to the choice of initial wind velocity.

A second instructive comparison is between `\textit{REF}' and the simulation that injects twice as much SN energy per unit stellar mass, `\textit{WML4}' (orange, long-dashed curves).  This simulation uses the same wind velocity as `\textit{REF}' and so the SN-driven winds become ineffective at the same mass.  This can be seen by comparing `\textit{REF}' (black, solid curve) to WML4 in panel (B), where above $M_{\rm tot}=10^{11.25}\,$\msun\ the two simulations converge towards each other with increasing mass, but below this mass, `\textit{WML4}' can suppress SF significantly more efficiently than `\textit{REF}'.  We can see this more quantitatively in panel (A), where we see that at low masses (where the feedback is effective), $f_{\rm star}$ is almost exactly a factor of two lower than in the `\textit{REF}' simulation.  This implies that the star formation is self-regulating: a galaxy's SFR increases until the rate of energy injection into the ISM reaches a critical value. This critical rate of energy injection depends on the halo mass and is presumably the rate for which outflows balance inflows (e.g.\ \citealt{owls}, \citealt{dave12}; see also 
\citealt{boothschaye10} for an analogous discussion of self-regulated black hole growth). Because the outflow rate depends not only on the rate of energy injection, but also on the initial wind velocity \citep[e.g.][]{dallavecchiaschaye08}, the same will be true for the critical SFR. Doubling the SN energy that is injected per unit stellar mass, while keeping the initial wind velocity constant, will half the SFR required to produce the same outflow rate and will thus cause the galaxy to form half as many stars.  This represents one of the fundamental conclusions of this paper, \emph{star formation is regulated by the interplay between the available fuel supply and feedback processes}. 

Because a given galaxy requires a fixed rate of energy injection to achieve a balance between inflows and outflows, increasing the efficiency of SN feedback causes the galaxies to decrease their gas reservoirs (panel D). Because the star formation law is non-linear and because the ISM of more massive galaxies tends to be denser, the decrease in the mass in the ISM is larger for lower mass galaxies. 

The gas consumption timescales of all models are very similar in the regime where SN feedback is effective (panel H) because the ratio between the amount of gas available for star formation and the rate at which stars are formed is mainly determined by the star formation law. At the point where feedback becomes incapable of lifting gas out of the galaxy, the gas in the galaxy is used up very efficiently. 

The kinetic feedback models discussed above introduce a slope in the relation between stellar mass and sSFR, that does not agree with observations. Models that do not use a constant wind velocity and mass loading result in different relations between these quantities as we will show below. 

For the `\textit{WDENS}' model, the energy injected in the wind per unit stellar mass is also the same as in the reference model, but the initial wind velocity scales with the local sound speed. The relation between halo mass and SFR is even shallower than it is for the run with $v_w =$ 848 km s$^{-1}$, indicating that the feedback is effective for all haloes. At high masses, this is our most effective wind model with constant energy.  This indicates that choosing the mass-loading and wind velocity carefully can make SN-driven winds capable of driving winds in all objects resolved in these simulations.  

As is clear from panel I of Fig.~\ref{fig:constwindE}, decreasing the slope of the low-mass end of the stellar mass function can be attained by increasing the mass loading factor in constant energy winds. The highest mass loading still gives a low-mass end slope that is somewhat steeper than power law fits to the low-mass end in the observations, although the discrepancy only becomes severe at masses lower than those observed. Such efficient feedback in low-mass objects also boosts the sSFRs in the higher-mass objects for which feedback from star formation is just becoming inefficient, as appears to be required by the observations (panel G). 

These simulations suggest that the fit to the observed mass function could be much improved by varying the mass loading factor with galaxy mass \citep[see also][]{oppenheimer10, okamoto10, bower12, puchweinspringel12}.

\subsection{Simulations with a top-heavy IMF at high pressures} \label{sec:dblimf}

\begin{figure*}
\centering
\includegraphics[width=0.33\linewidth]{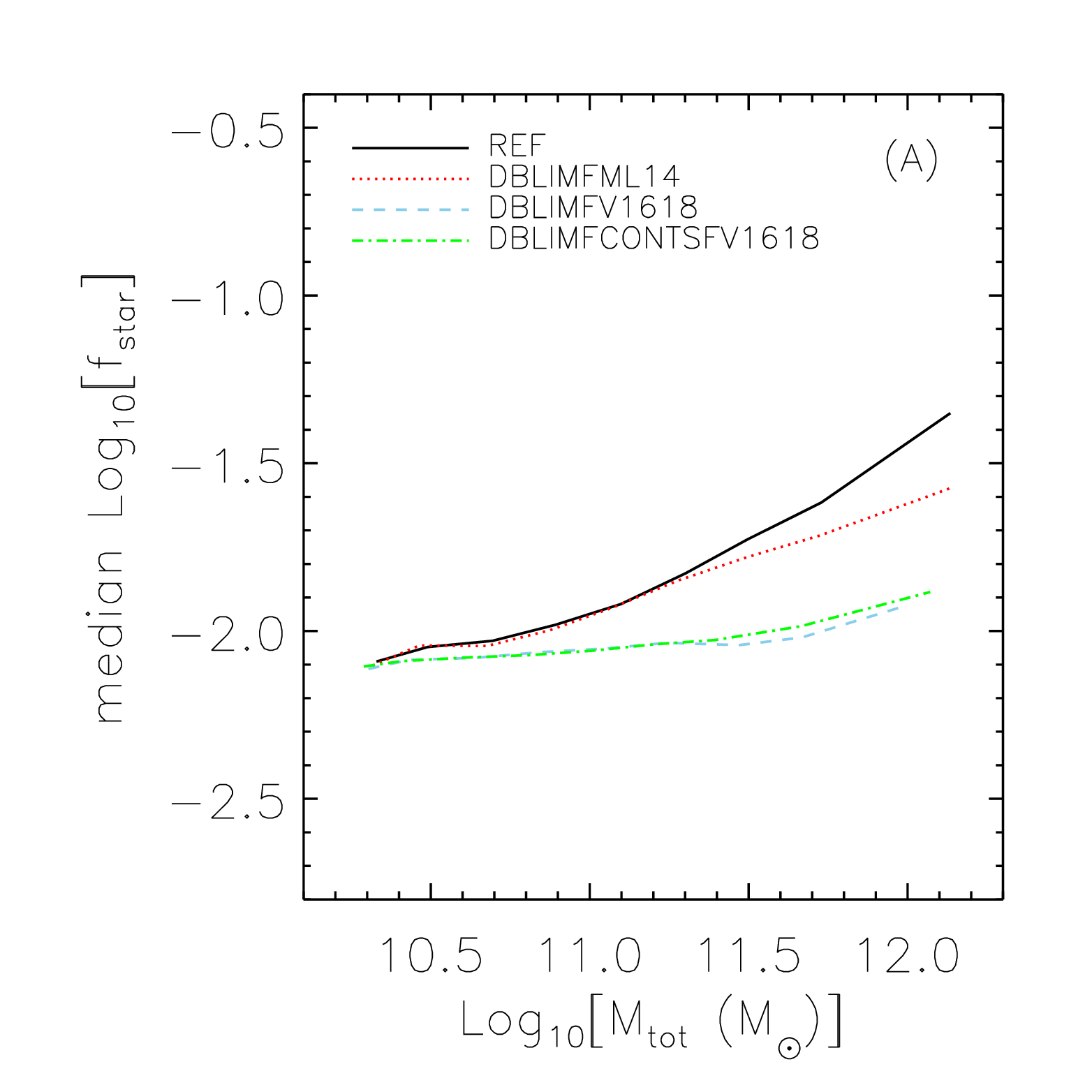}
\includegraphics[width=0.33\linewidth]{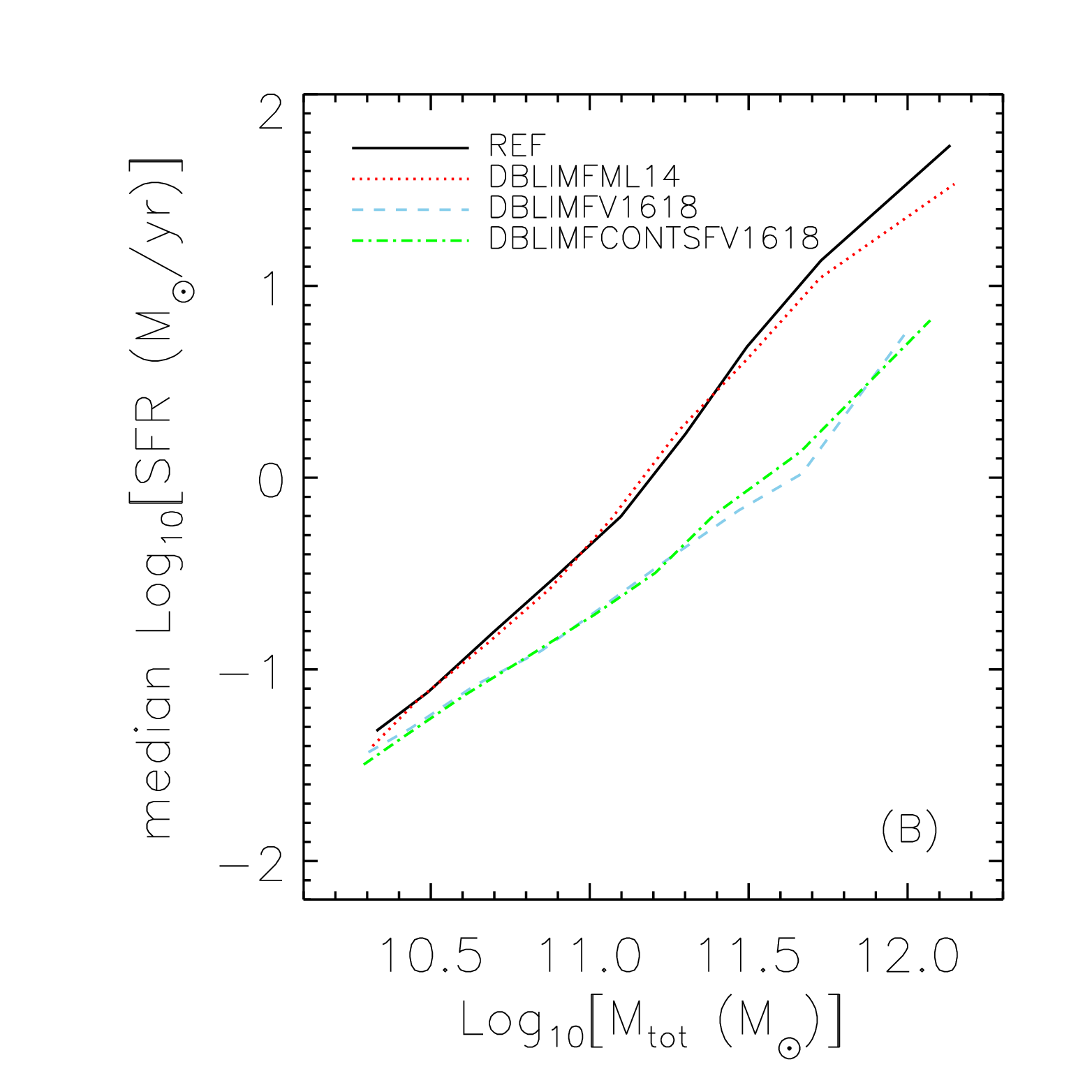}
\includegraphics[width=0.33\linewidth]{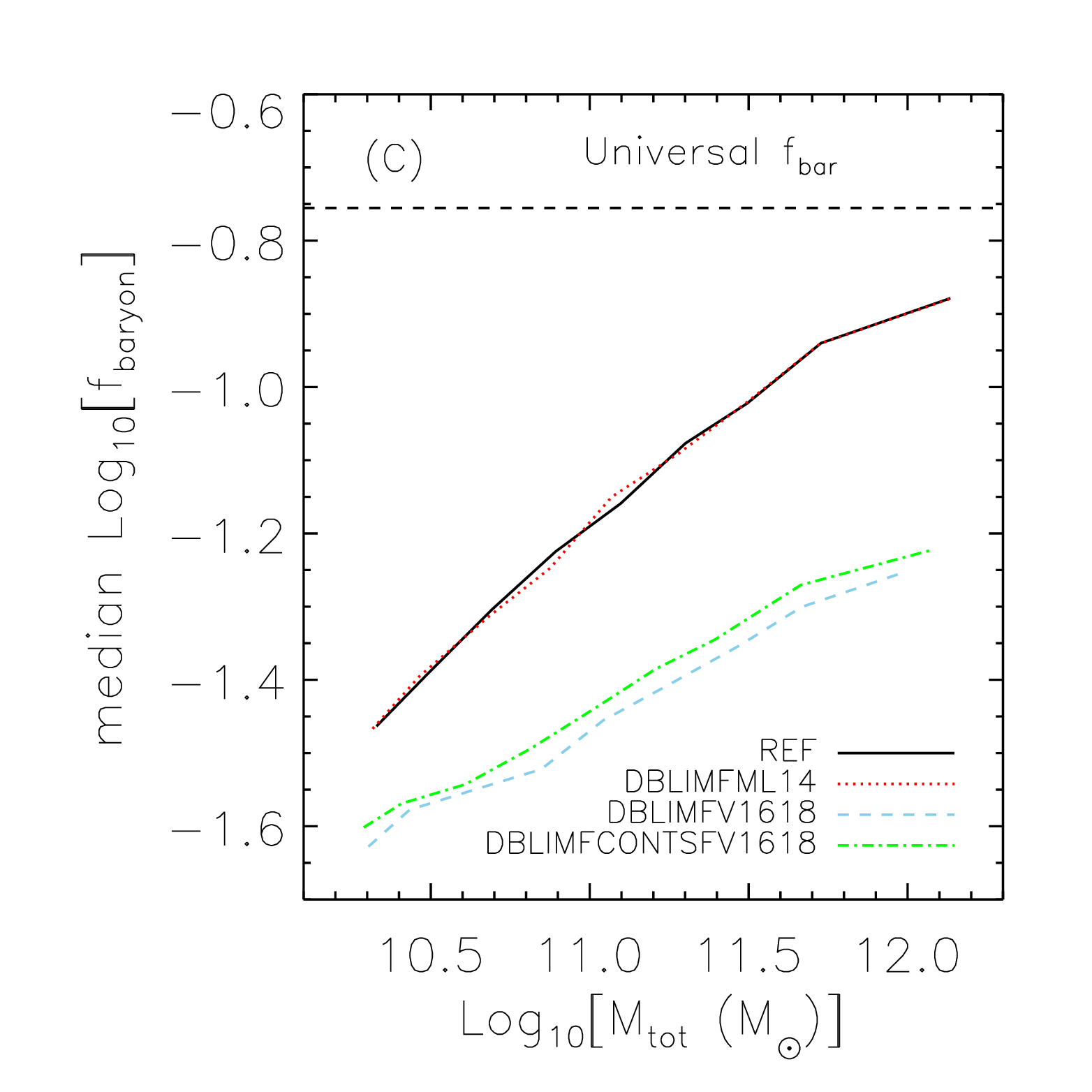} \\
\includegraphics[width=0.33\linewidth]{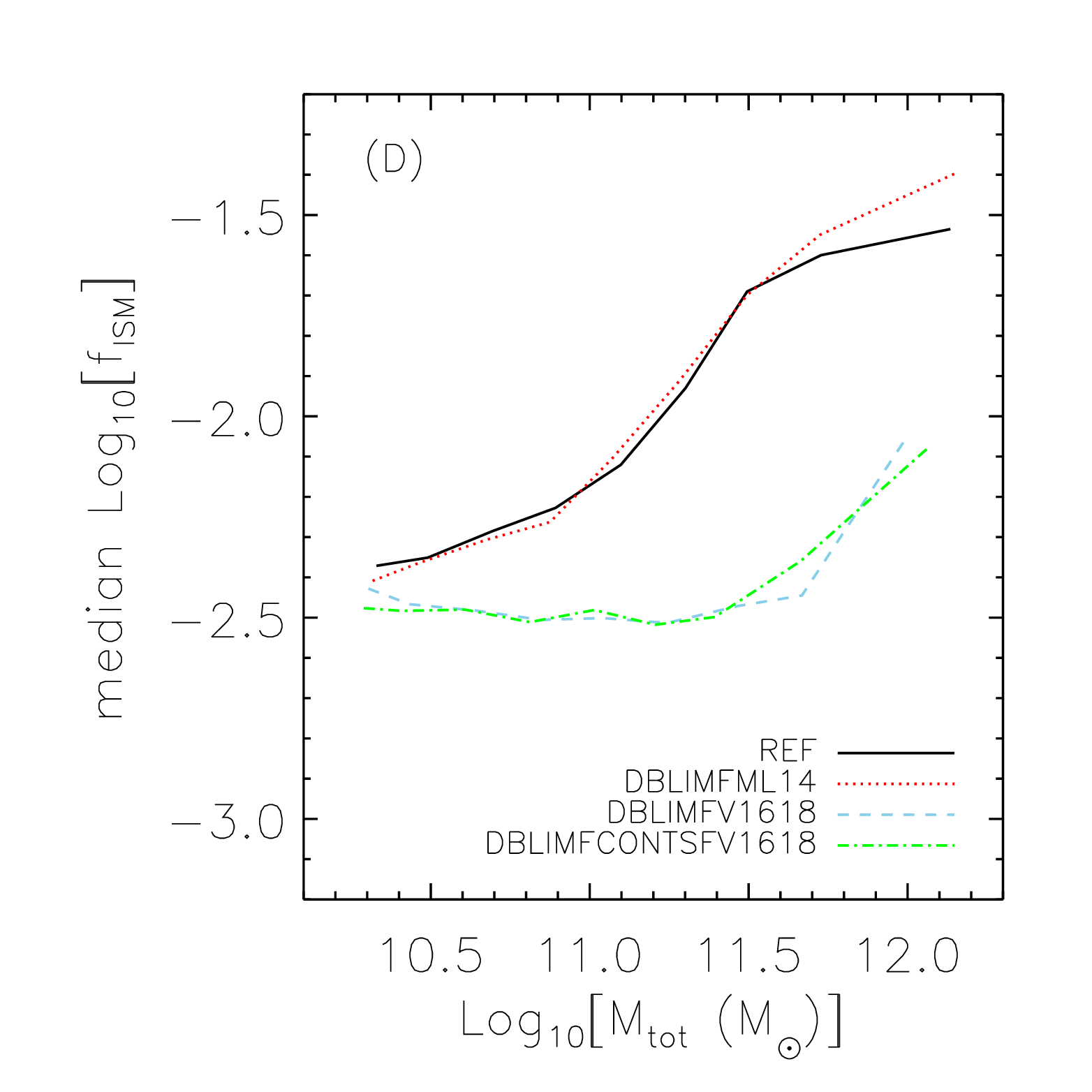}
\includegraphics[width=0.33\linewidth]{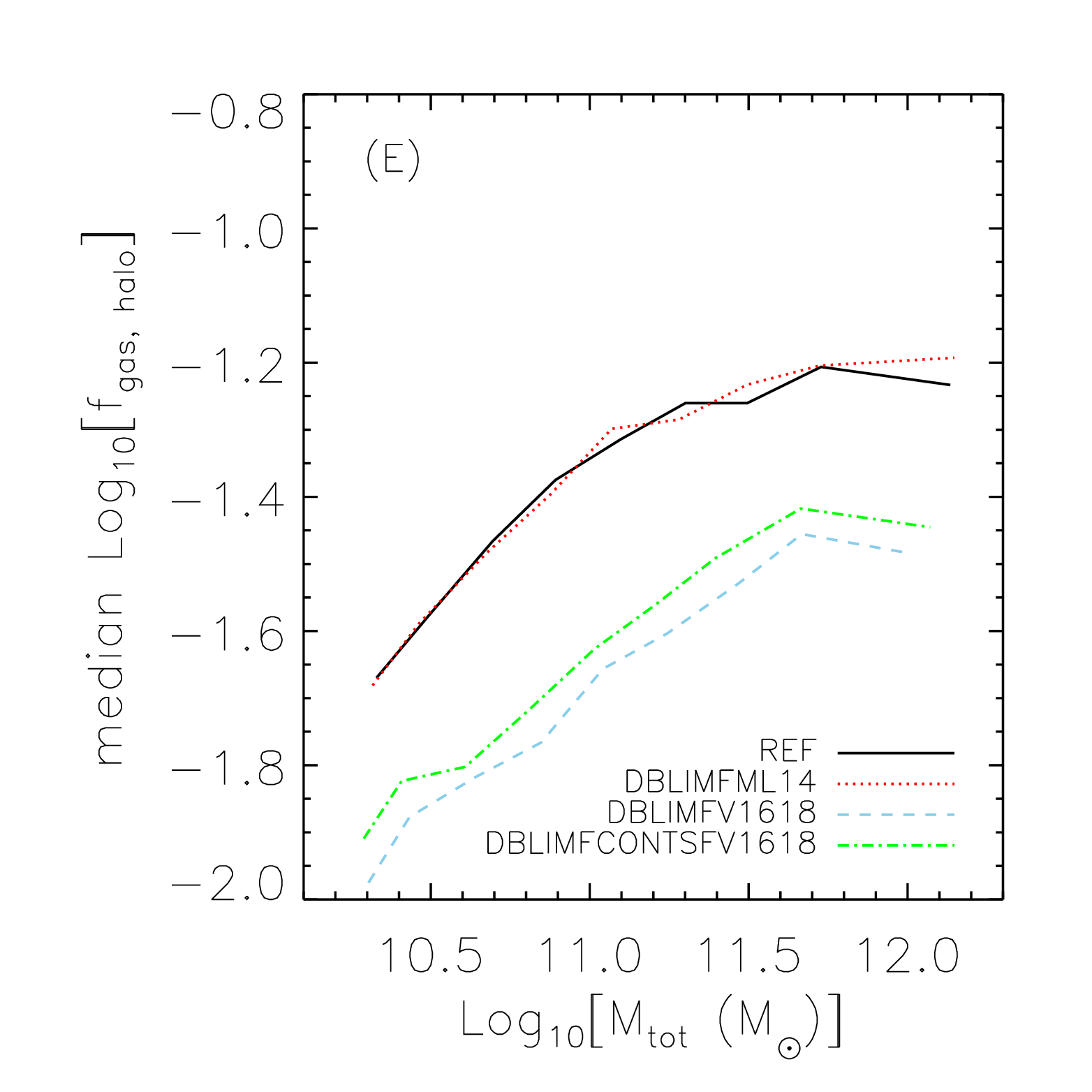}
\includegraphics[width=0.33\linewidth]{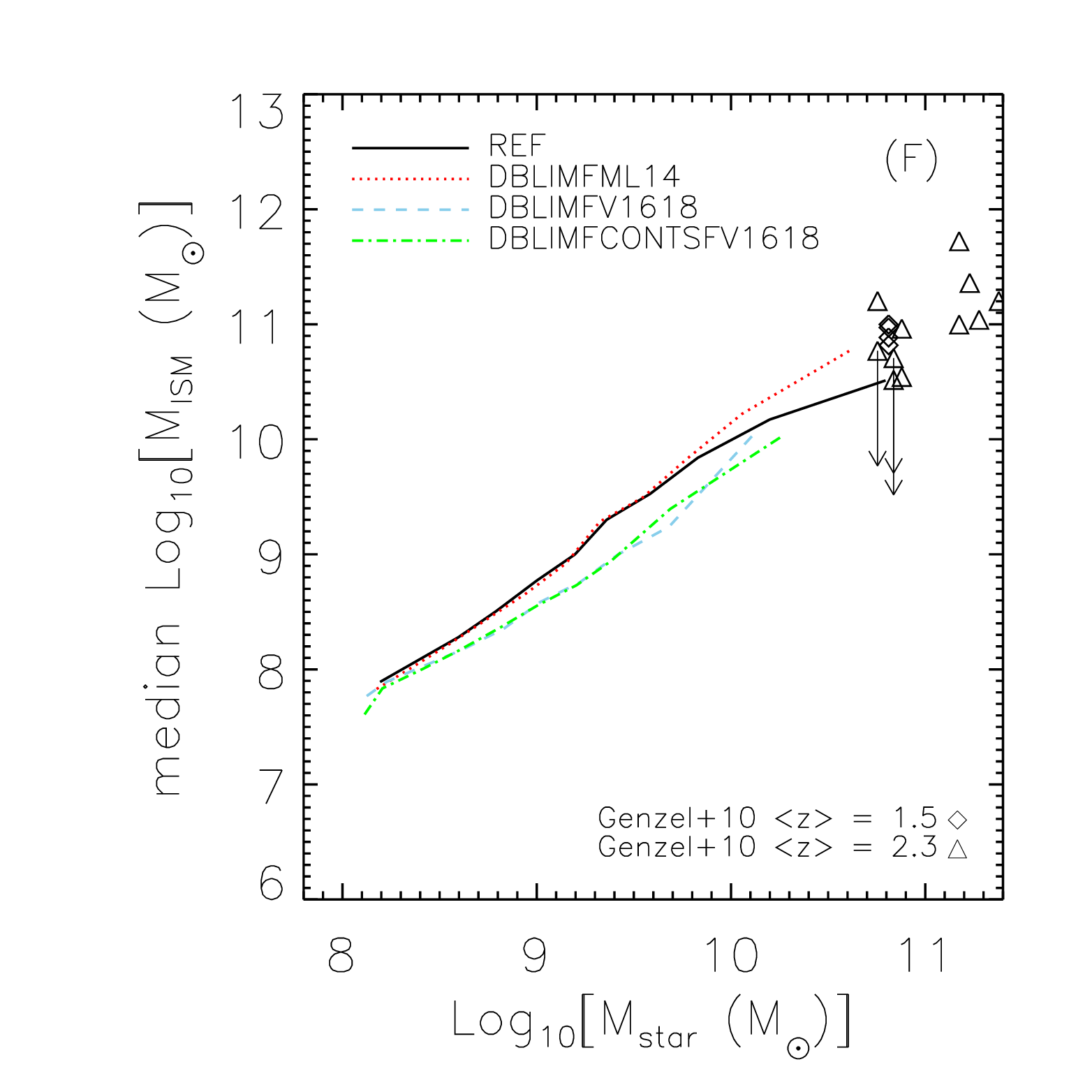} \\
\includegraphics[width=0.33\linewidth]{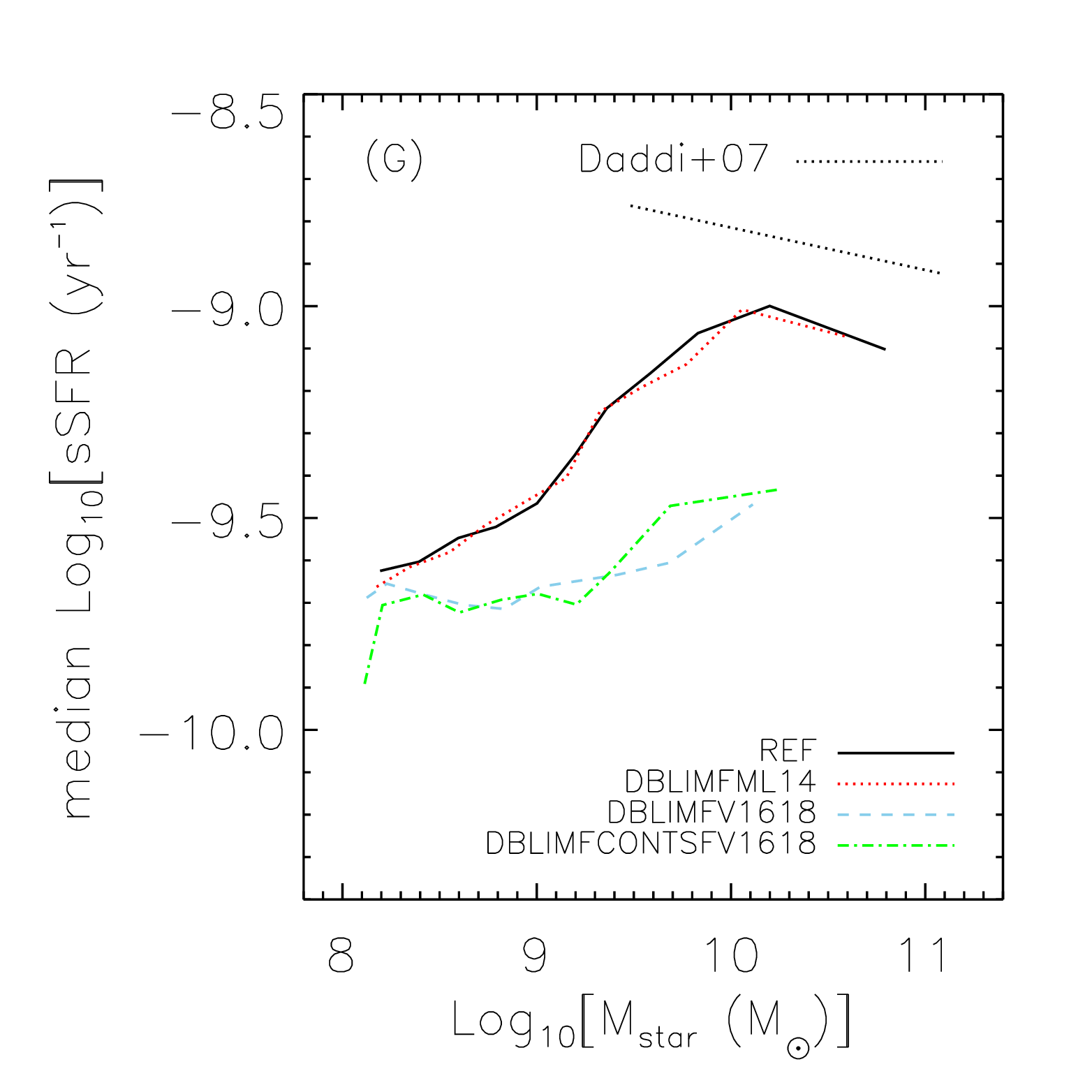}
\includegraphics[width=0.33\linewidth]{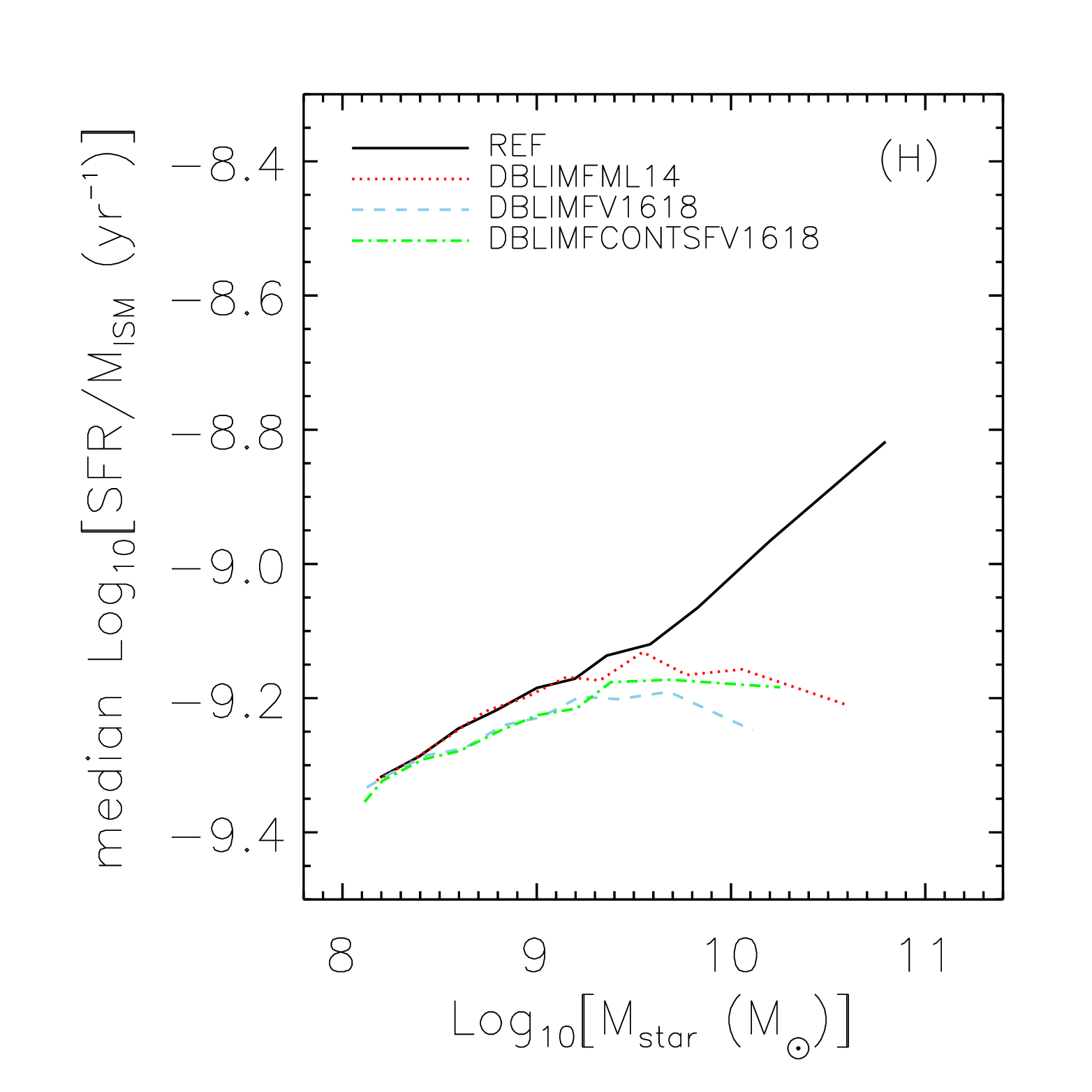}
\includegraphics[width=0.33\linewidth]{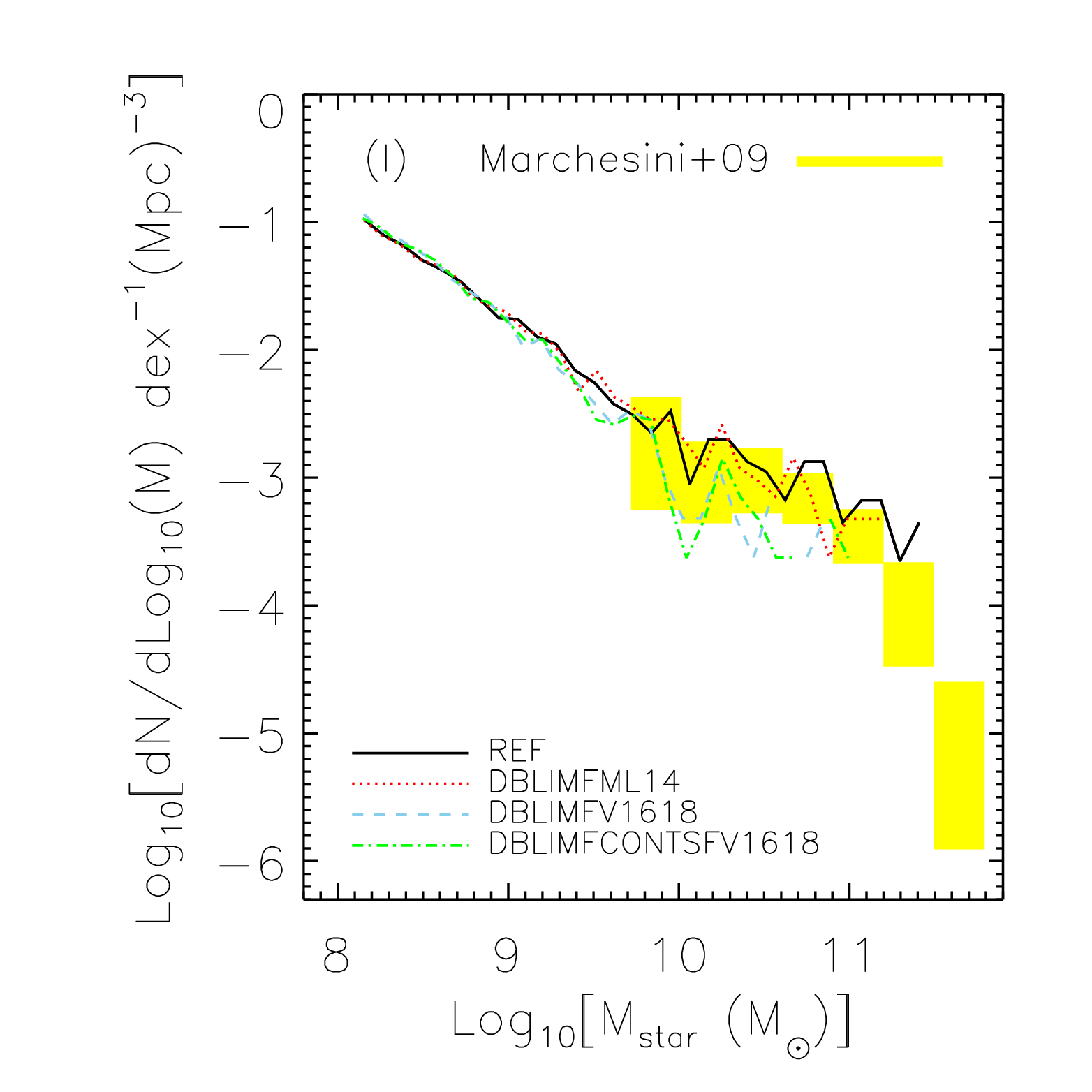} \\
\caption{As Fig.~\ref{fig:All_sims}, but showing only the set of simulations in which a top-heavy IMF is used for star formation at high pressure. The `\textit{REF}' simulation (without top-heavy IMF) is shown by the black, solid curve. The extra SN energy available from a top-heavy IMF can either go towards increasing the mass loading (`\textit{DBLIMFML14}'; red, dotted curve) or the initial velocity of the winds (`\textit{DBLIMFV1618}'; blue, dashed curve). In the presence of a sudden change in the IMF at some pressure, one can either allow the star formation rate or the rate of formation of massive stars (which is what is observed) to be a continuous function of the pressure (`\textit{DBLIMFCONSTSFV1618}'; green, dot-dashed curve).  Putting the extra SN energy into the mass-loading (`\textit{DBLIMFML14}') does not increase the level of suppression of the SFR, whereas increasing the wind velocity does allow more gas to escape from galaxies.} 
\label{fig:dblimf} 
\end{figure*}

\noindent Determining the stellar IMF represents a significant challenge outside of the solar neighbourhood.  The IMF determined locally is usually taken to be universal and is applied to all galaxies, at all redshifts.  There is, however, some evidence suggesting that the IMF may be \lq top-heavy\rq\, in extreme regions such as starbursts \citep[e.g.][]{baugh05,habergham10,weidner11}.  We performed a series of runs in which a different IMF is assumed in stars that formed in high-pressure environments.

In these simulations, stars form with an IMF $\d N/ \d M \propto M^{-1}$ (compared with $M^{-2.3}$ for the high-mass end of the Chabrier IMF) if the gas pressure exceeds $P/k = 2.0 \times 10^6$ cm$^{-3}$ K (evaluated at the resolution limit of the simulations). This process mimics the observation that stars that form in extreme environments (such as starbursts) appear to have an IMF that is flatter than the Chabrier IMF we assume in `\textit{REF}' \citep[e.g.][]{mccrady03, stolte05, maness07}.

For a top-heavy IMF, the available energy from SNe per unit stellar mass formed is higher by a factor of $\sim7$ than for the  Chabrier IMF. Numerically, we have some freedom in how this energy is distributed.  It can be used either to increase the wind mass loading or the wind velocity. We therefore ran two additional simulations with a top-heavy IMF at high pressures.  In the first simulation the wind velocity remains fixed at 600 km s$^{-1}$ (the `\textit{REF}' value), but the mass loading is increased to $\eta = 14$ (`\textit{DBLIMFML14}') for stars that form in high-pressure environments. In a second simulation, the mass loading is kept fixed at $\eta = 2$ (the `\textit{REF}' value) and the wind velocity is increased to $v_w = 1618$ km s$^{-1}$ (`\textit{DBLIMFV1618}') for stars that form at high pressures.

When changing the IMF suddenly as a function of the pressure, it is not immediately clear how to treat the star formation law. The Kennicutt-Schmidt law is inferred from observations that probe only massive stars. The actual SFR, therefore, depends on how many low-mass stars are formed per unit mass of massive stars. When changing the IMF, the star formation law can be changed in two ways:
\begin{enumerate}
\item From observations there is no indication of a discontinuity in the formation rate of massive stars with pressure. Although this is most likely the result of the IMF being a continuous function of SFR or pressure (if there is a relation at all), we nevertheless implemented a model that changes the normalization of the KS-law such that the formation of massive stars is continuous, resulting in a discontinuous SFR as a function of pressure (the KS-law normalization drops at the pressure above which the IMF is top-heavy). This is the procedure we follow in most of our simulations, and is the assumption made by both the simulations `\textit{DBLIMFML14}' and `\textit{DBLIMFV1618}'.
\item If the (total) SFR as a function of pressure is continuous, the formation of massive stars must be discontinuous, given that we assumed the IMF to change suddenly above some critical pressure. We run one additional model under this assumption, termed `\textit{DBLIMFCONTSFV1618}', which assumes that the total SFR as a function of pressure is continuous and the extra energy injected by a top-heavy IMF goes into increasing the wind velocity relative to the `\textit{REF}' simulation.
\end{enumerate}

When comparing simulations to observations, we do not correct the stellar mass of simulations with a double IMF. On average, only $\sim 10\%$ of the star particles in the simulation box formed with a top-heavy IMF (this depends slightly on resolution and weakly on whether the rate of formation of massive stars is a continuous function of the pressure). In \citet{owls} we showed that at late times, this correction can be significant, but that at $z=2$ the integrated SFR of the universe is the same, regardless of whether or not the SFRs of particles at pressures higher than the threshold pressure for the top-heavy IMF are corrected for another assumed IMF.  Also, we demonstrate in Paper II, by considering simulations in which the normalization and slope of the star-formation law are varied, that the form of the assumed SF law does not strongly influence the galaxy mass function or the SFR distributions in galaxies (it mainly affects the ISM fraction). At $z=2$ we therefore expect any differences between the runs that use a top-heavy IMF and the `\textit{REF}' simulation to be mostly due to the extra energy input from SN feedback and/or the increased rate of production of metals that results from a top-heavy IMF.
 
We show results from these simulations in Fig.~\ref{fig:dblimf}.  In all of the plots there are two \lq families\rq\, of curves. The simulations `\textit{REF}' (black, solid curves) and `\textit{DBLIMFML14}' are, in almost all plots, similar to one another, and the simulations `\textit{DBLIMFV1618}' (blue, dashed curves) and `\textit{DBLIMVCONTSFV1618}' (green, dot-dashed curves) are also very similar to one another.  The fact that `\textit{DBLIMFV1618}' and `\textit{DBLIMVCONTSFV1618}' are always very similar tells us that, if feedback is effective, whether or not we have a continuous SFR at the threshold pressure is only of minor importance when it comes to galaxy properties, and as such we will not consider `\textit{DBLIMVCONTSFV1618}' any further.

Next we compare `\textit{REF}' and `\textit{DBLIMFML14}'. As discussed in Sec.~\ref{sec:constwindE}, SNe are only capable of driving winds out of a galaxy when they are launched at sufficiently high velocities.  Therefore, `\textit{DBLIMFML14}' behaves similarly to `\textit{REF}' because the two models are capable of driving winds out of the same objects.  It is apparent that `\textit{DBLIMFML14}' does form fewer stars than `\textit{REF}' at high halo masses ($M_{\rm tot}\ga 10^{11.5}\,$\msun; panels A and B), even though the halo baryon fractions of the two simulations are very similar (panels C and E). That the two models differ only at high halo masses can be explained by the fact that the pressure in the ISM increases with the depth of the potential. Hence, only in massive haloes does a significant fraction of the ISM have pressures above the critical value above which we assume a top-heavy IMF.

The reason why `\textit{DBLIMFML14}' has lower stellar mass fraction in massive haloes than `\textit{REF}' can be seen by comparing the gas consumption timescale (panel H) and the mass of gas in the ISM (panels D and F) for the same two simulations.  Although there is significantly more gas in the ISM in `\textit{DBLIMFML14}' than in `\textit{REF}', the gas consumption timescale is much longer. This may reflect the drop in the normalisation of the KS-law in high-pressure gas in `\textit{DBLIMFML14}'. Because the gas consumption time scale becomes extremely long in the high-pressure gas ($\sim 10^{10}\,{\rm yr}$) and because the initial wind velocity is too low to be effective, self-regulation may be difficult to achieve.

Finally, we compare `\textit{REF}' and the top-heavy IMF simulation in which the initial wind velocity is increased (`\textit{DBLIMFV1618}').  It is immediately obvious that the effect of the top-heavy IMF is to strongly suppress star formation (panels A, B, F, G and H) as well as to effectively remove gas from galaxies (panel D) and from their host haloes (panels C and E). The effects increase with the mass, because the ISM pressure, and thus the fraction of stars forming with a top-heavy IMF, increase with the depth of the potential well.  The effect of the top-heavy IMF is very similar to increasing the wind velocity (see \S\ref{sec:constwindE}). In both simulations the SN-driven winds are able to drive out gas from all but the most massive objects, but the stellar fractions are lower for `\textit{DBLIMFV1618}' than for `\textit{WML1V848}'.  This is a result of the increase in the available SN energy per unit stellar mass, so a smaller star formation rate is required to drive winds from a galaxy and to effectively regulate star formation. Finally, we note that the simulations that include strong feedback from a top-heavy IMF suppress the amplitude of the galaxy stellar mass function below the observed points (panel (I)).  This behaviour is seen in the other models that contain very strong feedback (e.g.\ `\textit{WPOT}' and `\textit{WVCIRC}' in Sec.~\ref{sec:wmom} and `\textit{AGN}' in Sec.~\ref{sec:agn}).

We conclude that the effect of a top-heavy IMF at high pressures mainly reflects the increased efficiency of winds driven by massive stars. The extra energy available for driving winds is more important than the increase in the metal yields associated with a top-heavy IMF.

\subsection{`Momentum-driven' wind models} \label{sec:wmom}

\begin{figure*}
\centering
\includegraphics[width=0.33\linewidth]{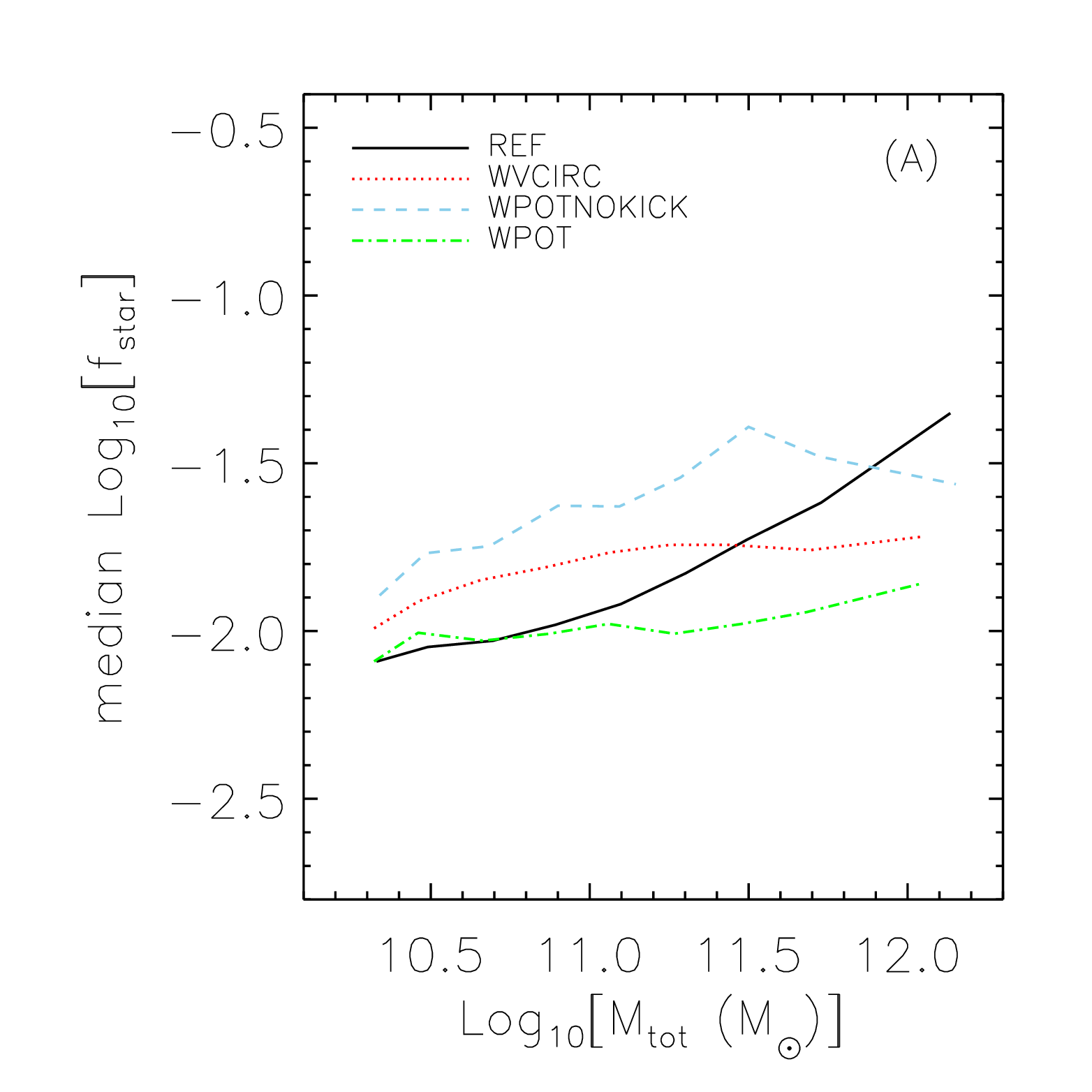}
\includegraphics[width=0.33\linewidth]{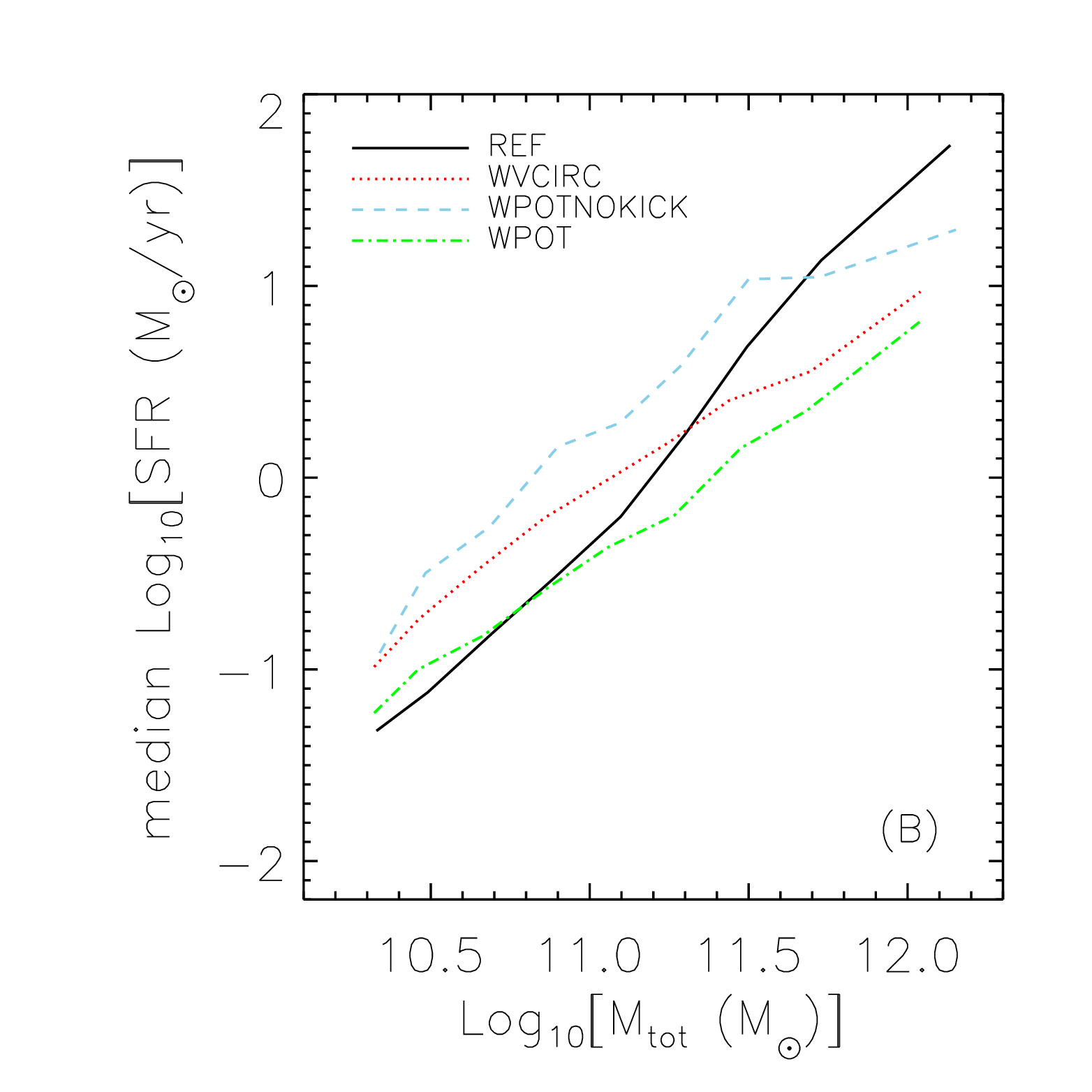}
\includegraphics[width=0.33\linewidth]{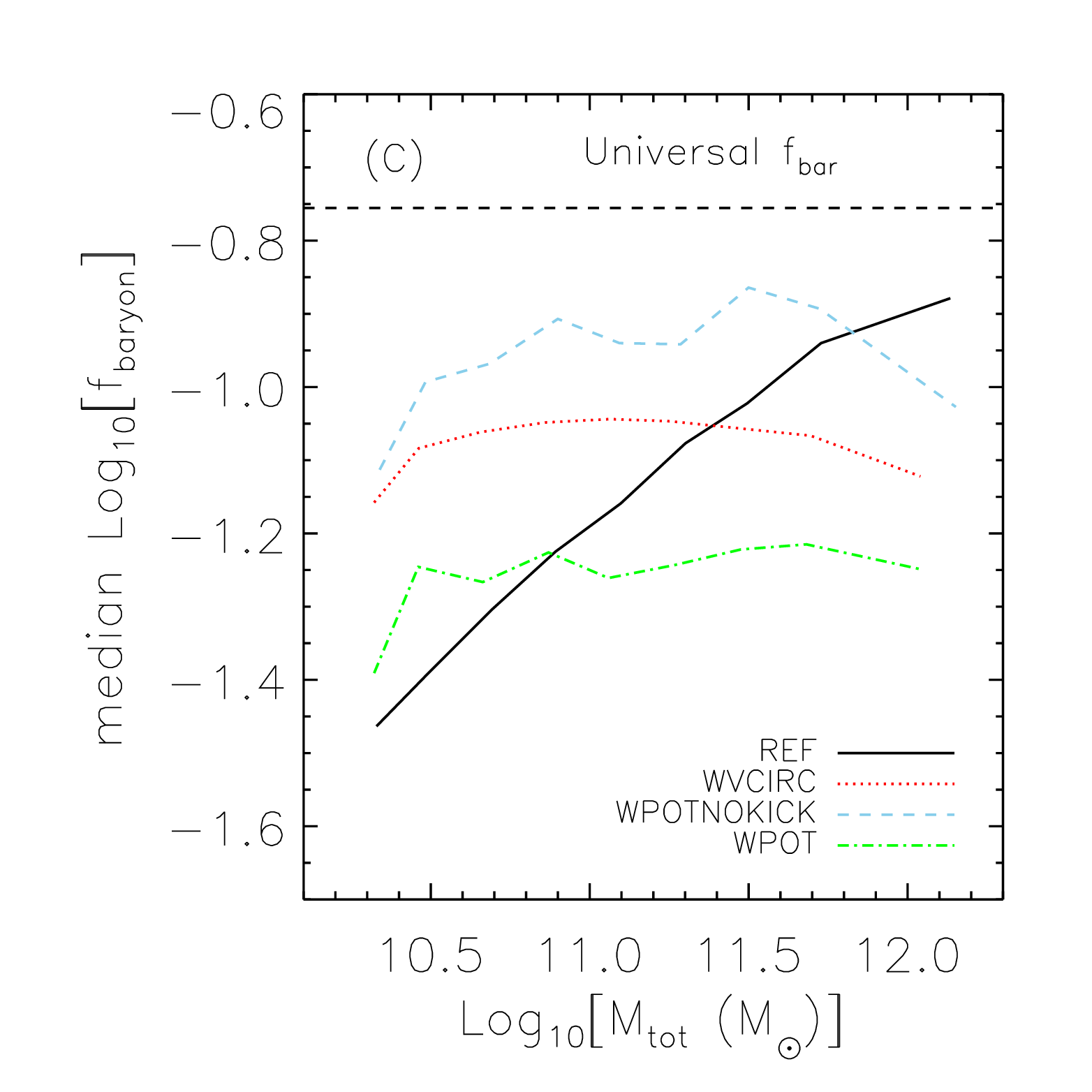} \\
\includegraphics[width=0.33\linewidth]{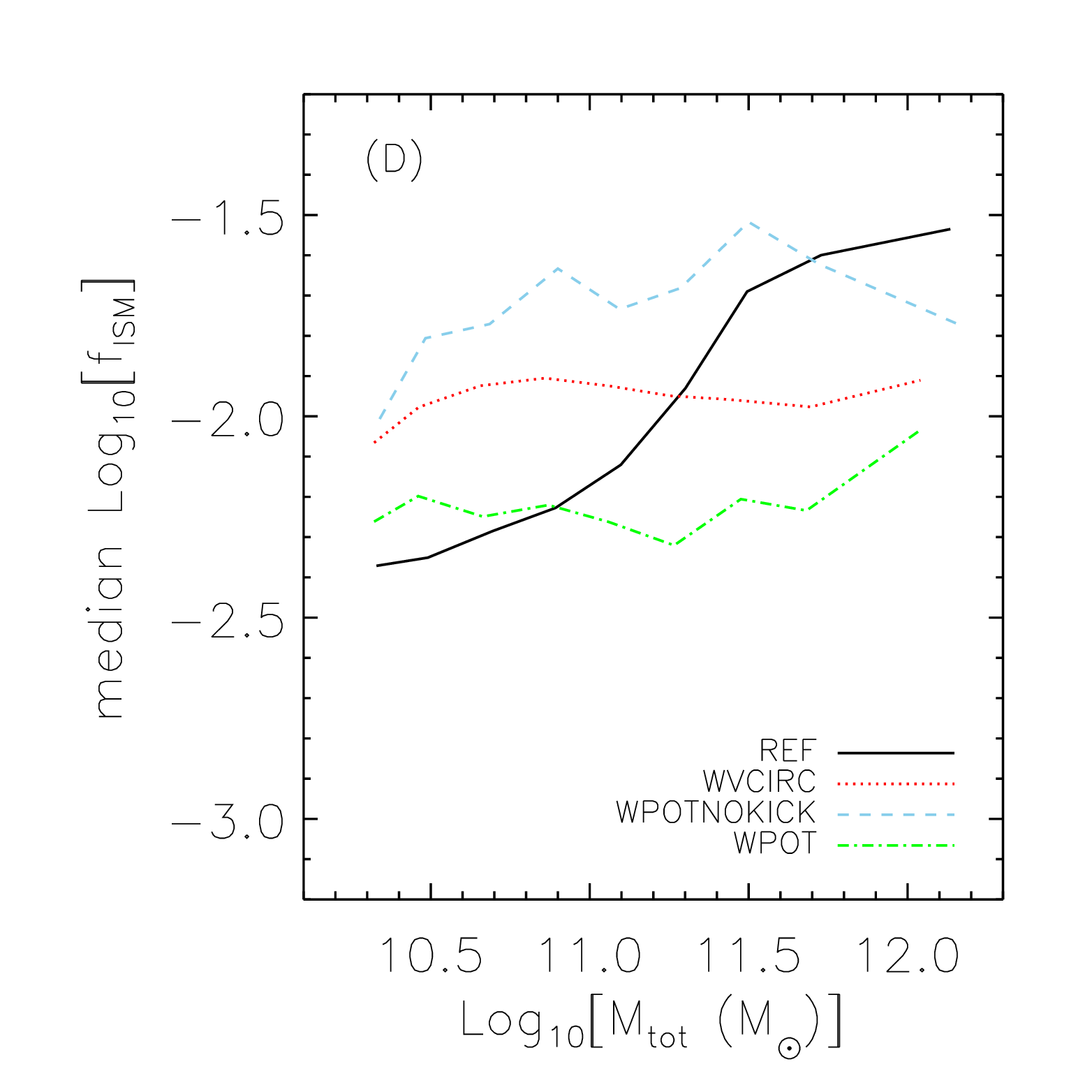}
\includegraphics[width=0.33\linewidth]{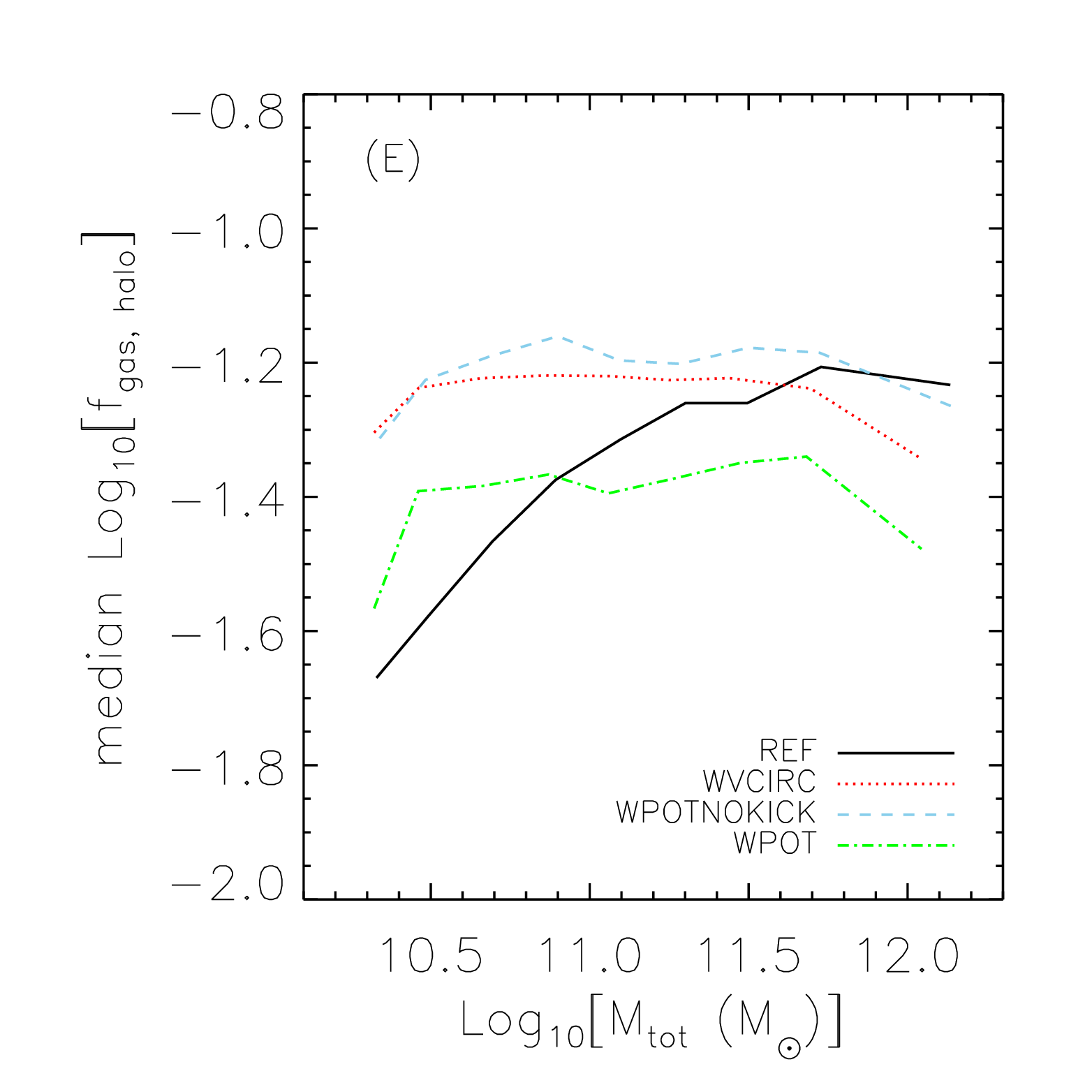}
\includegraphics[width=0.33\linewidth]{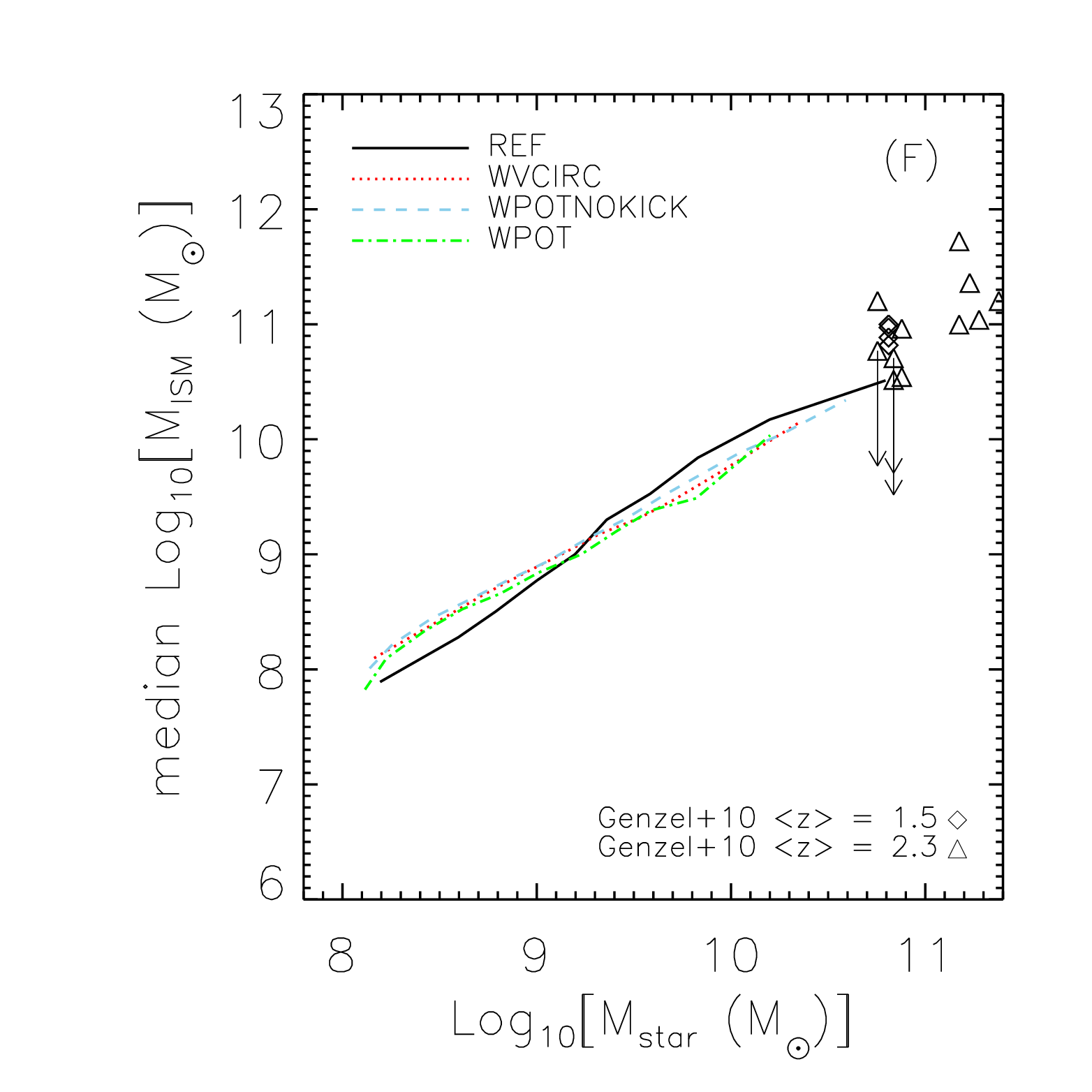} \\
\includegraphics[width=0.33\linewidth]{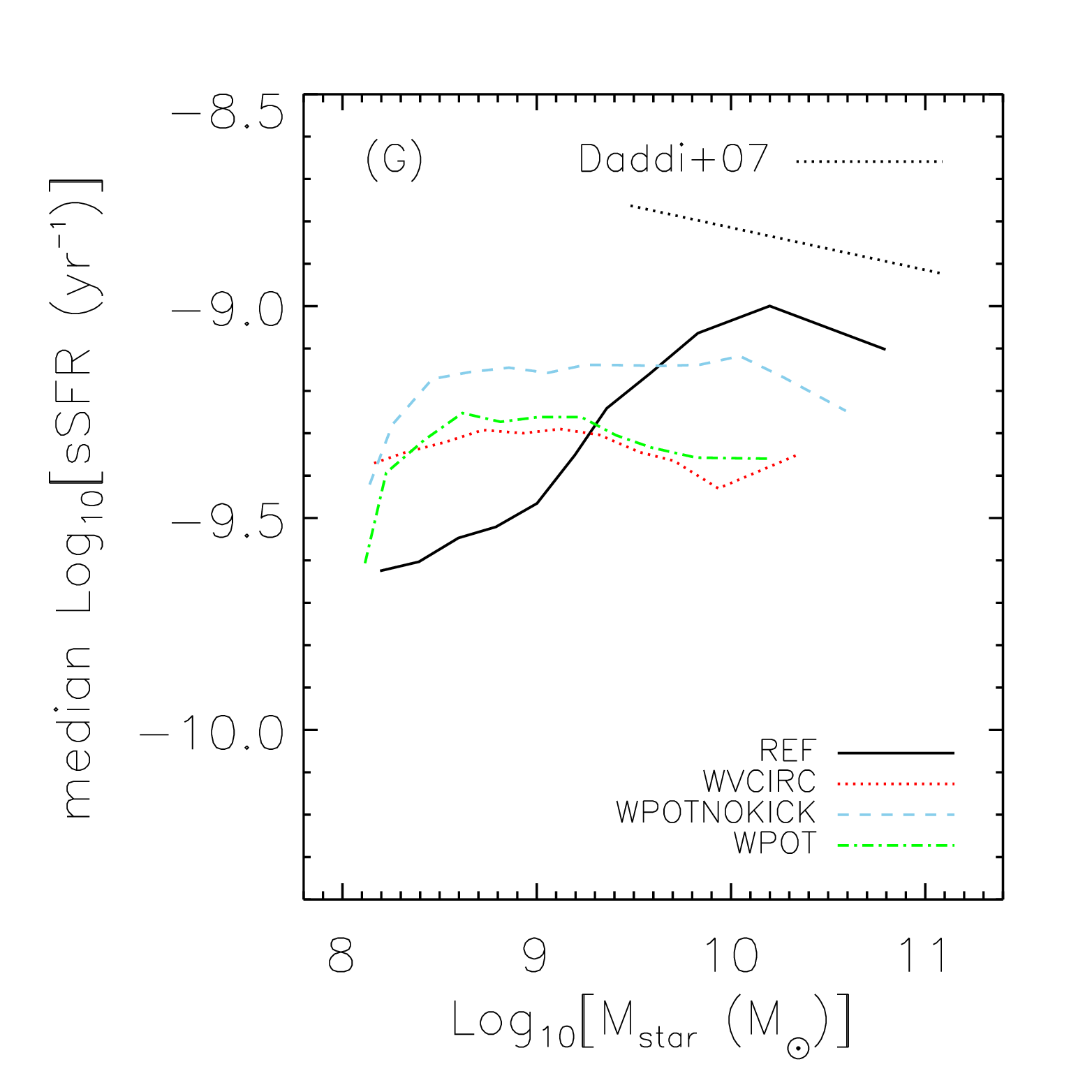}
\includegraphics[width=0.33\linewidth]{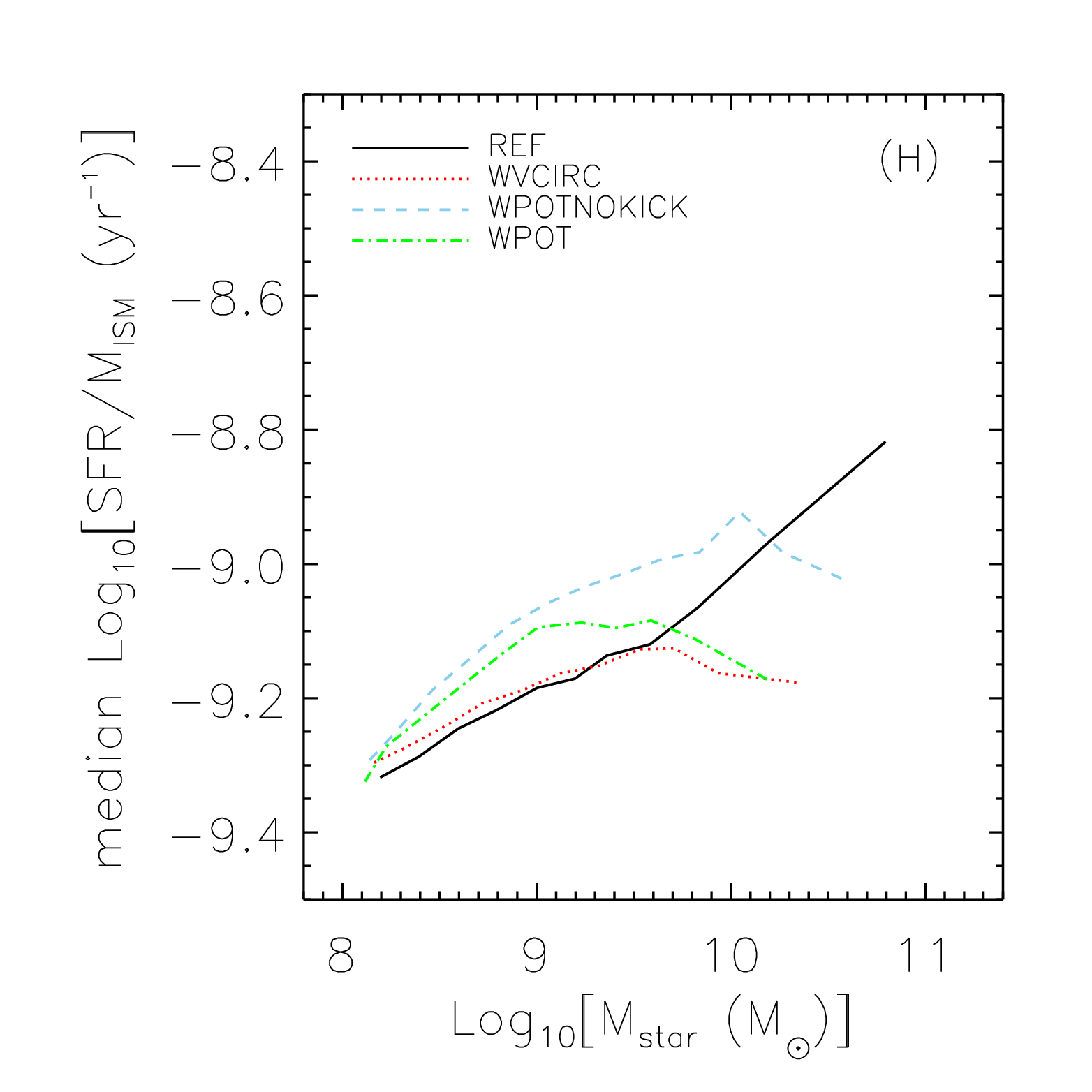}
\includegraphics[width=0.33\linewidth]{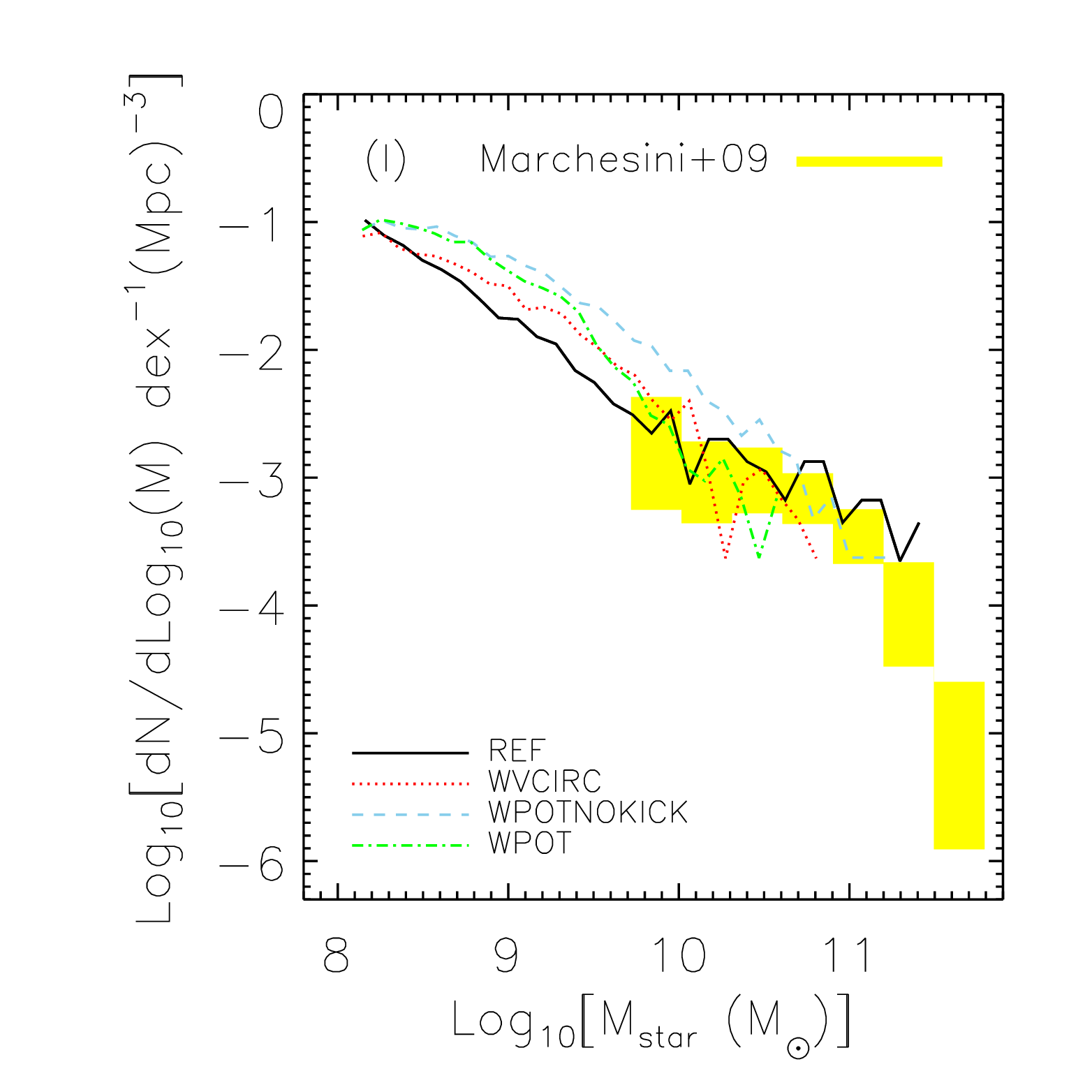} \\
\caption{As Fig.~\ref{fig:All_sims}, but showing only the reference simulation (`\textit{REF}'; black, solid curve) and the set of simulations in which \lq momentum-driven wind\rq\, scalings are employed. In \textit{`WVCIRC'} (red, dotted curve), the initial wind velocity scales with on the circular velocity of the halo the wind is launched from, while in \textit{`WPOTNOKICK'} and \textit{`WPOT'} the wind velocity is set by the the local gravitational potential (without and with an extra kick respectively, shown by the blue, dashed and green, dot-dashed curves respectively). In these models the energy injected into the wind per unit stellar mass is not constant and generally exceeds the energy in the reference simulation, as is evident from the increased suppression of star formation.} 
\label{fig:wmom} 
\end{figure*}

\noindent Galactic winds can be driven by radiation pressure on dust grains in the wind, which drag along the gas \citep[e.g.][]{murray05}. Here the driving force of the wind is the radiation pressure which injects momentum into the outflow. If the cooling time in the shock-heated gas is sufficiently short, then the wind will conserve its momentum and the wind is said to be `momentum-driven'. Although cosmological, hydrodynamical simulations have not yet tried to model this process directly by including radiation transport, simulations have been run that include the same type of kinetic feedback as used for SN-driven winds, but in which the initial wind velocity varies with local physical conditions or galaxy properties, and the initial mass loading varies as $\eta\propto v_{\rm w}^{-1}$, such that the rate with which momentum is injected per unit stellar mass, $\propto \eta v_{\rm w}$, is constant. This scaling should be contrasted with that used for the `constant energy' winds considered earlier, for which $\eta \propto v_{\rm w}^{-2}$.

We implemented several `momentum-driven wind' models that are similar to those used by \citet{oppenheimerdave06,oppenheimerdave08}. Here, the wind parameters vary with either the local potential (`\textit{WPOT}' and `\textit{WPOTNOKICK}'; \citealt{oppenheimerdave06}) or the circular velocity, $v_c = \sqrt{GM_{vir}/R_{vir}}$ of the halo from which the wind is launched (`\textit{WVCIRC}'; \citealt{oppenheimerdave08}). 
In `\textit{WVCIRC}' the wind velocity and mass loading are given by $v_w = (3 + n) v_c /\sqrt{2}$ and $\eta = \frac1{\sqrt{2}} \times (v_{c}/v_{\textrm{\scriptsize crit}})^{-1}$, where $n$ and $v_{\textrm{\scriptsize crit}}$ are parameters, set to 2 and 150 km s$^{-1}$, respectively. In `\textit{WPOTNOKICK}', the wind velocity is given by $v_w = 3\sigma$, where $\sigma$ is the velocity dispersion, calculated from the gravitational potential: $\sigma = \sqrt{-\Phi/2}$, and $\eta=150~{\rm km}\,{\rm s}^{-1}/\sigma$. In `\textit{WPOT}' an extra velocity kick of $2\times \sigma$ is added to each event. These models are not directly comparable to those of \citet{oppenheimerdave06,oppenheimerdave08,oppenheimerdave09} because these authors used different parameter values and different models in different papers (see \S4.9 of \citealt{owls}). Moreover, contrary to Oppenheimer \& Dav\'e, we do not decouple the wind particles from the hydrodynamics, which likely makes our implementations less efficient. As was shown by \citet{dallavecchiaschaye08} and \citet{owls}, and as we will show in Sec.~\ref{sec:whydrodec}, temporarily turning off pressure forces for wind particles makes the winds more efficient and has a large effect on galaxy properties.

One important difference between the models discussed here and those
in Sec.~\ref{sec:constwindE} is that here the total amount of energy
put into the wind per unit stellar mass formed is not constant, but
depends on the mass of the halo in which it takes place ($E \propto
M_{\rm tot}^{1/3}$ in `\textit{WVCIRC}'), and we caution that the total energy
used in feedback exceeds the total available energy from SNe for the
most massive galaxies\footnote{This occurs for $M_{\rm tot} \gtrsim 10^{12.5}$
\msun, although the exact mass of equality is redshift dependent, due
to the redshift dependence of the virial radius.}. Most worryingly, the amount of
momentum in the winds also exceeds the momentum available from
SNe and radiation, assuming that every photon is absorbed once
to drive the outflow (so no boost from optical thickness in the IR),
by a factor of about 7 (\citealt{haasphd}, see also the discussion in \citealt{dave11}). 
It should therefore be kept in mind that these models may inject an unphysically large amount of momentum. Inspection of Fig.~\ref{fig:prettypics} reveals that these wind prescriptions are capable of almost completely disrupting the galaxy disk. 

We compare the effects of the different momentum-driven wind models in Fig.~\ref{fig:wmom}. All the simulations studied in this section predict relatively flat relations between $M_{\rm tot}$ and $f_{\rm star}$ (panel A), yielding a flatter galaxy mass function at the low-mass end (panel I). None of the models show a characteristic halo mass above which feedback becomes ineffective (panels B -- F), indicating that the feedback remains effective at all masses.  In all of these simulations, the combination of a large mass-loading in low-mass haloes and a high wind-velocity in high-mass haloes means that feedback from star formation is always able to drive strong winds and suppress SF.

\subsection{Hydrodynamically decoupled winds} \label{sec:whydrodec}
\begin{figure*}
\centering
\includegraphics[width=0.33\linewidth]{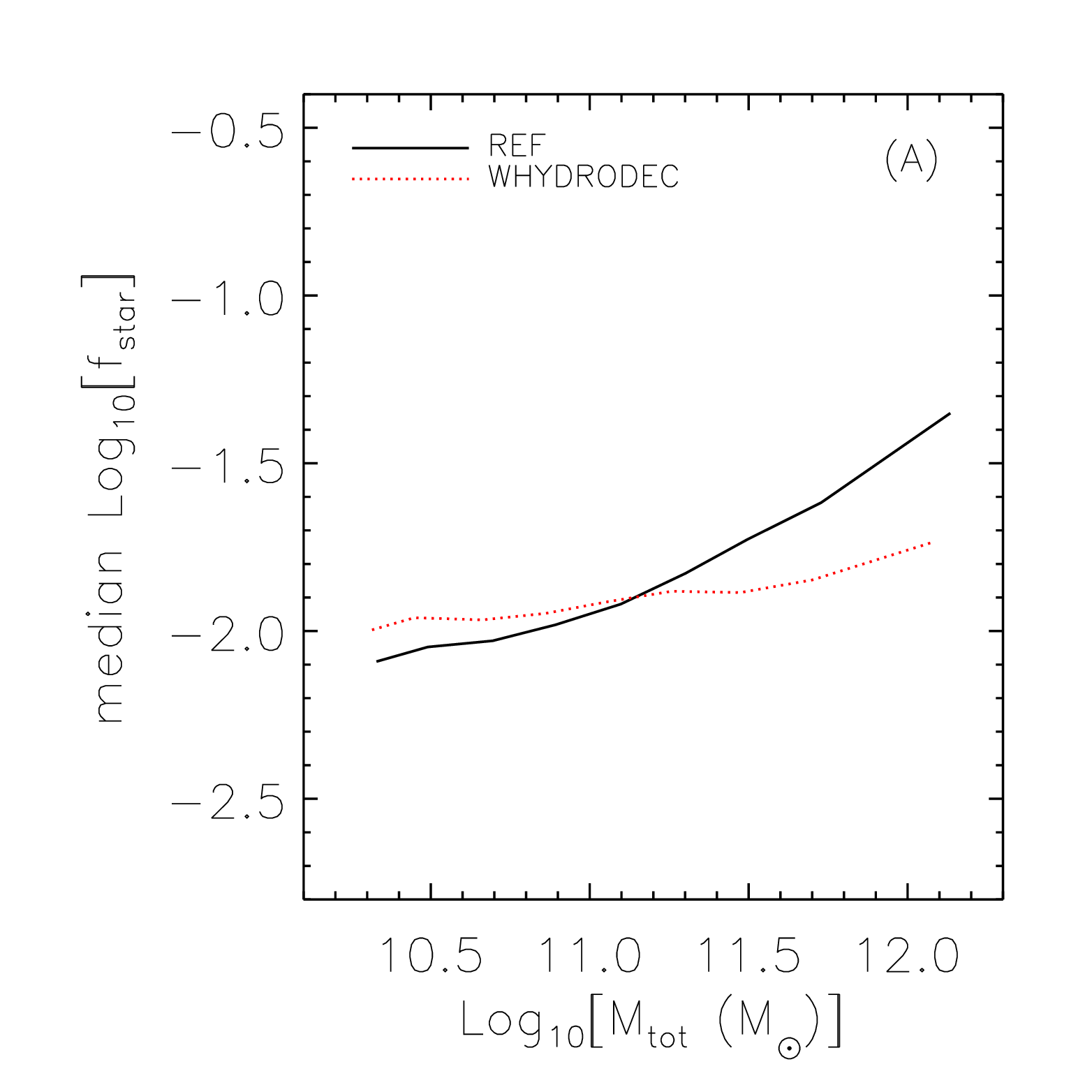}
\includegraphics[width=0.33\linewidth]{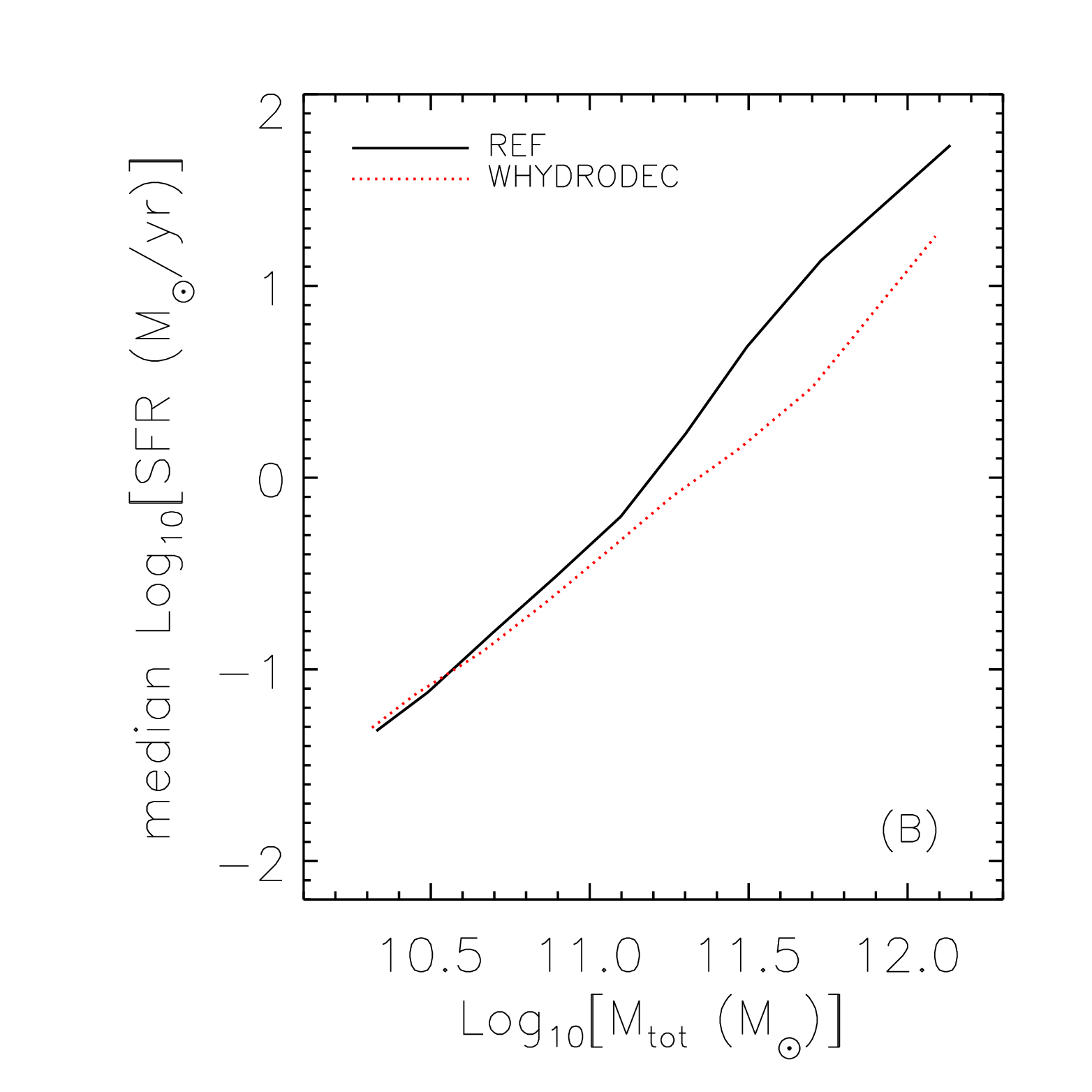}
\includegraphics[width=0.33\linewidth]{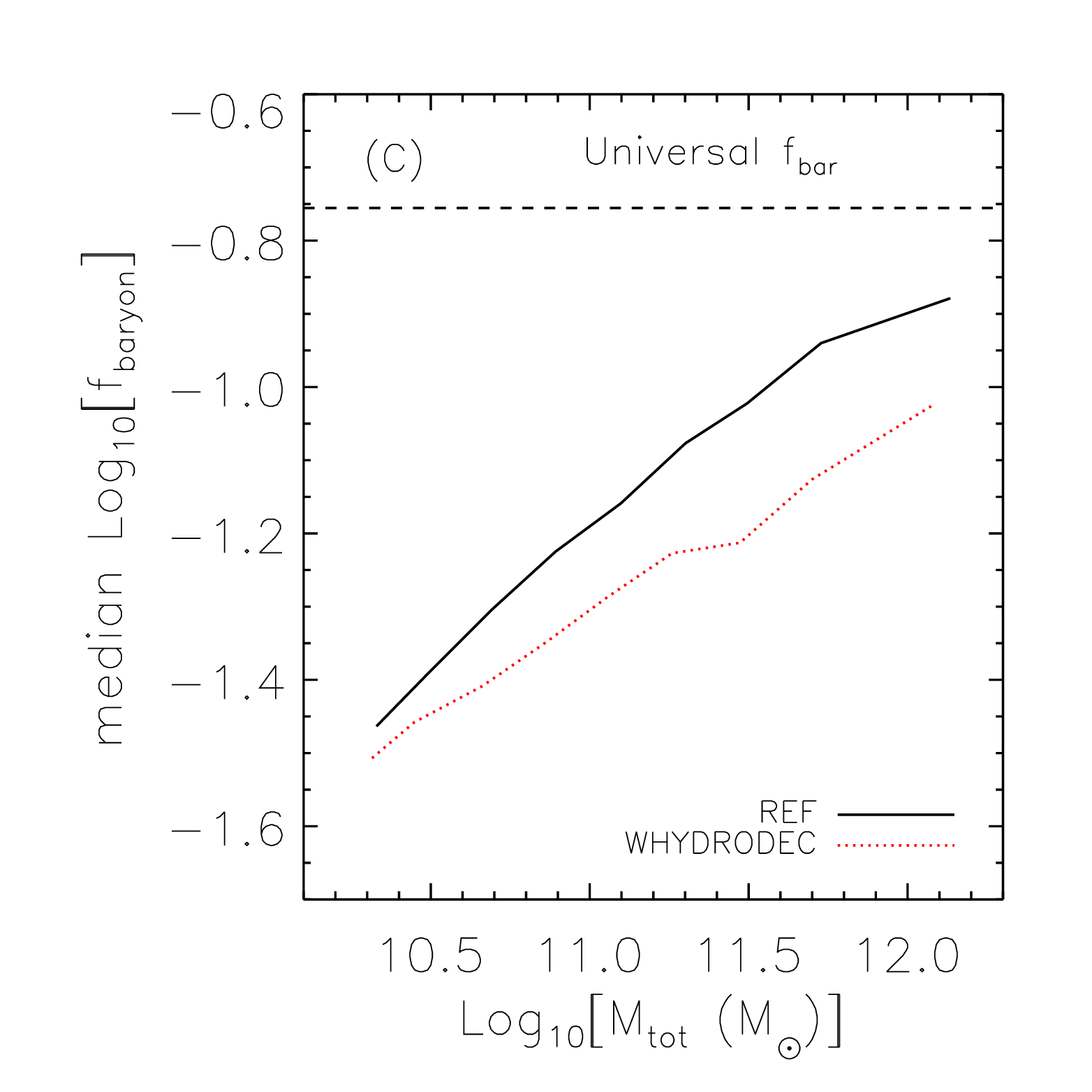} \\
\includegraphics[width=0.33\linewidth]{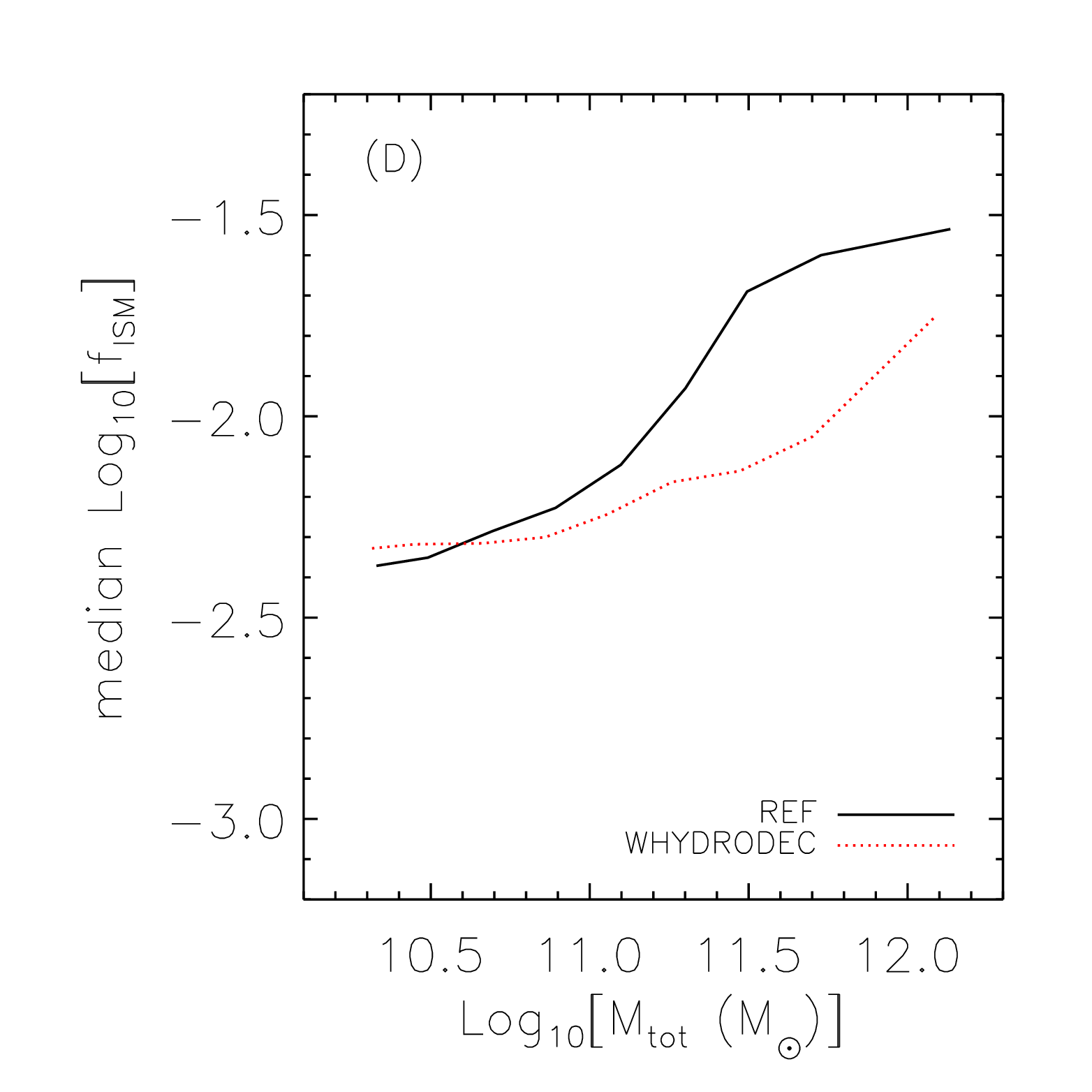}
\includegraphics[width=0.33\linewidth]{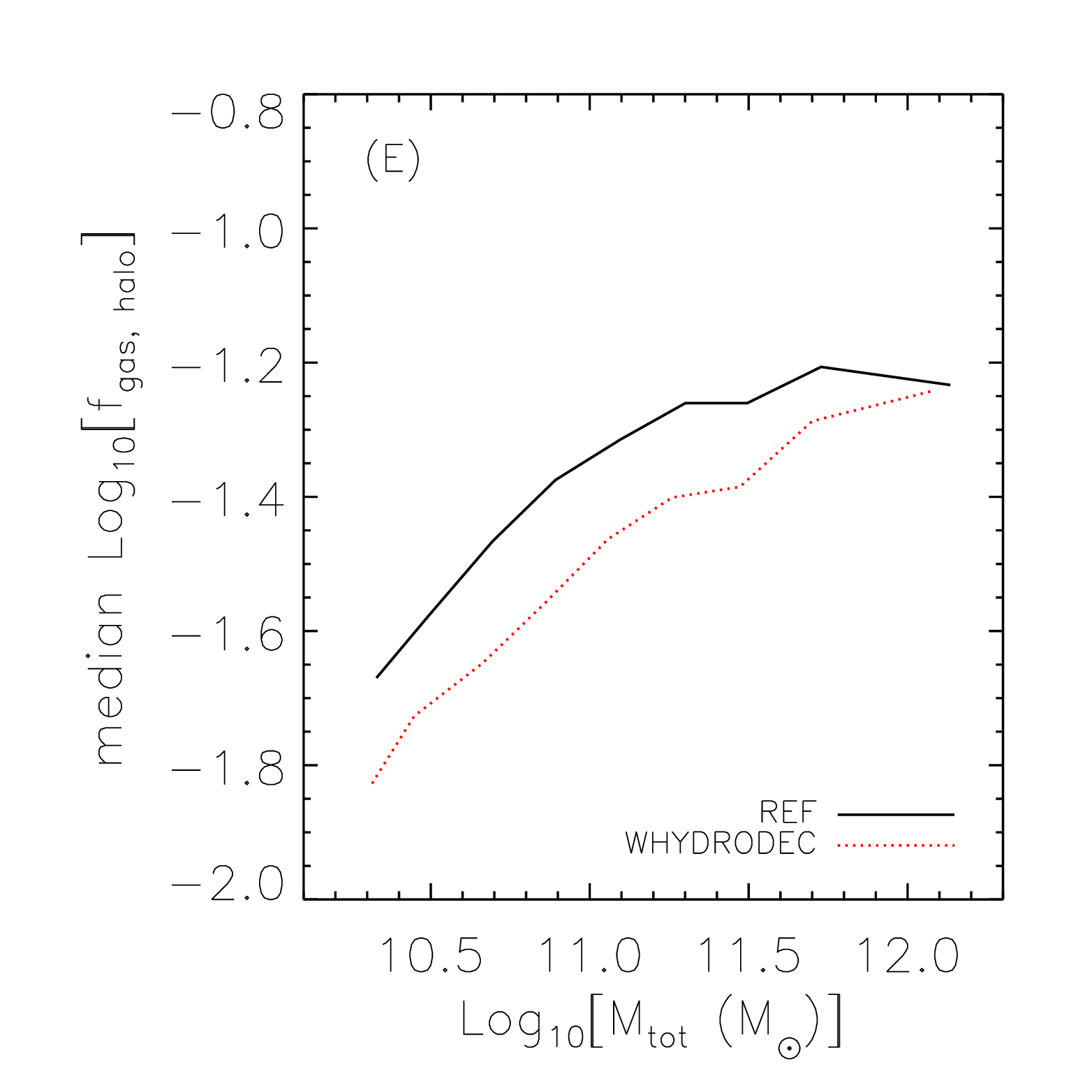}
\includegraphics[width=0.33\linewidth]{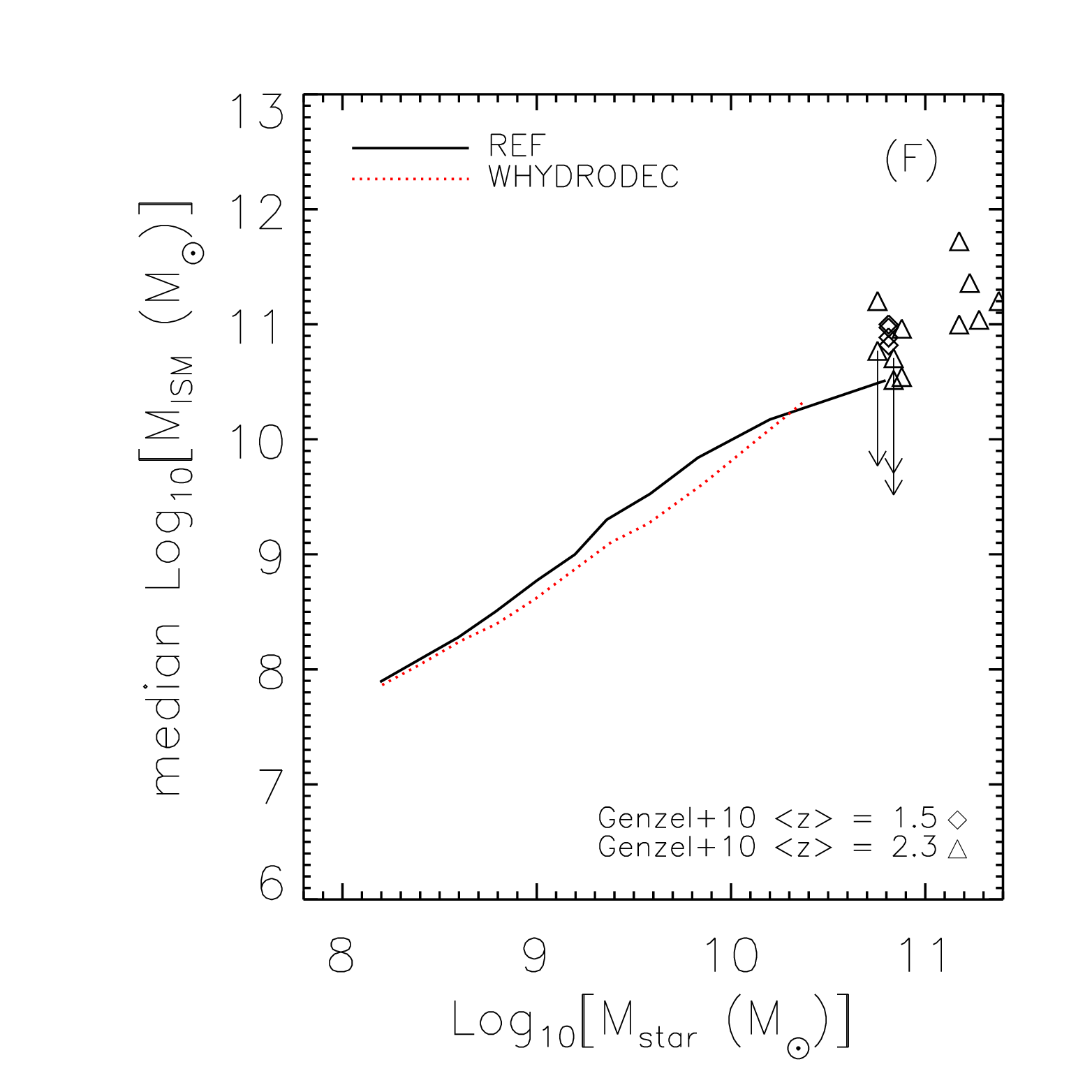} \\
\includegraphics[width=0.33\linewidth]{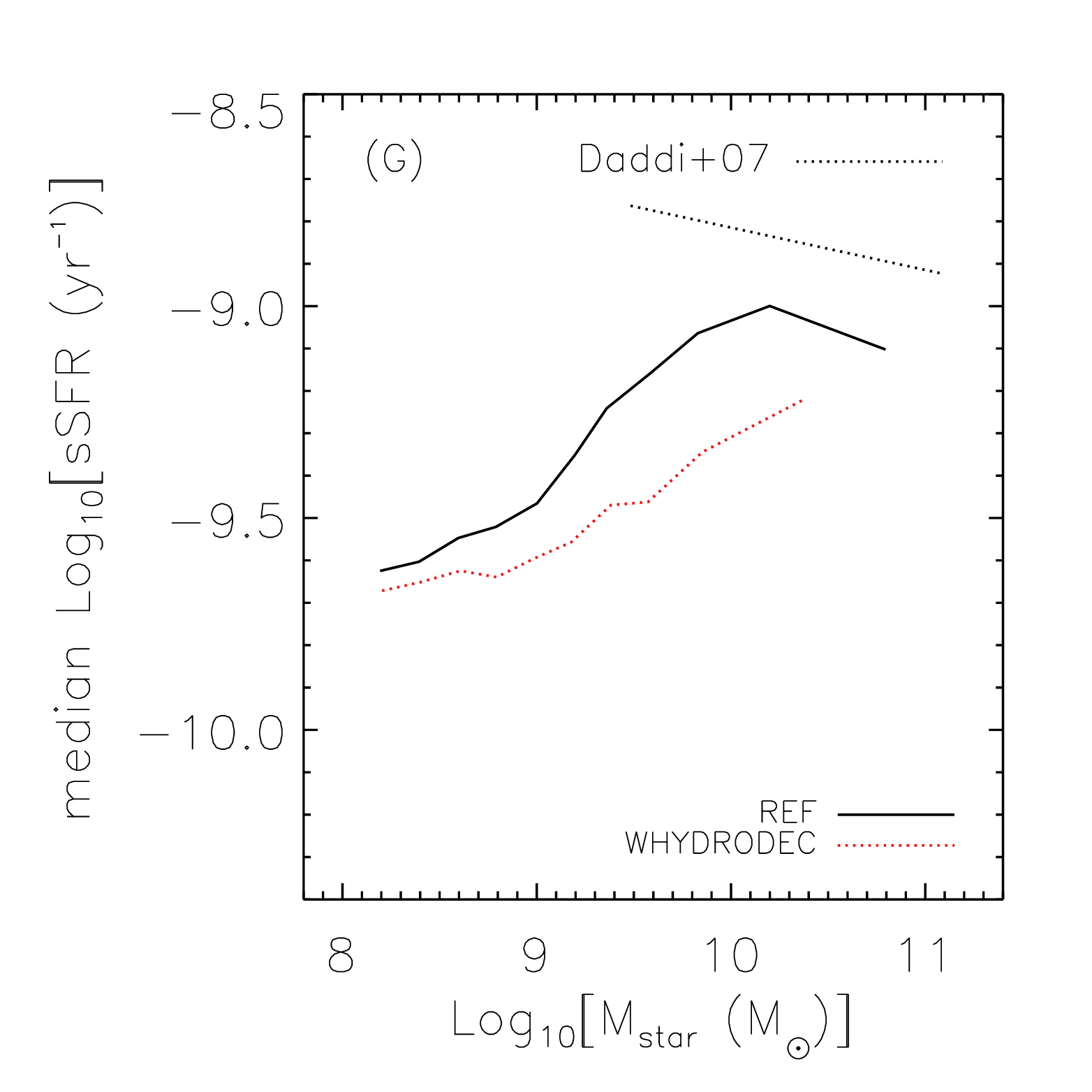}
\includegraphics[width=0.33\linewidth]{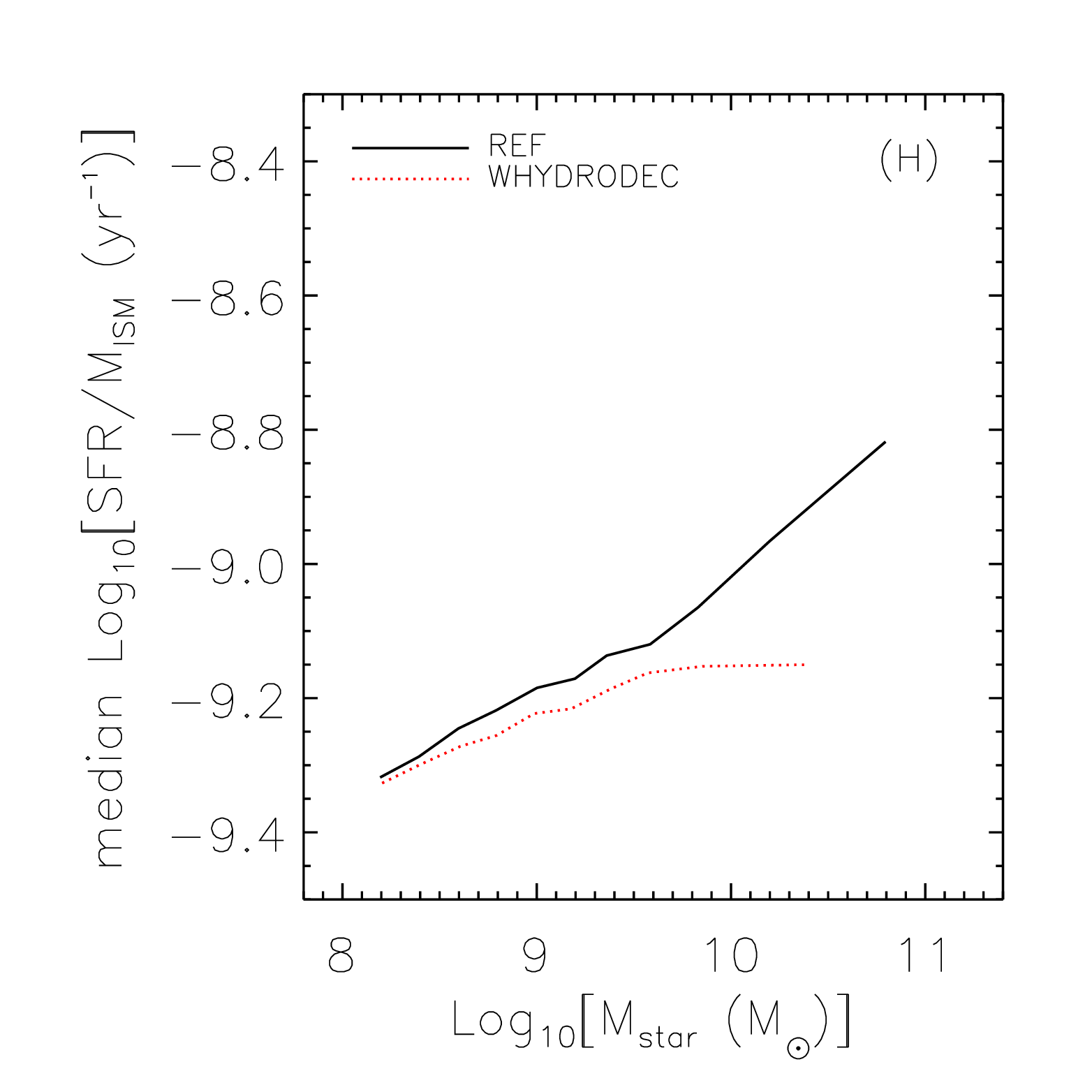}
\includegraphics[width=0.33\linewidth]{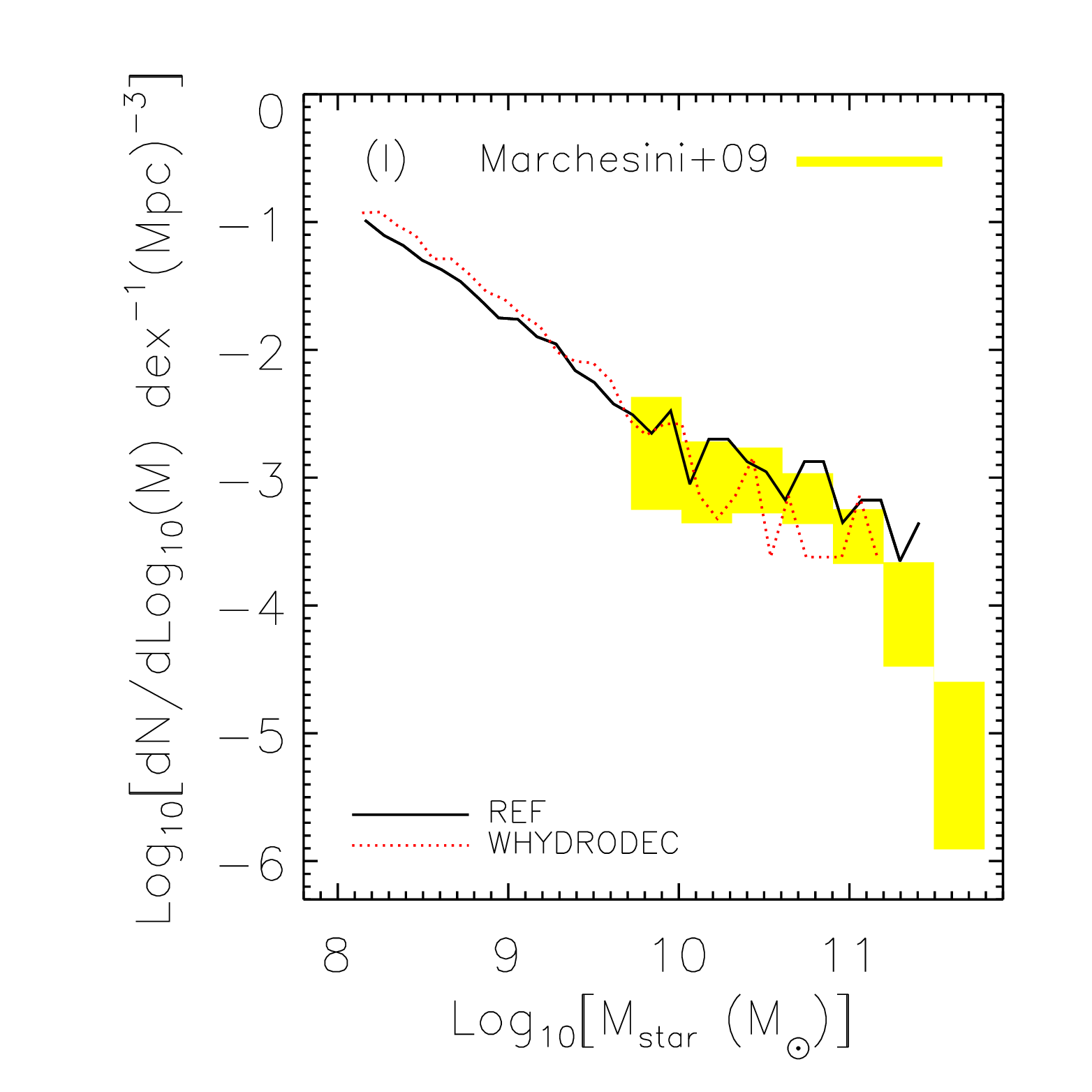} \\
\caption{As Fig.~\ref{fig:All_sims}, but showing only the `\textit{REF}' (black, solid curve) and the simulation in which the wind particles are temporarily decoupled from the hydrodynamics (`\textit{WHYDRODEC}'; red, dotted curve).  Despite using the same wind parameters in both of these simulations, artificially decoupling the SN-driven winds from the galaxies makes the winds very efficient at suppressing star formation.} 
\label{fig:whydrodec} 
\end{figure*}

\noindent 
Many SPH simulations that use kinetic SN feedback employ a method known as ’decoupling’ that allows SN-driven winds to effectively escape galaxies before they begin to interact with the halo gas. In these simulations, wind particles, once launched, are temporarily decoupled from the hydrodynamics until they escape the ISM. During decoupling a gas particle experiences gravity, but feels no hydrodynamic drag. Decoupling prevents shock-heating and hence the radiative losses which may otherwise quench the outflow in high-pressure gas \citep{dallavecchiaschaye12}, and it prevents the wind from entraining any surrounding ISM. To some extent, decoupling wind particles from the hydrodynamics mimics the existence of unresolved `chimneys' of low density in the ISM, through which winds can easily escape without entraining much other gas. However, decoupling does not account for the energy required to create such chimneys. For a detailed study of the effect of decoupling for the case of isolated disk galaxy simulations, see \citet{dallavecchiaschaye08}.

To explore the effect of decoupling, one of the OWLS simulations, `\textit{WHYDRODEC}', employs decoupling, following the recipe of \citet{springelhernquist03}. Every particle that is kicked into the wind feels no hydrodynamic forces for either a fixed amount of time (50 Myr), or until the density of the wind particle falls below some value (10\% of the star formation density threshold, i.e.\ when $n_\textrm{\scriptsize H} < 10^{-2}$ cm$^{-3}$). If the wind would retain its original velocity of 600 km s$^{-1}$, 50 Myr would correspond to a travelling distance of roughly 30 kpc, so it ensures that SN-driven winds escape the galaxy.

Inspection of Fig.~\ref{fig:prettypics} reveals that decoupling the SN-driven winds dramatically increases the density of gas around this galaxy as every wind particle is able to escape from the galaxy \citep[see also][]{dallavecchiaschaye08}. Decoupling the winds means that SN feedback remains capable of suppressing SF to significantly higher halo masses, until the wind velocity falls below the escape speed from the halo (see also \citealt{springelhernquist03new}).  This is visible in Fig.~\ref{fig:whydrodec} where the galaxies in the `\textit{REF}' simulation show sharp upturns in $M_{\rm star}$ (panel A), SFR (panel B), $f_{\rm ISM}$ (panel D) at a halo mass $\sim 10^{11.25}\,$\msun, which are not present in the `\textit{WHYDRODEC}' simulation, indicating that the SN feedback remains capable of quenching SF if winds are artificially decoupled from the hydrodynamics. The decoupled winds are capable of ejecting more gas entirely from both the galaxy (panels D and F) and the halo (panels C and E).

Interestingly, for the lowest halo masses the SFR is higher for `\textit{WHYDRODEC}' than for `\textit{REF}' (panel B). This is expected, because wind particles will drag other particles into the wind, thereby increasing the effective mass loading, provided the initial wind velocity is sufficiently high for the winds to leave the galaxy.

The differences between `\textit{WHYDRODEC}' and `\textit{REF}' are very significant, up to 0.5~dex in the ISM fraction (panel D).  From this we can conclude that gravity (which acts on the winds in both simulations) is not the process that makes the winds ineffective.  It is rather the hydrodynamic drag and the associated radiative losses that make the winds less able to escape from high-mass haloes \citep[see ][]{dallavecchiaschaye08, dallavecchiaschaye12, creasey12}.

\subsection{Thermal SN feedback} \label{sec:wthermal}

\begin{figure*}
\centering
\includegraphics[width=0.33\linewidth]{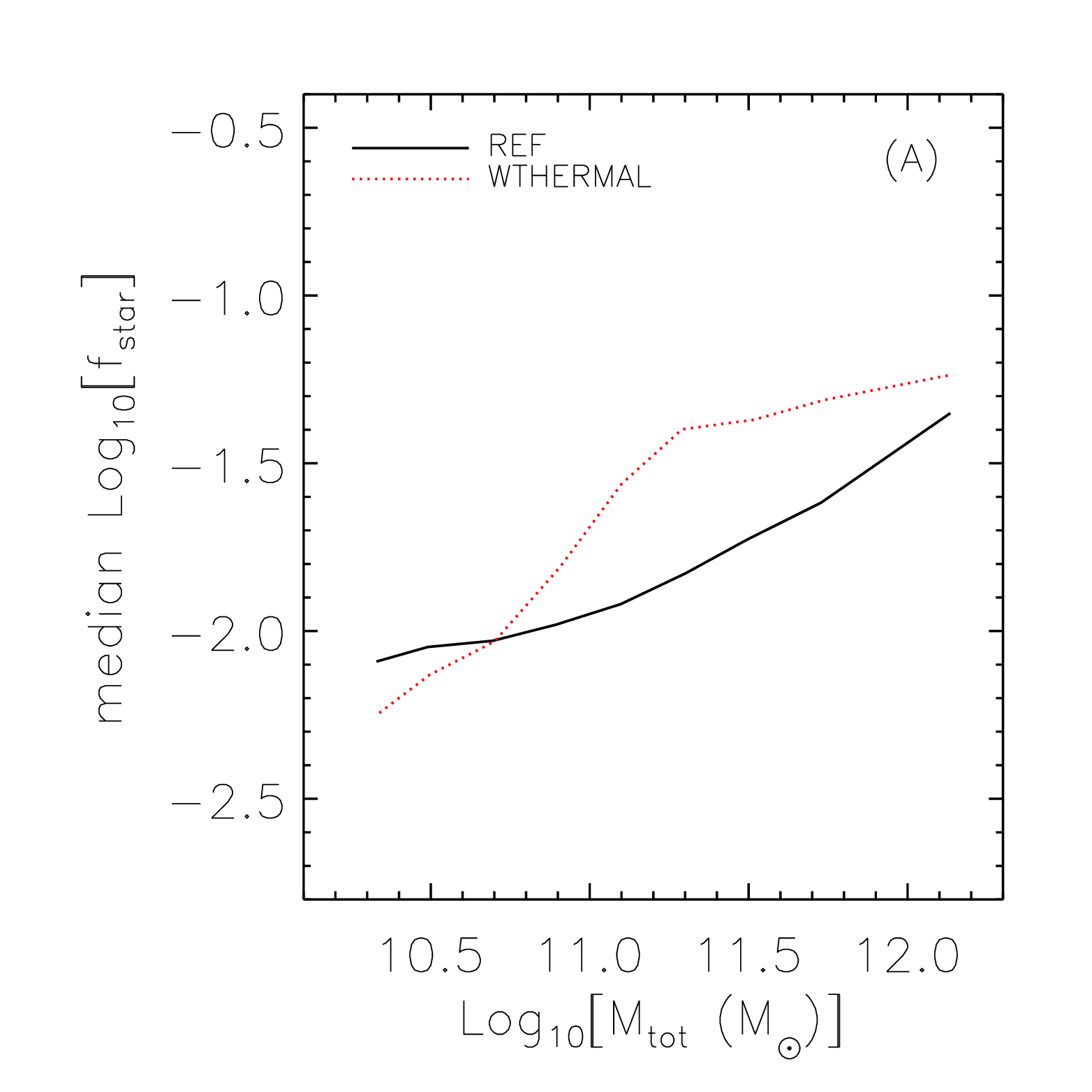}
\includegraphics[width=0.33\linewidth]{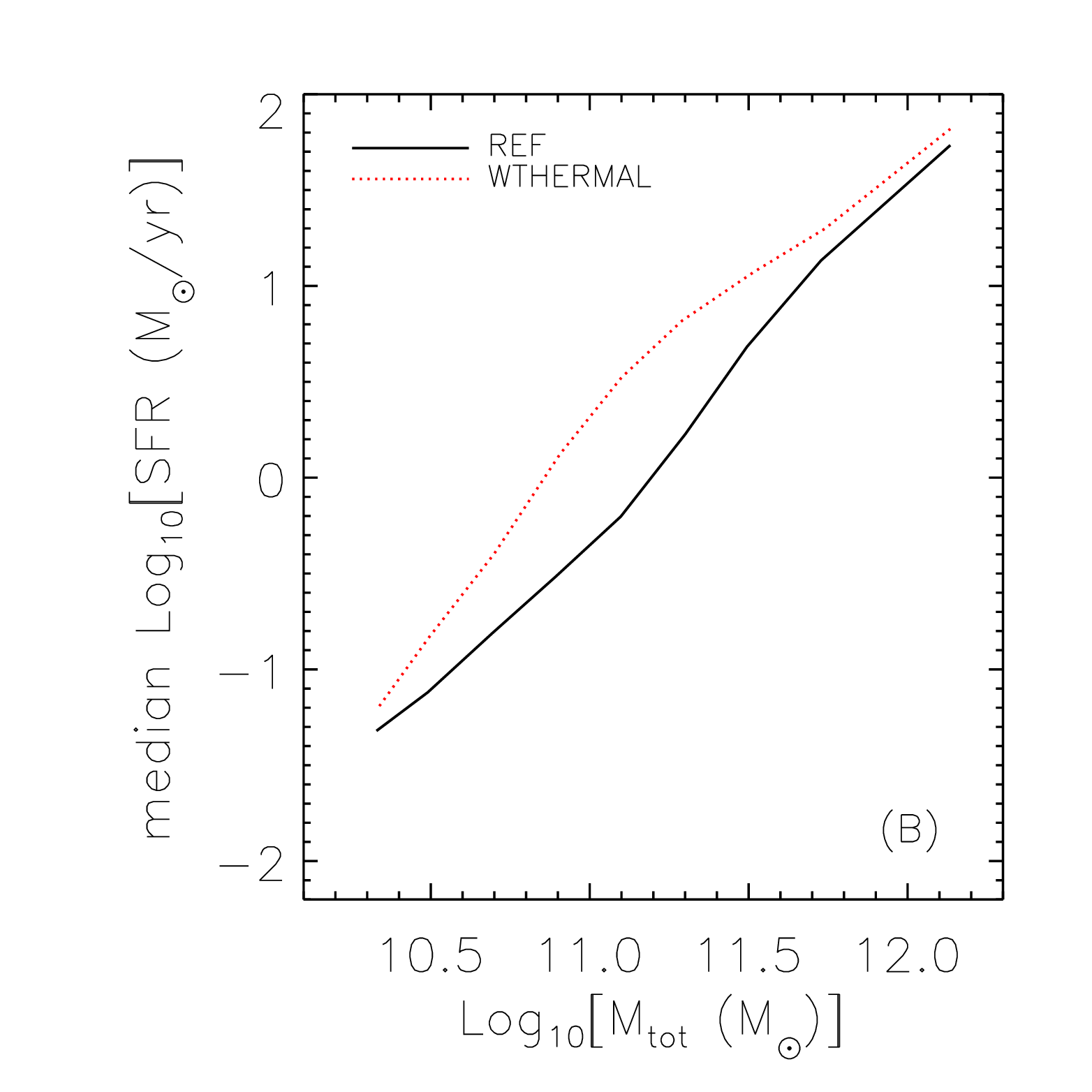}
\includegraphics[width=0.33\linewidth]{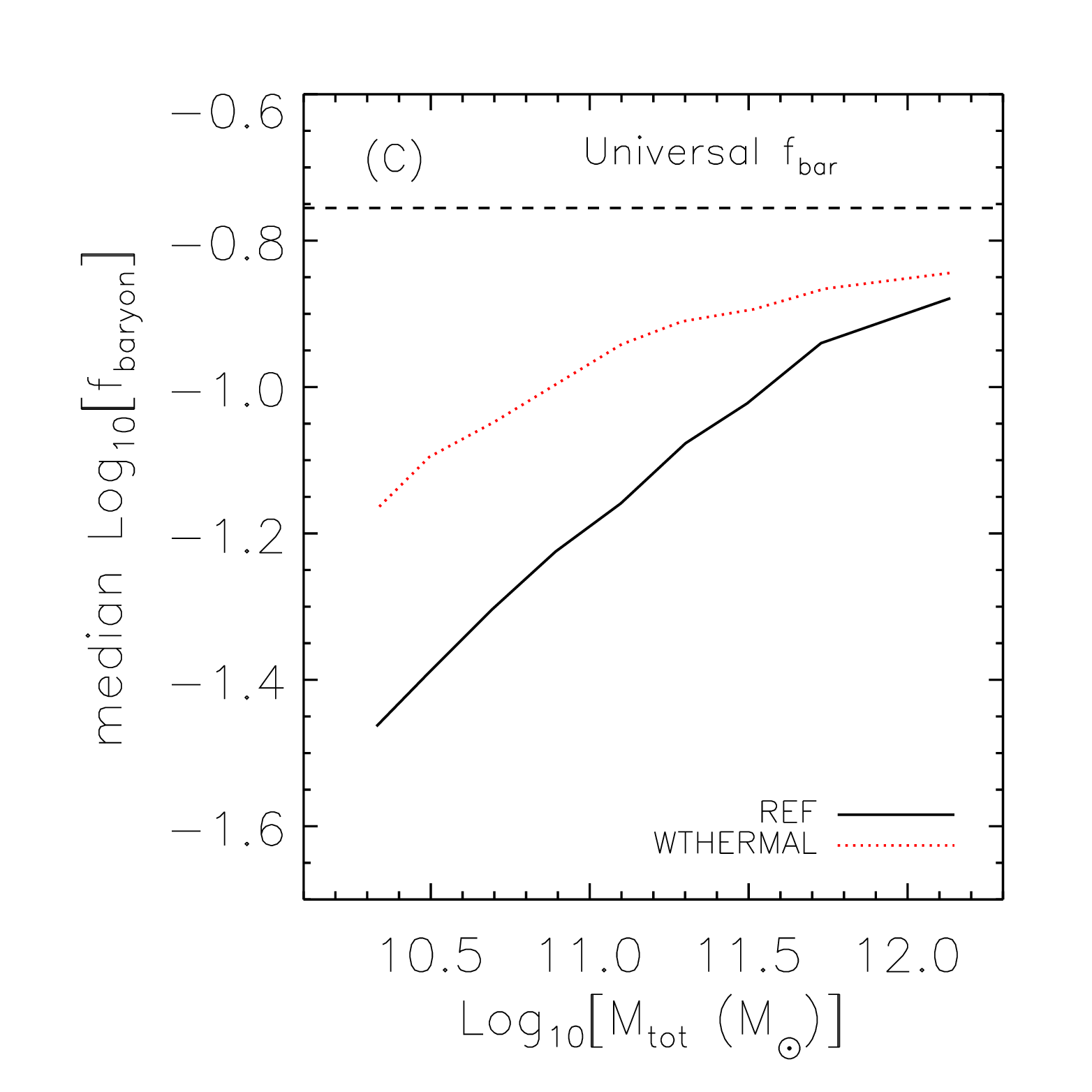} \\
\includegraphics[width=0.33\linewidth]{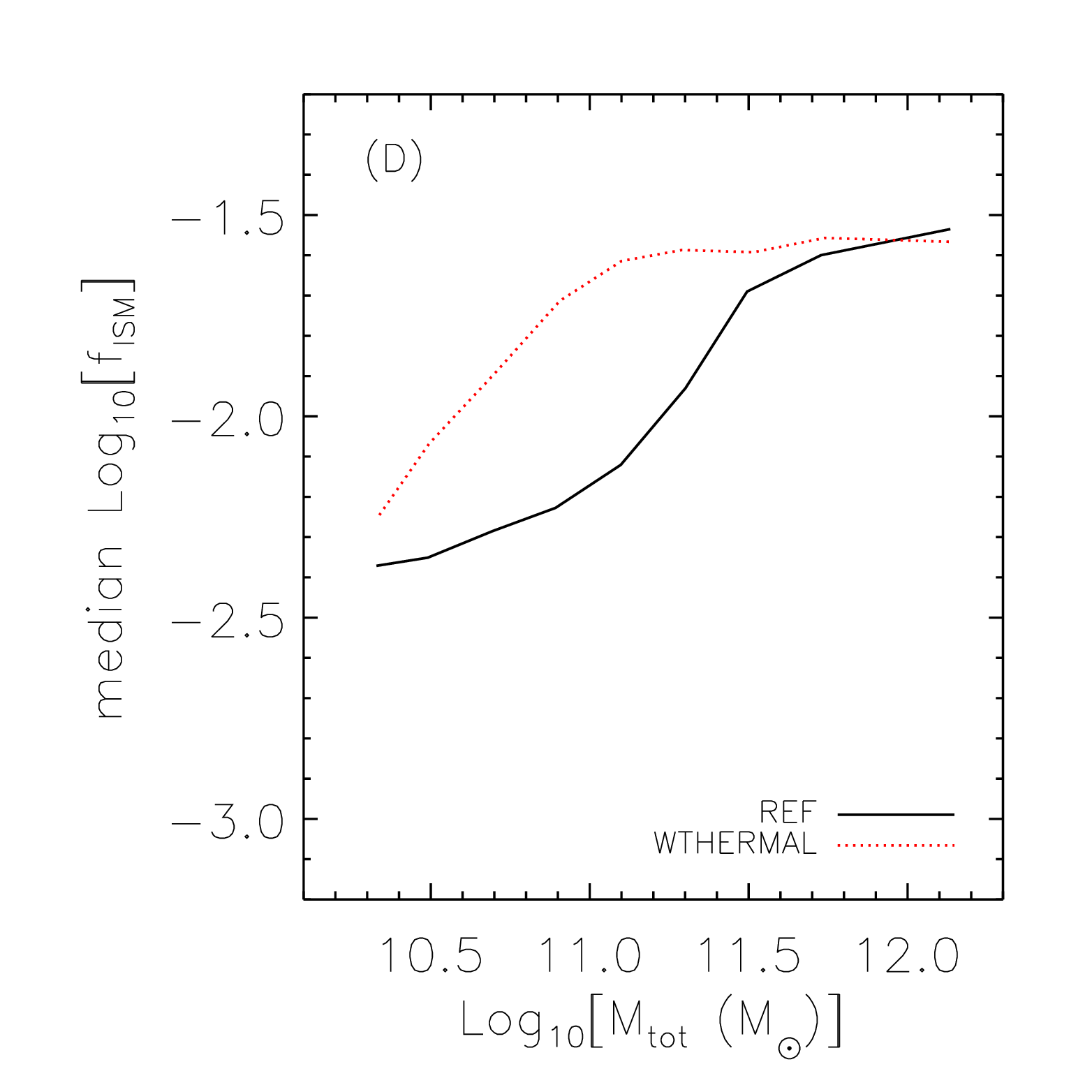}
\includegraphics[width=0.33\linewidth]{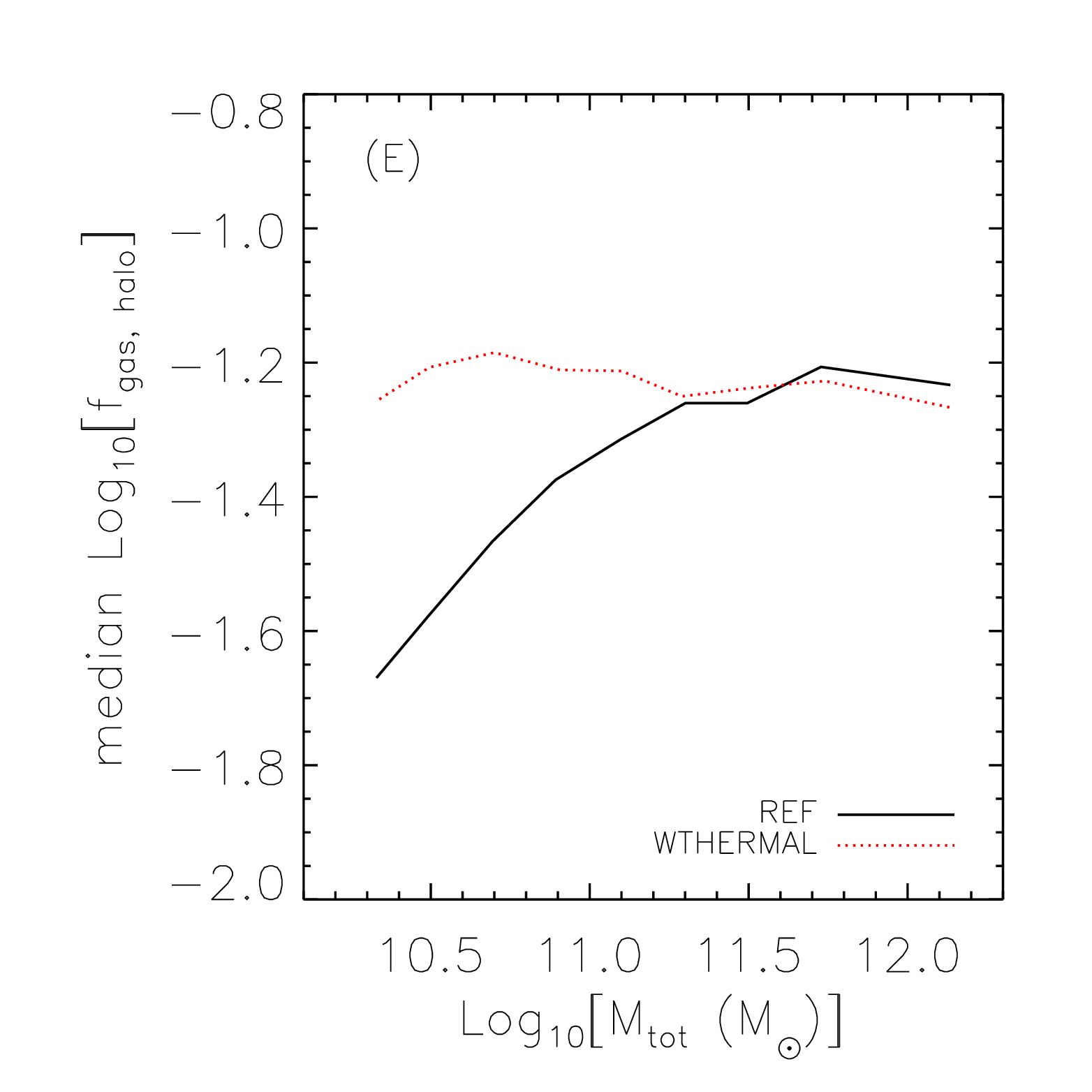}
\includegraphics[width=0.33\linewidth]{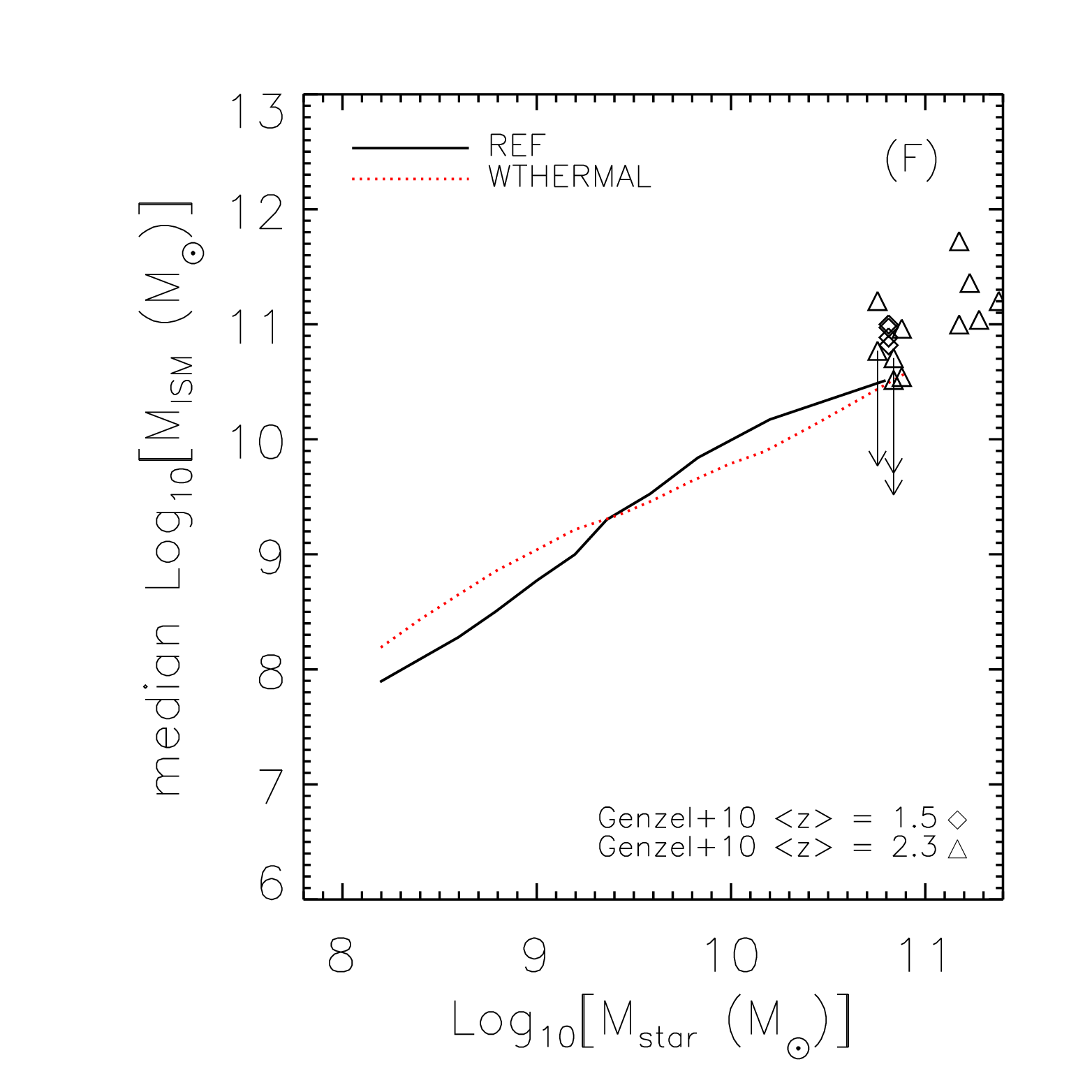} \\
\includegraphics[width=0.33\linewidth]{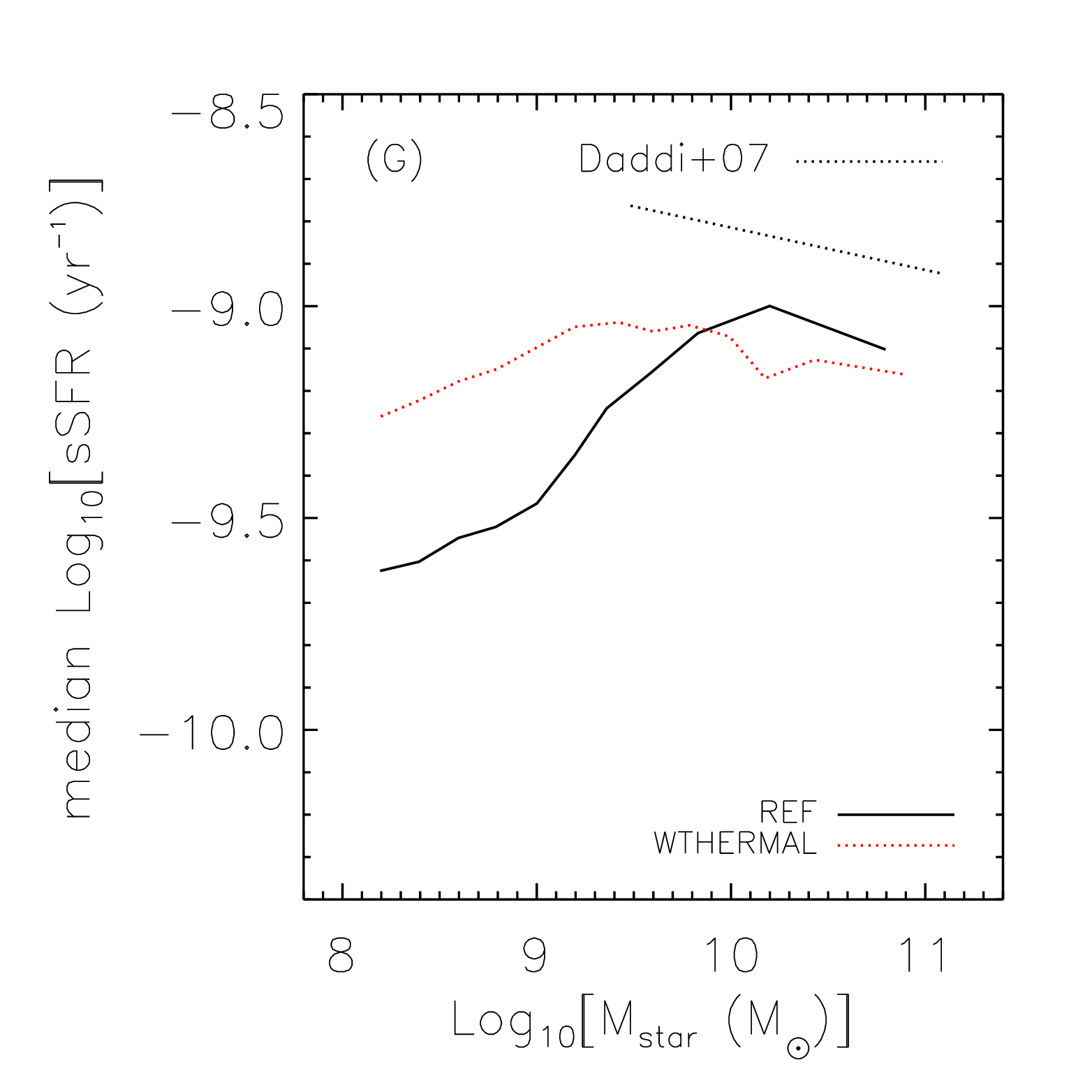}
\includegraphics[width=0.33\linewidth]{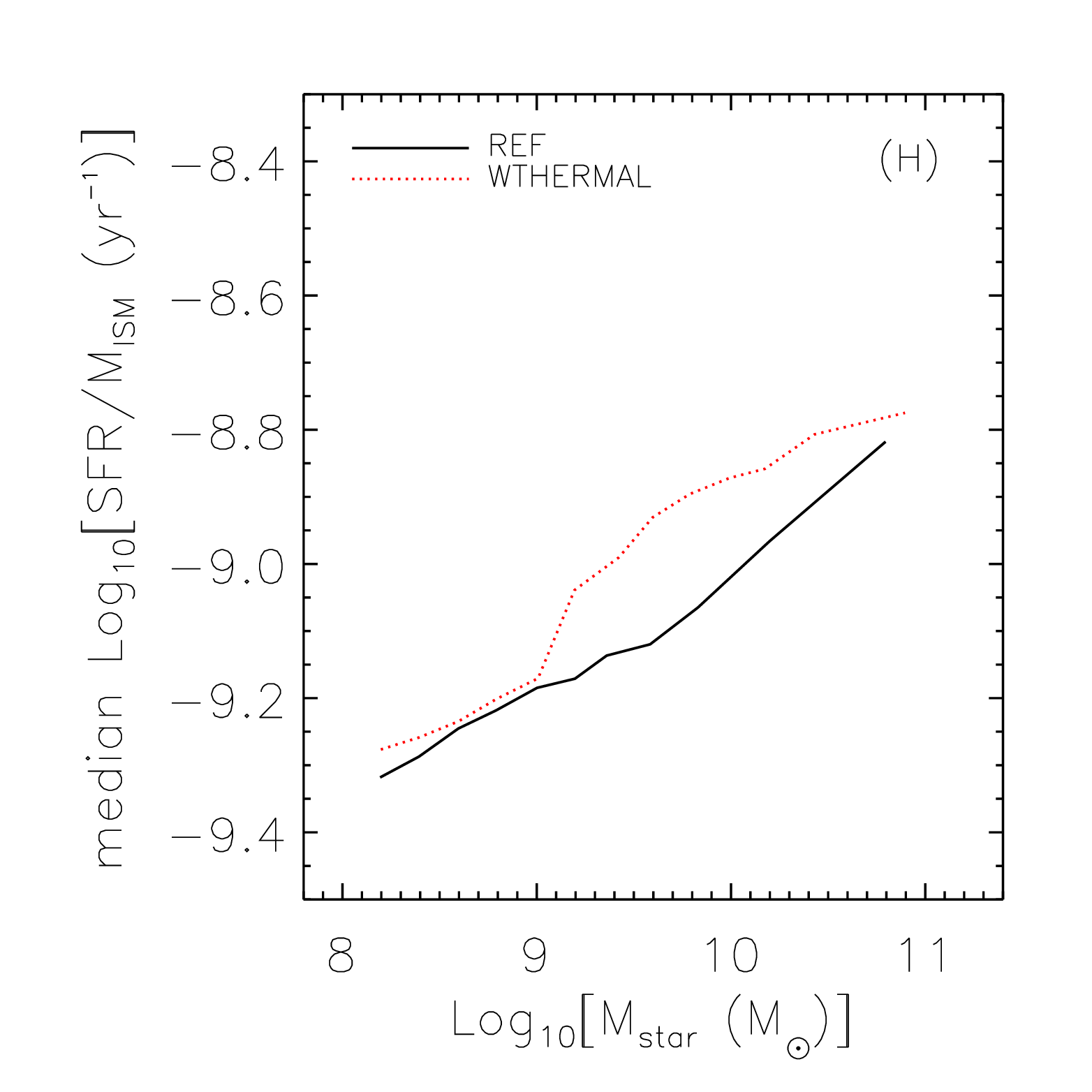}
\includegraphics[width=0.33\linewidth]{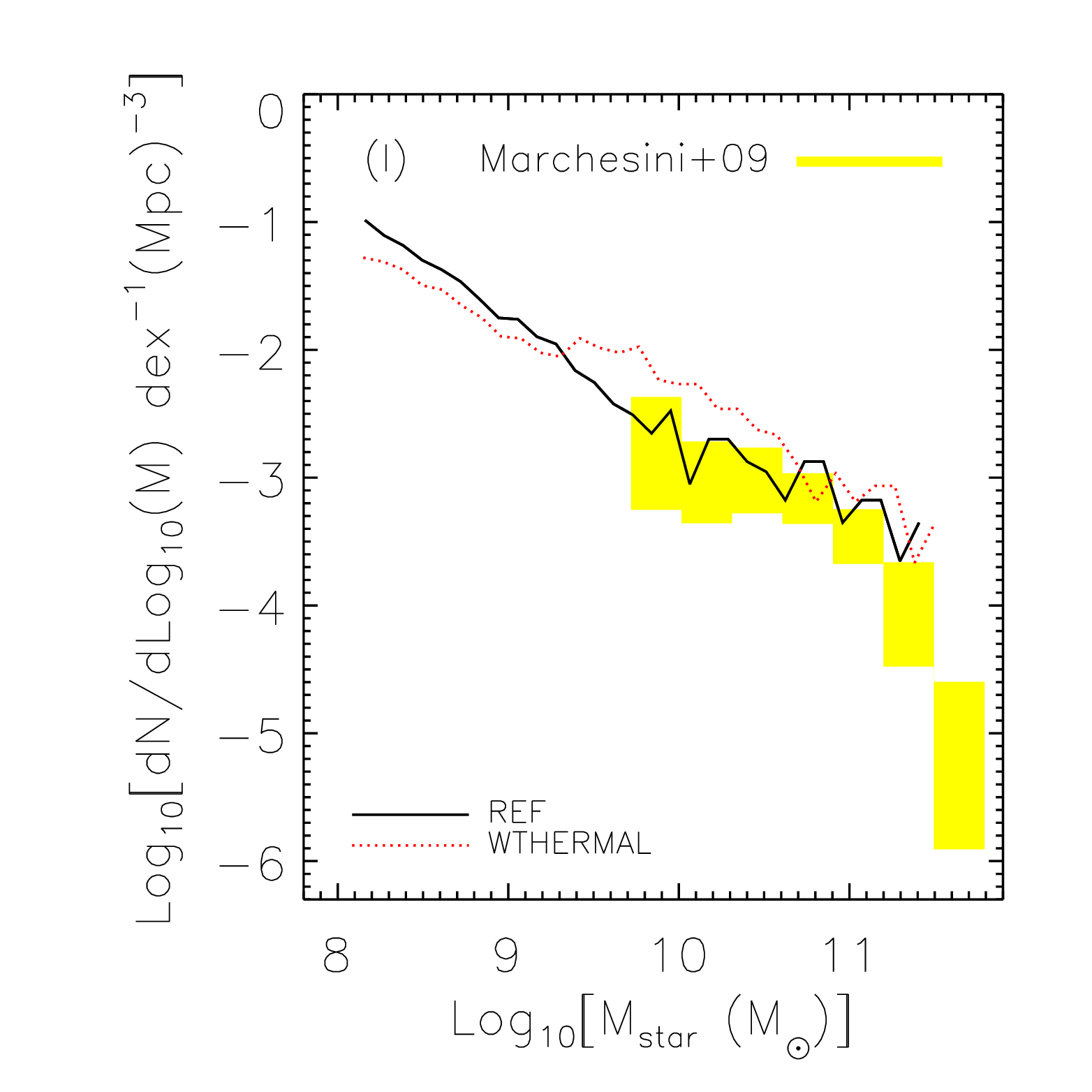} \\
\caption{As Fig.~\ref{fig:All_sims}, but showing only the simulation in which SN feedback is implemented thermally (`\textit{WTHERM}'; red, dotted curve) and `\textit{REF}' (black, solid curve).  Both thermal and kinetic feedback suppress star formation, but for the models presented here, thermal feedback is somewhat weaker, except for the lowest, and in these simulations poorly resolved, halo masses.} 
\label{fig:wthermal} 
\end{figure*}

\noindent 
All the models that we considered up till now, implemented feedback from star formation kinetically. In this section we investigate what happens if, instead of launching the wind by injecting kinetic energy, we inject thermal energy into the gas surrounding each newly formed star particle. As described in \citet{dallavecchiaschaye12}, the \lq thermal feedback\rq\, is implemented stochastically. If the SN energy is distributed amongst all of a star particle's neighbours, then the rise in the temperature of each particle is so low that cooling times remain very short and the particle immediately re-radiates all of the energy. In this case, the feedback will have little effect, unless the cooling is temporarily suppressed \citep{mori97, thackercouchman00, kay02, sommerlarsen03, brook04, stinson06}. Therefore, we choose to inject the thermal energy into neighbouring gas particles stochastically by specifying a temperature jump, $\Delta T=10^{7.5}\,$K, and then calculating for each neighbouring gas particle the probability that it is heated such that the expectation value for the total injected energy agrees with the amount of feedback energy that is available. This method has some similarity with the ‘promotion feedback’ model of \citet{scannapieco06}. We do not turn off radiative cooling at any time. The simulation that employs this model is termed `\textit{WTHERMAL}'.  Given our choices for $\Delta T$ and IMF, the expectation value for the number of heated particles per star particle is about 1.34 for a fully ionised plasma (equation 8 of \citealt{dallavecchiaschaye12}).  We inject $\sim40\%$ of the available SN energy in order to facilitate comparison with the other models.

`\textit{WTHERMAL}' is compared to `\textit{REF}' in Fig.~\ref{fig:wthermal}. Although the two models inject the same amount of SN energy per unit stellar mass formed, `\textit{WTHERMAL}' suppresses star-formation less effectively than `\textit{REF}' at most masses.  This is visible in panels A--D, where the thermal feedback simulation lies systematically above the kinetic feedback model. 
The differences are small at high masses, because there both types of feedback are effective. However, at the lowest halo masses ($M_{\rm tot}\la 10^{10.5}\,$\msun, close to our resolution limit), `\textit{WTHERMAL}' predicts lower stellar masses, but still slightly higher star formation rates, than `\textit{REF}'. This suggests that the thermal feedback is more efficient than the kinetic feedback in the poorly resolved galaxies that are not plotted here. 

Our finding that the thermal feedback is less efficient than kinetic feedback is consistent with \citet{dallavecchiaschaye12}, who predict that for $\Delta T = 10^{7.5}\,{\rm K}$, radiative losses should become significant for densities $n_{\rm H} \gtrsim 1~{\rm cm}^{-3}$ (see their equation 18), which are certainly reached in all but the lowest-mass galaxies in our simulations. 


\subsection{AGN feedback} \label{sec:agn}

\begin{figure*}
\centering
\includegraphics[width=0.33\linewidth]{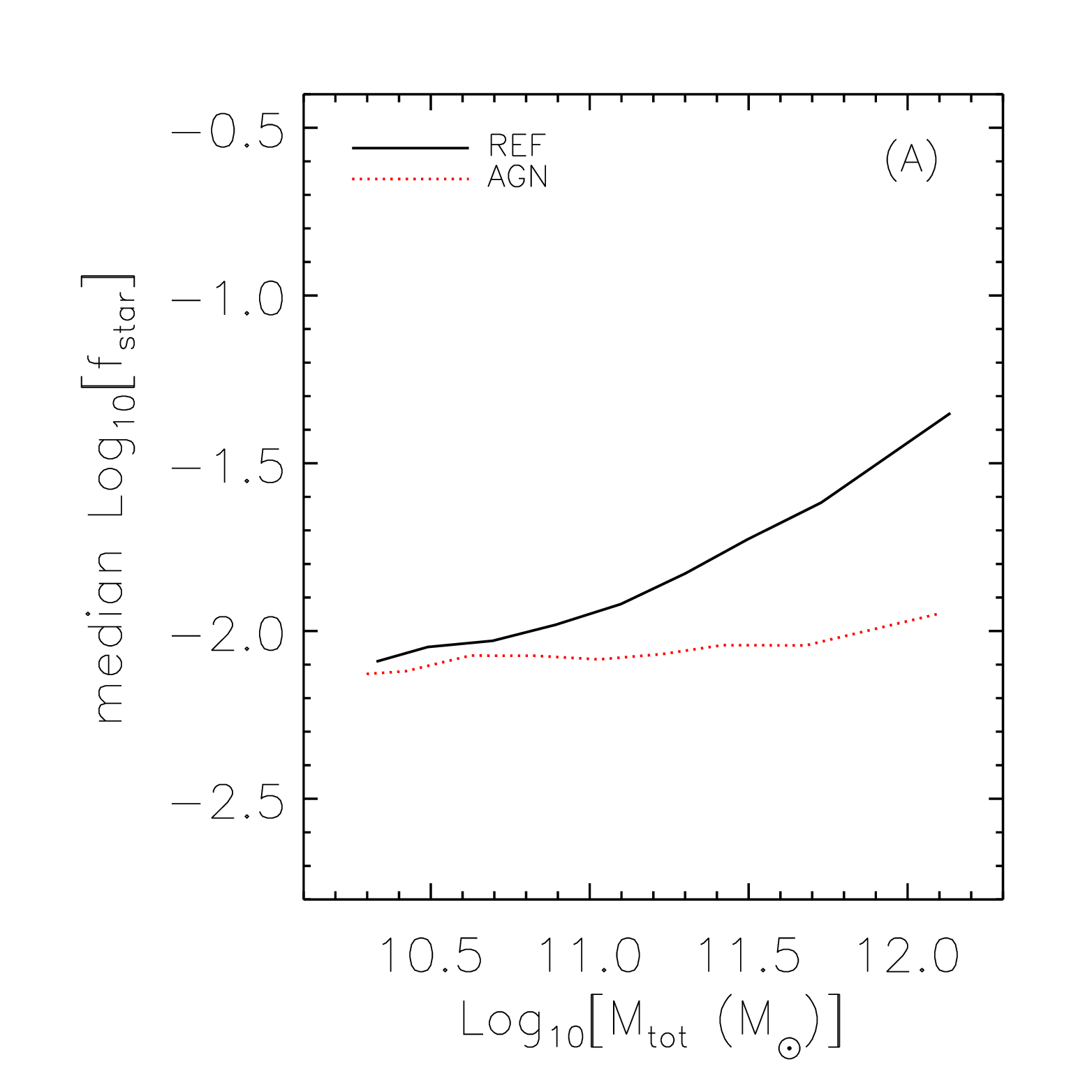}
\includegraphics[width=0.33\linewidth]{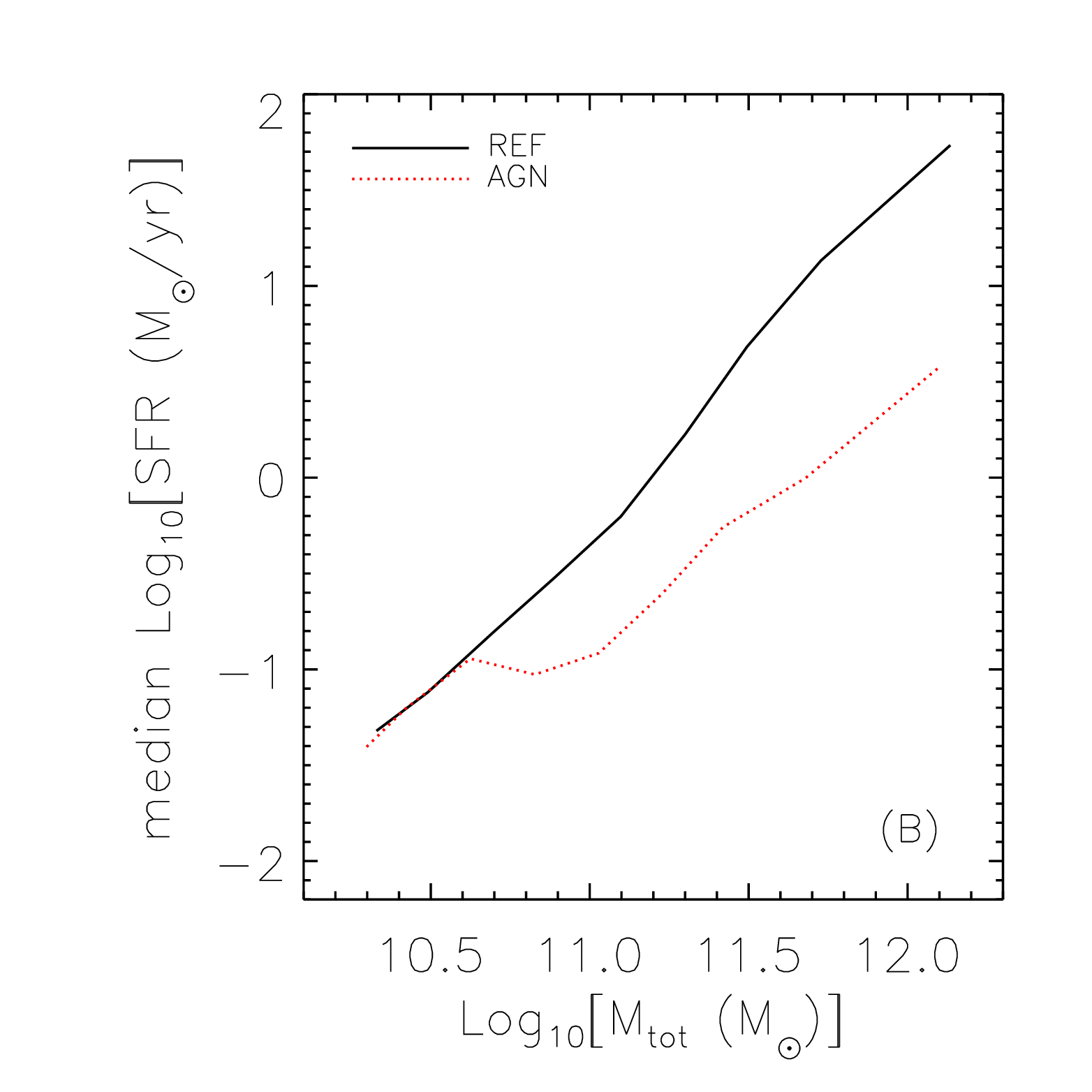}
\includegraphics[width=0.33\linewidth]{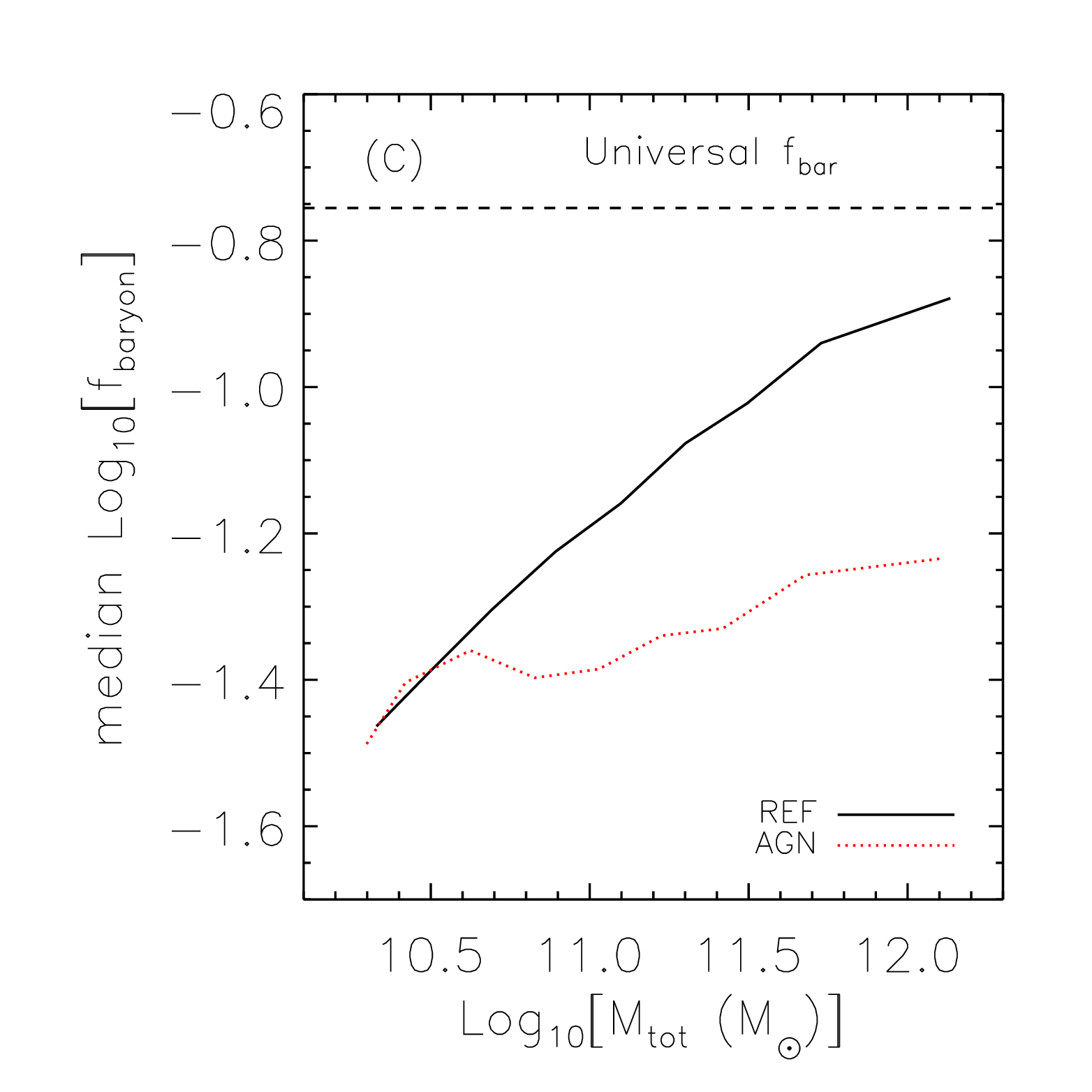} \\
\includegraphics[width=0.33\linewidth]{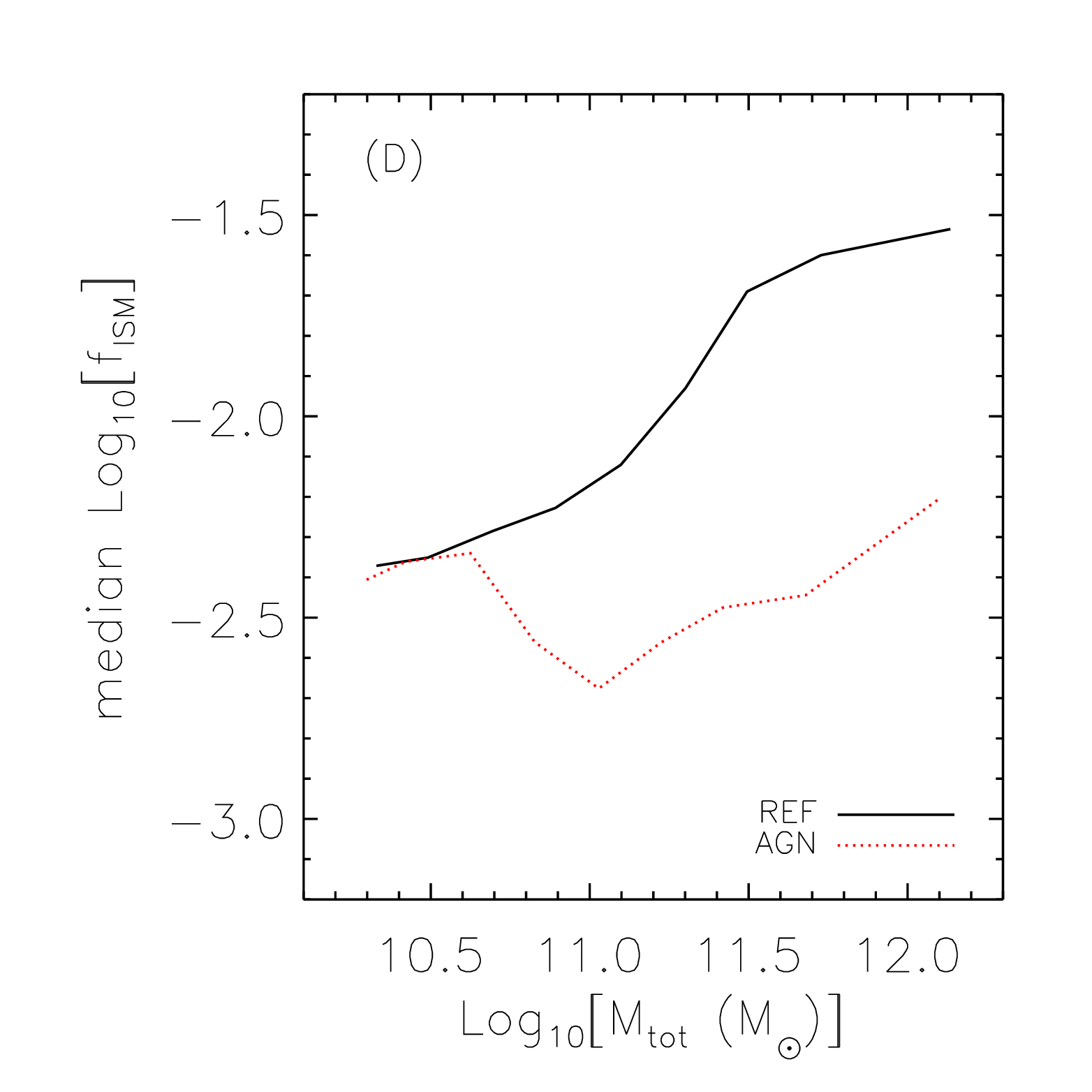}
\includegraphics[width=0.33\linewidth]{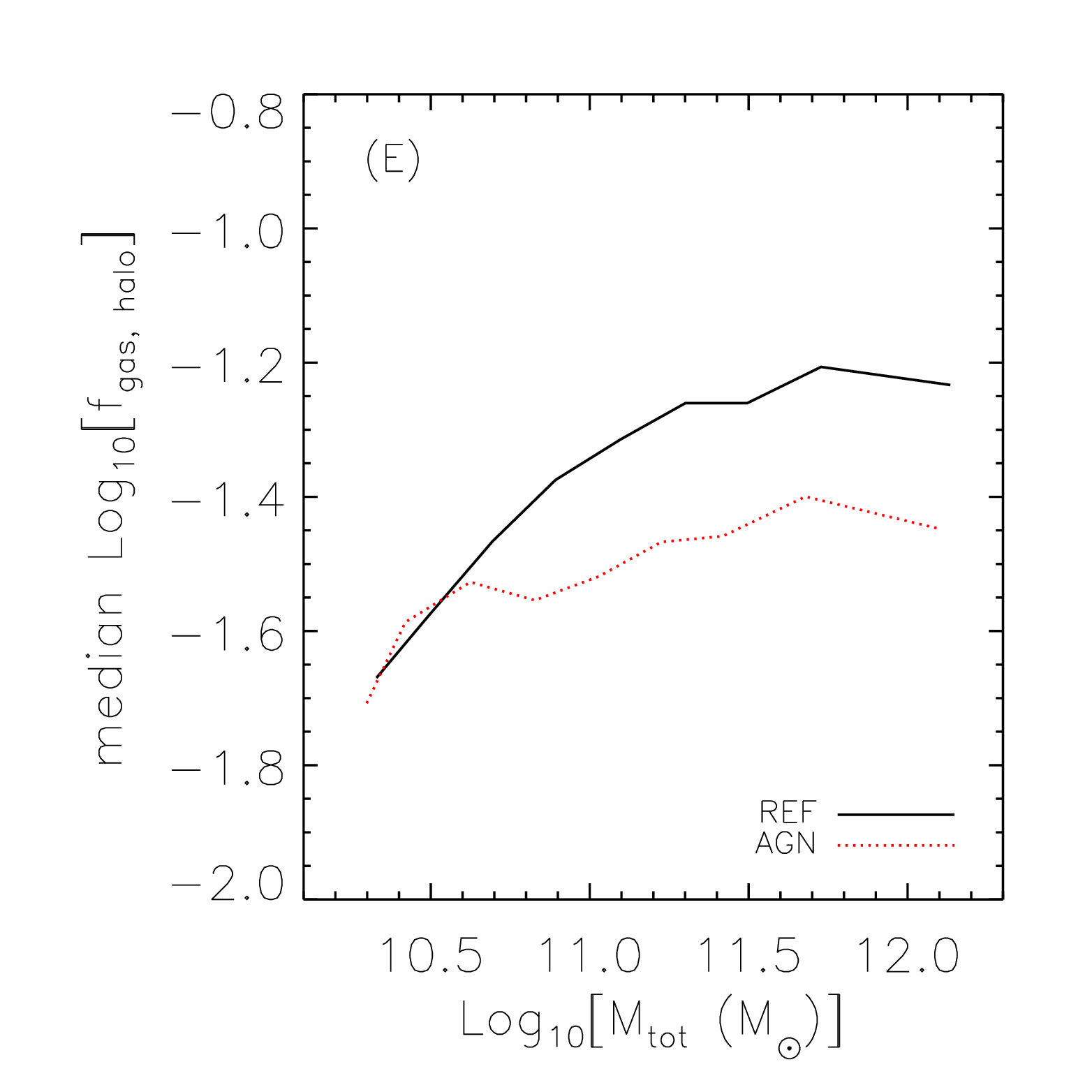}
\includegraphics[width=0.33\linewidth]{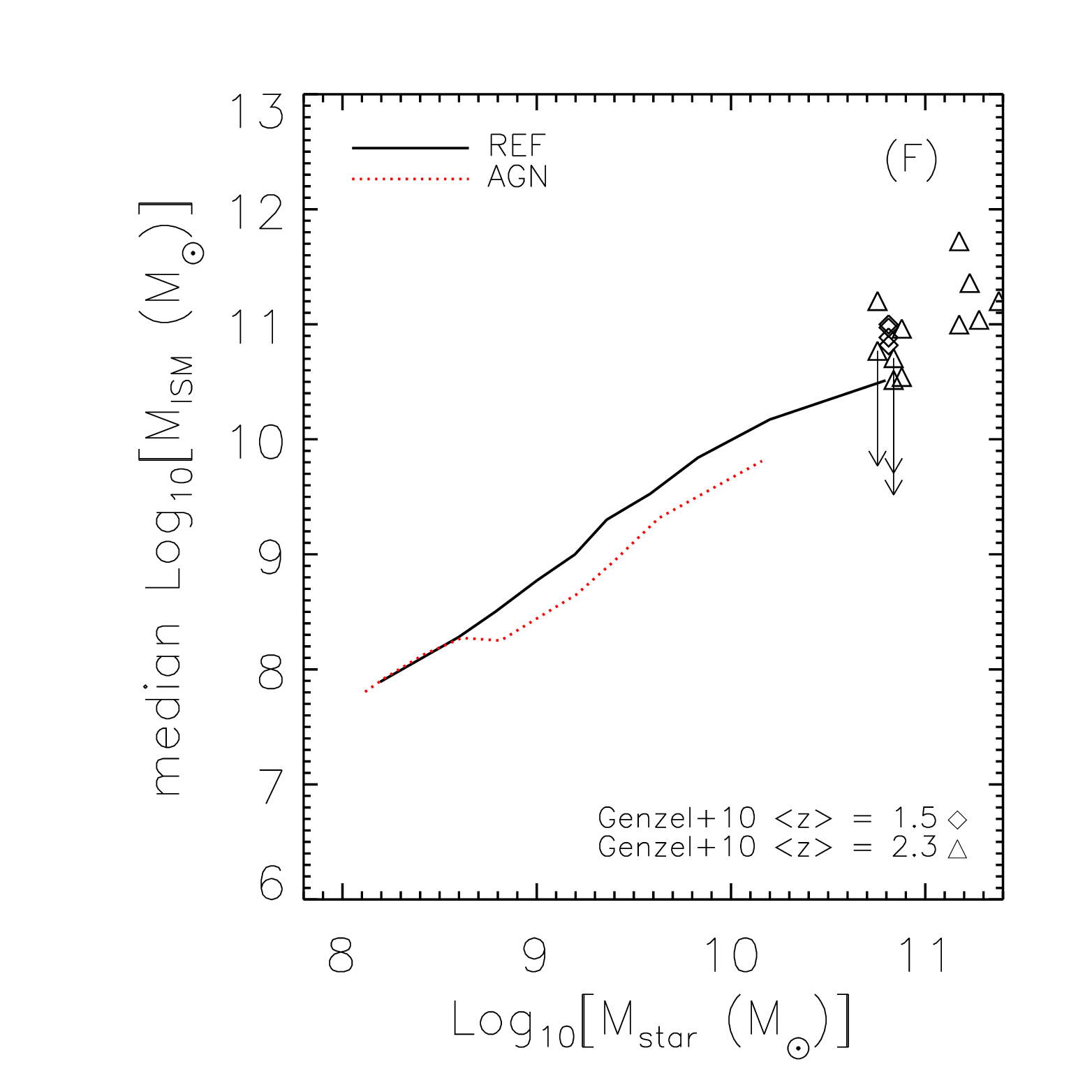} \\
\includegraphics[width=0.33\linewidth]{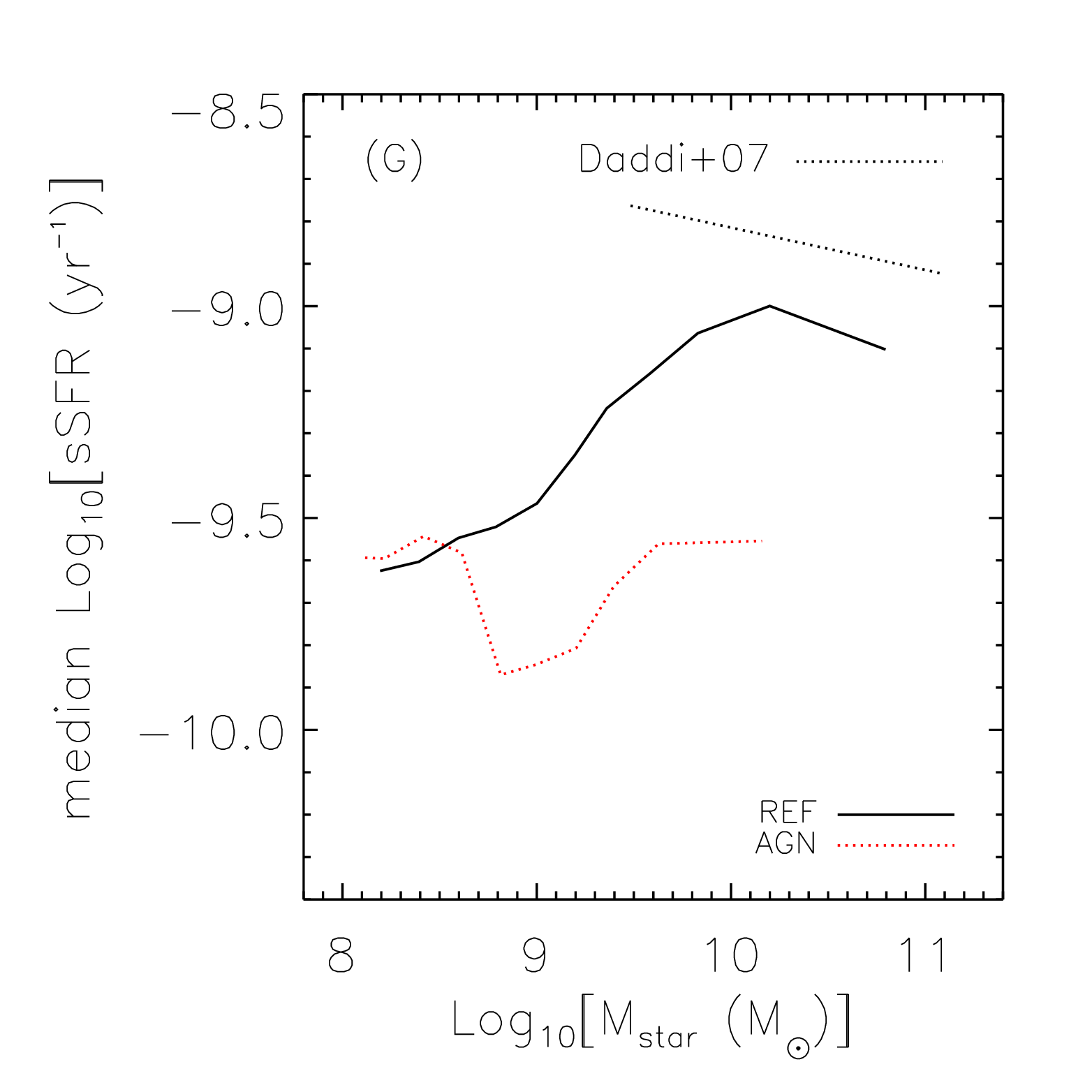}
\includegraphics[width=0.33\linewidth]{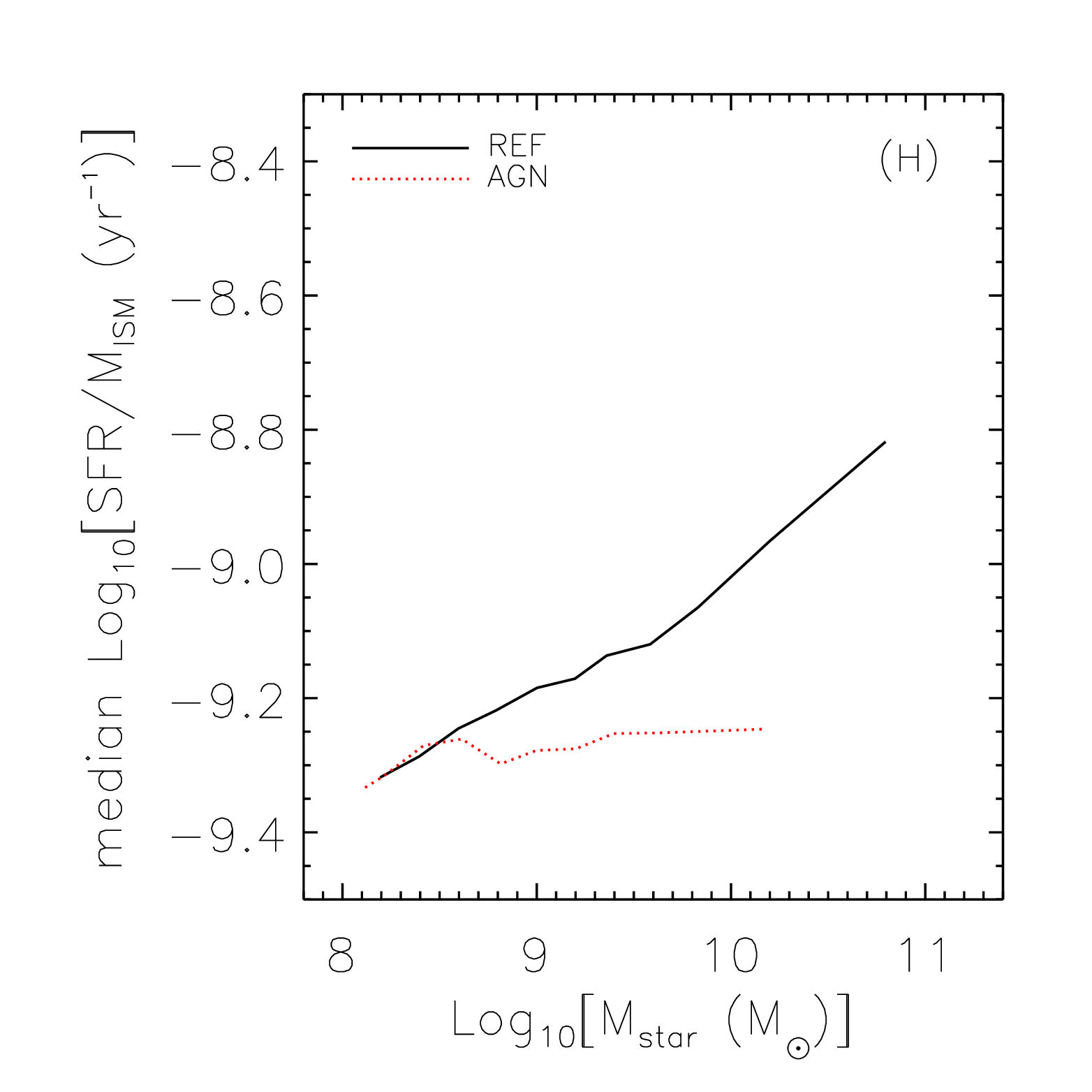}
\includegraphics[width=0.33\linewidth]{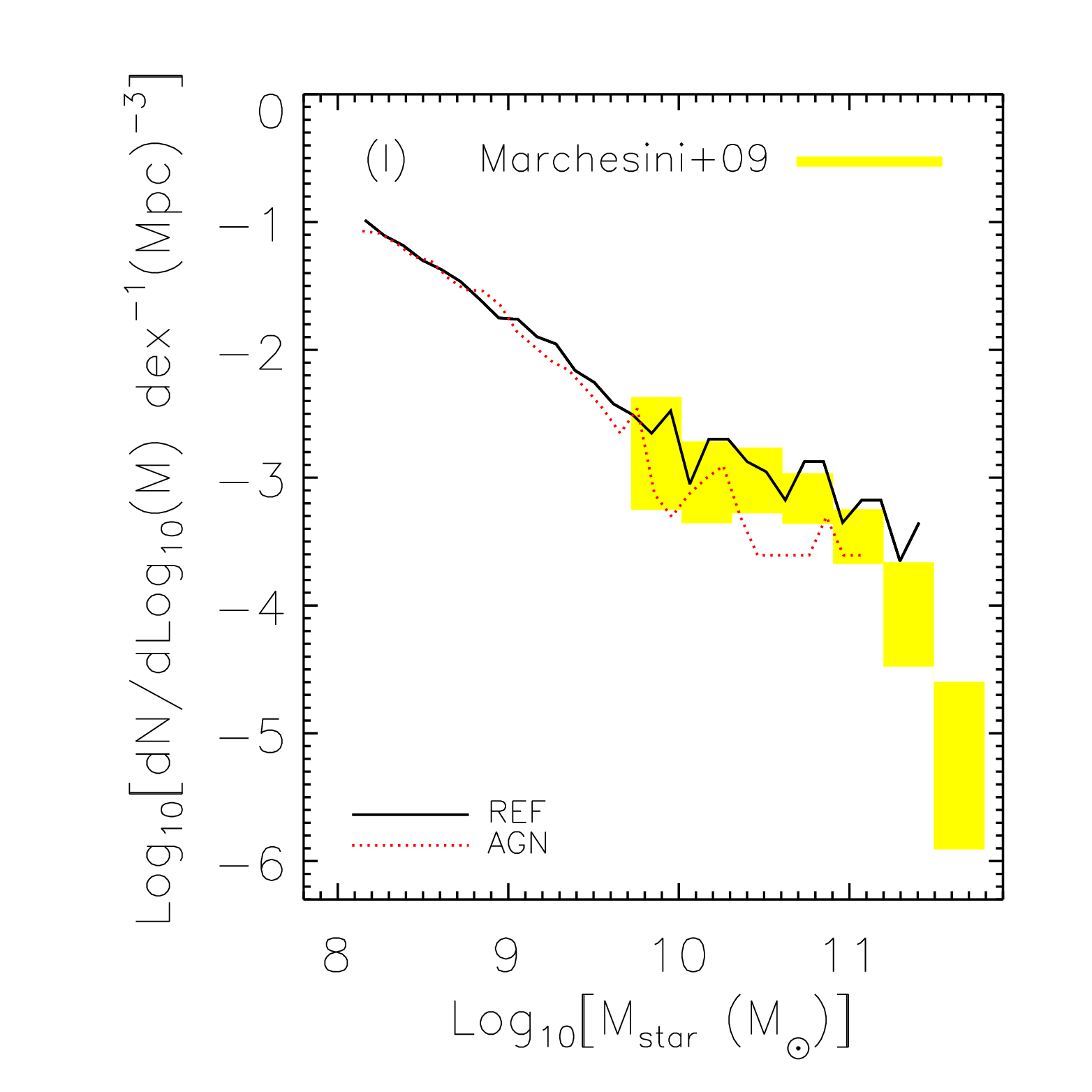} \\
\caption{As Fig.~\ref{fig:All_sims}, but comparing only the simulation with AGN feedback (`\textit{AGN}'; red, dotted curve) to `\textit{REF}' (black solid line).  AGN feedback effectively ejects gas from galaxies in all haloes where black hole seeds have been placed ($M_{\rm tot}> 4 \times 10^{10} \, h^{-1}$\msun, thus strongly suppressing the star formation rate.} 
\label{fig:agn} 
\end{figure*}

\noindent AGN feedback is implemented using the method of \citet{boothschaye09} which is, in turn, a modification of that of \citet{springel05}. We frequently run a halo finder and insert a seed mass black hole (BH) ($m_{\rm seed}=10^5\, h^{-1}$\msun) into every halo with mass $>4\times 10^{10} \, h^{-1}$\msun\ that does not yet contain a BH.  These seed black holes then grow both through merging and gas accretion (which is limited to the Eddington rate). Accretion rates in low-density gas ($n_\textrm{H} < 10^{-1}$cm$^{-3}$) are assumed to be equal to the Bondi-Hoyle rate.  For higher-density, star-forming gas the Bondi-Hoyle rate is boosted by a factor $(n_{\rm H}/10^{-1}\,{\rm cm}^{−3})^2$ to compensate for the lack of a cold, interstellar gas phase and the finite resolution \citep[see][for a full discussion]{boothschaye09}. The BH growth rate is related to its accretion rate by  $\dot{m}_{\rm BH} = (1-\epsilon_{\rm r})\dot{m}_{\rm accr}$, where $\epsilon_{\rm r}=0.1$ is the radiative efficiency of the BH.  The amount of energy coupled to the surroundings of the BH is then given by $\dot{E}=\epsilon_{\rm f}\epsilon_{\rm r}\dot{m}_{\rm accr}c^2$, where $c$ is the speed of light and $\epsilon_{\rm f}$ is a free parameter, the \lq feedback efficiency\rq, that is tuned to reproduce the global BH density at $z=0$.  In the fiducial runs $\epsilon_{\rm f}=0.15$.  This model reproduces the observed black hole scaling relations \citep{boothschaye09}, and the observed X-ray and optical properties of low-redshift galaxy groups \citep{mccarthy10}. Note, however, that our simulations have a substantially higher resolution than the $100 h^{-1}$ Mpc, $512^3$ runs used to compare with low-redshift observations. As shown by \citet{boothschaye09}, the model is not fully converged for intermediate masses with higher resolution, resulting in more efficient BH growth until the growth becomes self-regulating. This model does not include `radio-mode' feedback (which may be necessary to get a sharp exponential drop-off in stellar mass function) and we have not varied the parameters of this model in this paper.

Fig.~\ref{fig:prettypics} shows that AGN feedback is the most destructive form of feedback, and in a massive galaxy at $z=2$ the AGN destroys the gaseous disk. This effect is also visible in panels C - E of Fig.~\ref{fig:agn}, where it is apparent that the ISM and halo gas content (and thus the baryonic content, which is dominated by gas in the halo) of galaxies is strongly suppressed relative to the `\textit{REF}' simulation.

In the very lowest-mass resolved haloes ($M_{\rm tot} < 10^{10.6}\,$\msun), `\textit{AGN}' and `\textit{REF}' show virtually identical results because in these haloes BHs have not yet been seeded.  However, even slightly above these halo masses, the BH accretion prescription used in these simulations is very efficient at this resolution and the BHs grow and effectively suppress star formation (panels A, B, G and H of Fig.~\ref{fig:agn}).   It has been argued that AGN feedback is necessary to suppress rapid cooling of hot halo gas and suppress star formation in high-mass haloes \citep[e.g.][]{dimatteo05, croton06, bower08, boothschaye09, mccarthy10}. In panel (B) of Fig.~\ref{fig:agn} we show the halo SFR as a function of mass for `\textit{AGN}'. The effect of AGN is indeed very strong at high masses.  Although the galaxy sSFRs are almost an order of magnitude below the observations (panel G) this may not be a problem.  The observations are of galaxies that are actively star forming. While in the `\textit{REF}' simulation all galaxies are active, in the `\textit{AGN}' simulation there are both active and passive galaxies at all masses, leading to a median sSFR that lies below the relation obtained for active galaxies alone.  The stellar mass function (panel I), slightly undershoots the observed stellar mass function in the range where both observations and simulations exist. This may occur because the SN feedback in the `\textit{REF}' simulation was tuned to reproduce the peak in the global star formation rate density of the Universe. Including AGN feedback will then under-produce the amount of stars.


\section{Conclusions} \label{sec:conclusions}
We have analyzed a large set of high-resolution cosmological simulations from the \owls\ project \citep{owls}, focusing on the baryonic properties of the galaxy population at redshift 2. In particular, we studied the effects of variations in the input physics on the stellar mass, SFR, baryon fraction, ISM fraction, and gas fraction as a function of halo mass, on the sSFR, ISM mass and the gas consumption time scale as a function of stellar mass, and on the galaxy stellar mass function. In this paper we focused on variations of the parameters of the sub-grid models for radiative cooling, feedback from star formation and AGN, as well as the box size and the resolution, as shown in Appendix A. In Paper II we concentrate on the remaining sub-grid prescriptions (e.g.\ those for star formation), which tend to be less important for galaxy properties other than the mass and density distribution of the ISM.  

A central conclusion from this work is that the SFR in star forming galaxies is self-regulated by galactic winds driven by massive stars. Many of our results can be understood if the SFR adjusts so that the (time averaged) rate at which energy and momentum are injected is sufficient to drive a large-scale outflow that, on average, balances the gas accretion rate \citep[e.g.][]{owls,dave12}. Chemical feedback is also important, because,
for a fixed redshift and halo mass, the accretion rate is sensitive to the radiative cooling rate and hence to the metallicity.

Our results for variations of the kinetic feedback parameters and radiative cooling rates can be summarised as follows:
\begin{itemize}
\item Feedback from star formation is only effective if the initial wind velocity used for kinetic feedback is sufficiently high. The required wind velocity increases with the mass of the galaxy and also if metal-line cooling is included. If wind particles are temporarily decoupled from the hydrodynamics, then the feedback remains effective up to much higher masses. All these results can be understood if radiative losses inside the ISM are responsible for the quenching of the winds \citep[][see also \citealt{creasey12}]{dallavecchiaschaye12}. 
\item If the winds are not quenched, then the SFR is inversely proportional to the amount of energy that is injected per unit stellar mass. For a fixed initial wind velocity, this rate of energy injection is determined by the initial wind mass loading.
\item If, on the other hand, the initial wind velocity is too low for the winds to escape the galaxies, then the behavior of the simulations tends to that of a simulation without any feedback at all: catastrophic cooling resulting in excessive star formation.
\end{itemize}
The main conclusions drawn from other variations are:
\begin{itemize}
\item Using a top-heavy IMF for star formation at high pressure mainly influences the simulated halos through the extra amount of SN energy available per unit stellar mass. 
\item Feedback can remain highly efficient at all masses if the initial wind velocity is increased with the galaxy mass while keeping the momentum that is injected per unit stellar mass constant, as motivated by models of winds driven by radiation pressure (although the momentum-driven winds used here and in the literature may use much more momentum than is actually available). 
\item AGN feedback is very efficient at reducing the star formation rate and gas content, especially at high masses.
\end{itemize}

We compared our predictions to three different observational results: the molecular gas mass as a function of stellar mass, the sSFR as a function of stellar mass, and the galaxy stellar mass function. The latter can be thought of as a convolution of the halo mass function and the stellar mass as a function of halo mass. The comparison with observations revealed that:
\begin{itemize} 
\item For most simulations, the mass in the ISM in simulated galaxies is slightly lower than the observed molecular gas masses as a function of stellar mass. Note though, that the observed gas masses may have been overestimated \citep{narayanan12}.
\item Except for models with inefficient feedback, the galaxy stellar mass function is close to the observed one over much of the observed mass range. The shape is different though, with most simulations predicting a steeper low-mass end than (Schechter-like extrapolations of) observational results. Models with higher initial wind mass loading factors predict shallower faint-end slopes, as appears to be required by the observations \citep[see also][]{bower12}. None of the simulations predict a clear exponential cut-off at the high-mass end, but this could be due to our limited box size or lack of `radio-mode' AGN feedback.
\item The predicted sSFRs as a function of stellar mass tend to be lower than observed. For the observed mass range the discrepancy is worst if the feedback is efficient. The high observed sSFR can only be matched if the feedback is inefficient at the observed mass, but highly efficient at lower masses. The observed negative slope in the relation between the sSFR and stellar mass is only reproduced for galaxies for which feedback is inefficient. 
\end{itemize}

Thus, there is tension between the comparison of simulated and observed galaxy stellar mass functions and the comparison between simulated and observed sSFRs. The high observed sSFRs are difficult to match unless feedback suddenly becomes inefficient at the stellar masses for which observations are available ($M_\ast > 10^{9.5}\,$M$_\odot$). It can, however, not remain inefficient as the stellar mass increases or else the galaxy mass function would become too high at the high-mass end. Even though there may be other explanations, one possibility is that our simulations form too many stars in low mass systems at higher redshift. That would result in a steep low-mass end of the mass function, a lower gas reservoir in galaxies at $z=2$ and consequently, too low sSFRs.

We have shown that winds driven by feedback from star formation determine the main properties of galaxies residing in haloes of a given mass (the scatter among the red lines in Fig.~\ref{fig:All_sims} is much larger than for the blue lines that represent the simulations
appearing in Paper II). Even for a fixed amount of feedback energy per unit stellar mass, variations in the sub-grid implementation, e.g.\ different wind velocities and mass loadings (that can be either constant or fixed functions of local physical conditions), provide us with considerable freedom to \lq tune\rq\, galaxy properties. This freedom can be exploited to match observations spanning a wide range of masses, which would provide the simulations with some of the attractions of semi-analytic models. However, this potential success comes also with the disadvantages of such models: the underlying physics may remain poorly understood. As higher resolution simulations become feasible, the freedom provided by subgrid models to generate galactic outflows in cosmological simulations will be reduced, as the results become less sensitive to the manner in which the energy is injected \citep{dallavecchiaschaye12}.

Further improvement in our understanding of the physics that determines the global properties of galaxies will likely come from theoretical models and observations focusing on galactic winds.
The physics of star formation is less crucial as self-regulation implies that the time-averaged, galaxy-wide SFRs are determined by the large-scale inflows and the efficiency with which star formation drives galactic winds.


\section*{Acknowledgements}
The authors kindly thank the anonymous referee for a thoughtful report that helped improve the paper. Furthermore, we thank Danilo Marchesini for useful discussions about the observations of the stellar mass function and for providing his results in digital form. Furthermore, we are grateful to the members of the \owls\ collaboration for useful discussions about, and comments on, this work. The simulations presented here were run on Stella, the LOFAR BlueGene/L system in Groningen, on the Cosmology Machine at the Institute for Computational Cosmology in Durham (which is part of the DiRAC Facility jointly funded by STFC, the Large Facilities Capital Fund of BIS, and Durham University) as part of the Virgo Consortium research programme, and on Darwin in Cambridge. This work was sponsored by the National Computing Facilities Foundation (NCF) for the use
of supercomputer facilities, with financial support from the
Netherlands Organization for Scientific Research (NWO),
also through a VIDI grant. The research leading to these results has received funding from the European Research
Council under the European Union's Seventh Framework Programme (FP7/2007-2013) / ERC Grant agreement 278594-GasAroundGalaxies
and from the Marie Curie Training Network CosmoComp (PITN-GA-2009-238356). VS acknowledges support
through SFB 881, ‘The Milky Way System’, of the DFG.

\bibliographystyle{mn2e}
\bibliography{fb}


\appendix
\section{Numerical convergence} \label{sec:physprop_convergence}

\begin{figure}
   \begin{center}
   \resizebox{\hsize}{!}{\includegraphics{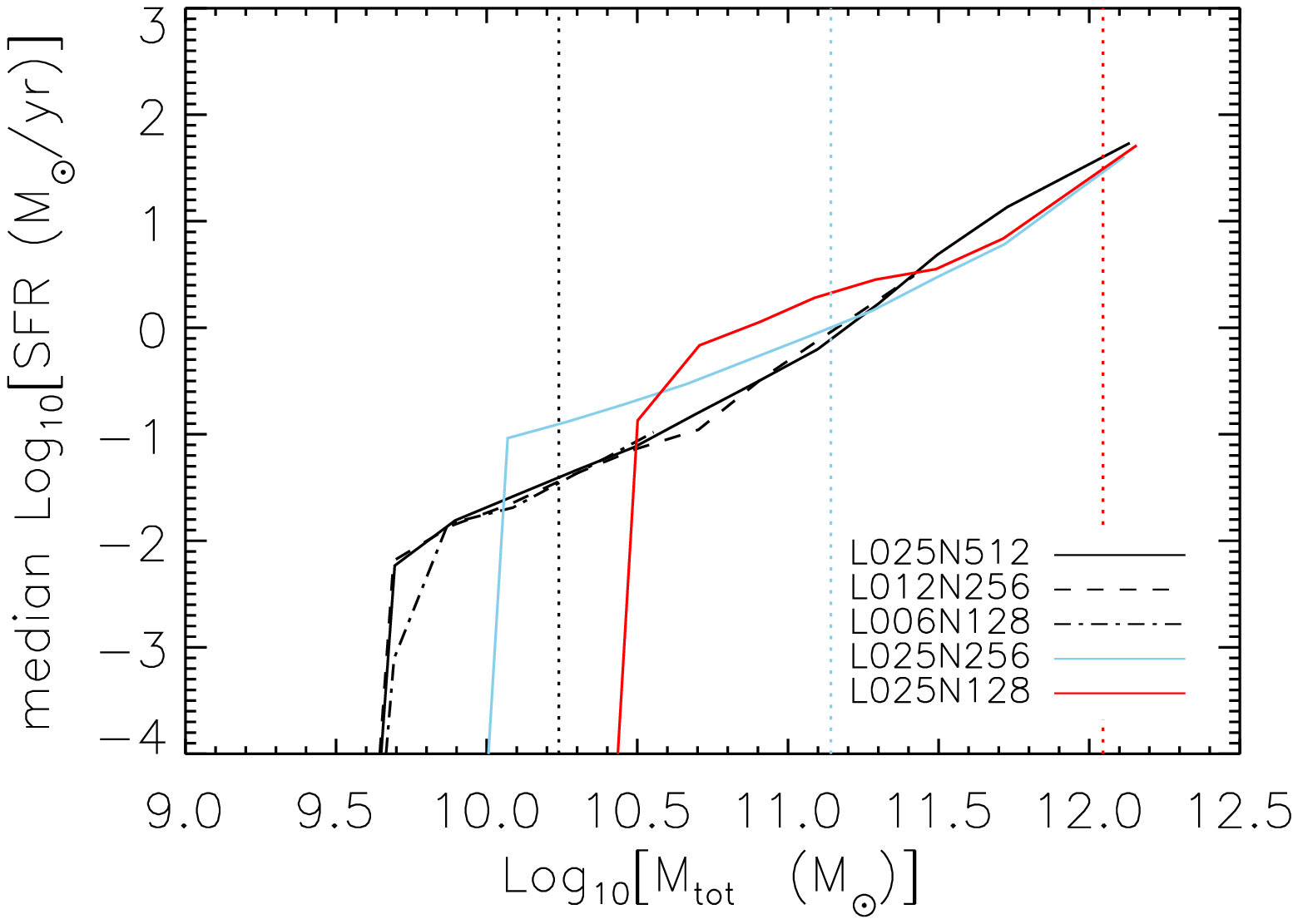}} \end{center}
    \caption{Median SFR as a function of halo mass at $z=2$ for 5 simulations with different particle numbers and/or box sizes as indicated in the legend. The vertical dotted lines indicate the mass of 2000 dark matter particles in the simulations shown by the curves in the corresponding colours. At the low-mass end, the median SFR falls to zero, as more than half of the haloes in a bin do not have gas particles with a density above the star formation threshold. Above a mass corresponding to 2000 dark matter particles per halo, the SFR as a function of halo mass is reasonably well resolved.}{\label{fig:conv_halomass_sfr}
    }     
\end{figure}

\begin{figure}
   \begin{center}
   \resizebox{\hsize}{!}{\includegraphics{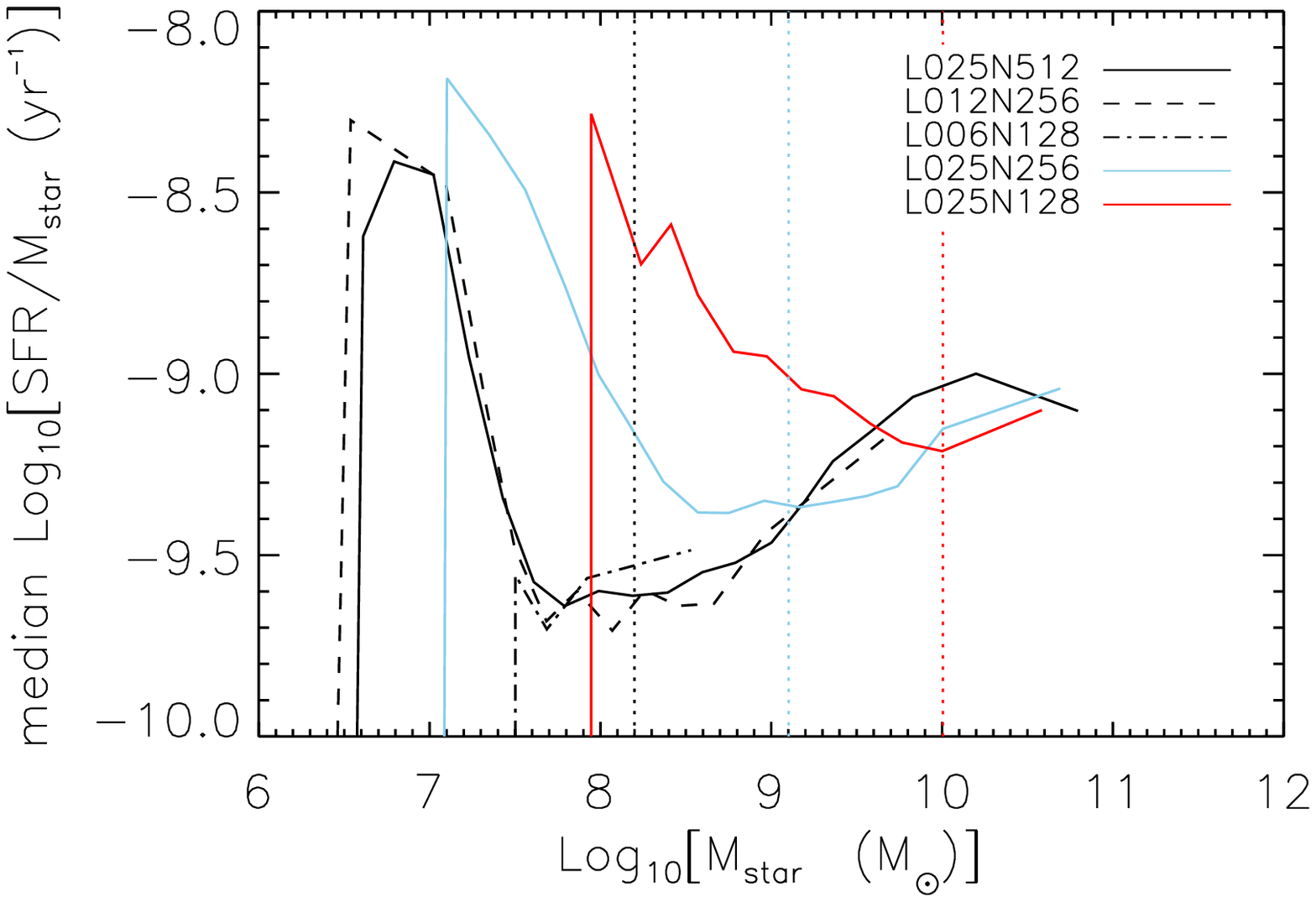}}\end{center}
    \caption{The median sSFRs of haloes as a function of their stellar mass at $z=2$ for 5 simulations with different particle numbers and/or box sizes as indicated in the legend. The vertical dotted lines indicate the mass corresponding to 100 star particles in the simulations shown by the curves in the corresponding colours. The sharp cut-off at low masses again stems from the fact that there is a minimum to the (non-zero) SFR. Right of the vertical dotted lines the sSFRs are reasonably well converged.}{\label{fig:conv_stellarmass_ssfr}
    }     
\end{figure}

\begin{figure}
   \begin{center}
   \resizebox{\hsize}{!}{\includegraphics{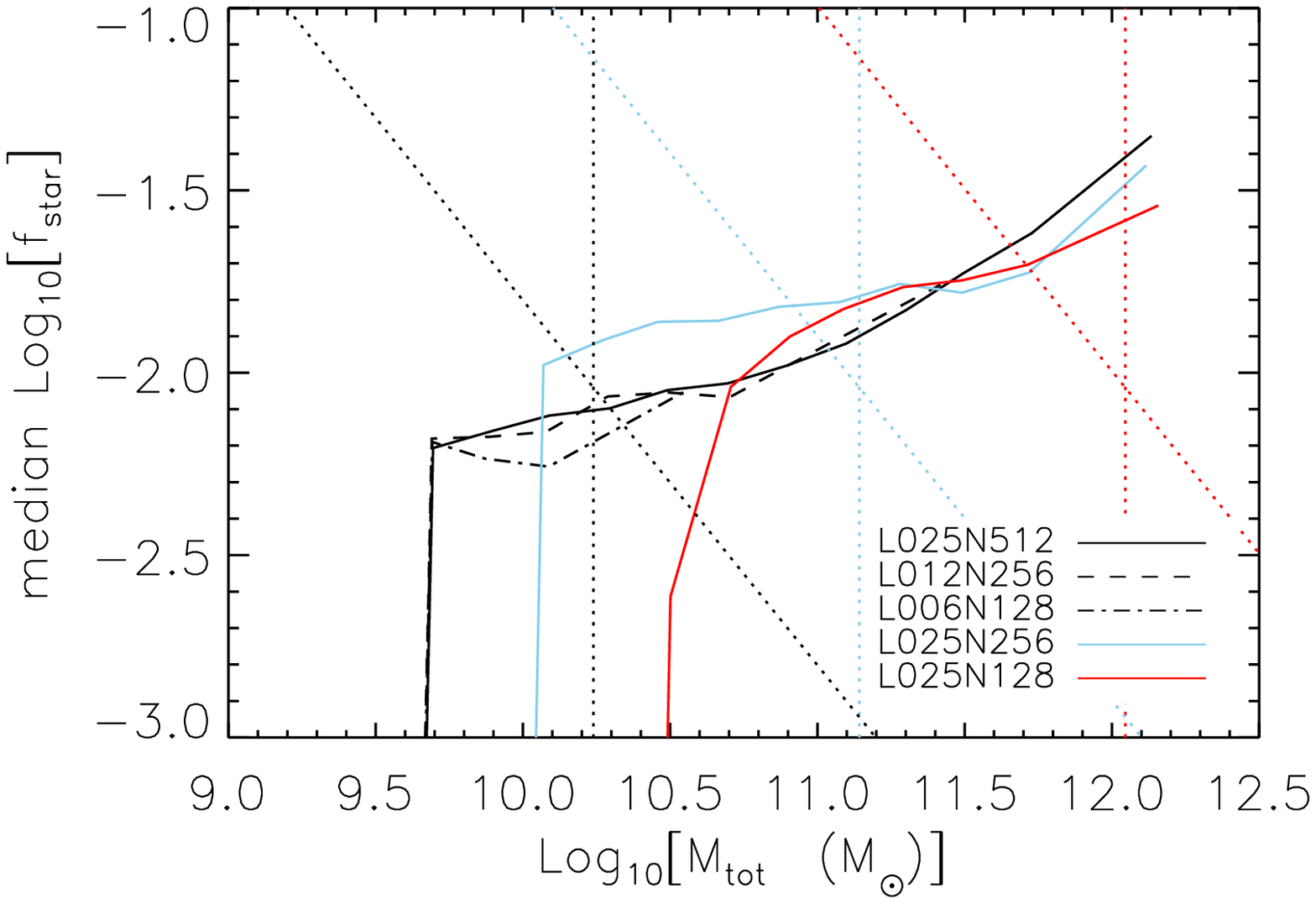}} \end{center}
    \caption{Median stellar mass fraction as a function of halo mass at $z=2$ for 5 simulations with different particle numbers and/or box sizes as indicated in the legend. The vertical dotted lines indicate the mass corresponding to  2000 dark matter particles in the simulations shown by the curves in the corresponding colours. The diagonal black dotted line indicates the relation for haloes with 100 star particles (the cut that is made in the rest of the paper where relations with stellar mass are shown). As can be seen, in the highest resolution simulation, the cuts made throughout this paper in minimum number of dark matter particles and the minimum number of star particles roughly correspond to each other. At lower resolutions, the cut in dark matter particle number is more stringent.}{\label{fig:conv_halomass_stellarfraction}
    }     
\end{figure}

Here we explore the sensitivity of our results to numerical resolution and simulation box size, and assess down to what mass limits we can consider our results numerically converged.  All simulations used in this section employ the same physical model as the `\textit{REF}' simulation, but are run with different particle numbers and box sizes.   We denote the simulations with `\textit{LXXXNYYY}', where \textit{XXX} is the size of the simulation box in co-moving $h^{-1}$Mpc and \textit{YYY} is the number of particles per spatial dimension (for both dark matter and baryons we use \textit{YYY}$^3$ particles). In this nomenclature, the `\textit{REF}' simulation is `\textit{L025N512}'.  In order to independently examine the effects of box size and resolution, we compare two sets of simulations:
\begin{itemize}
\item \textit{L025N512, L012N256} and \textit{L006N128}: These simulations have the same resolution, but the box size is varied in steps of a factor of two from 25 Mpc/$h$ to 6.25 Mpc/$h$. These three runs are shown as black curves with different line styles (solid, dashed and dot-dashed, respectively) in all figures in this appendix
\item \textit{L025N512, L025N256} and \textit{L025N128}: These simulations have the same box size, but different resolutions. The dark matter particle mass in these simulations is $6.3\times10^6$\,\msun/$h$, $5.1\times10^7$\,\msun/$h$ and $4.1\times10^8$\,\msun/$h$, respectively and the maximum proper gravitational softening is 0.5 kpc$/h$, 1 kpc$/h$ and 2 kpc$/h$, respectively.  In all figures in this appendix, these three simulations are shown as solid black, blue and red curves, respectively.
\end{itemize}

All plots in this appendix include all haloes identified by the FoF algorithm (the lowest mass haloes contain 20 dark matter particles).  Fig.~\ref{fig:conv_halomass_sfr} shows the median SFR as a function of halo mass.  It is immediately clear that simulation box size has no influence on the SFRs, as the lines with different line styles (\textit{L025N512}, \textit{L012N256} and \textit{L006N128}) show almost perfect agreement. The only effect of box size is that a larger box allows one to sample more massive, rarer objects.

The simulations are less well converged with respect to resolution (compare the solid black, red and blue curves in Fig.~\ref{fig:conv_halomass_sfr}).  The vertical, dotted lines denote 2000 times the dark matter particle mass in the simulations of the same colour. The halo SFRs are reasonably converged above these halo masses, so we employ 2000 dark matter particles as our resolution limit when comparing galaxy properties as a function of halo mass. The difference between the `\textit{L025N512}' and `\textit{L025N256}' runs are $\lesssim 0.5$dex. It is clear from Fig.~\ref{fig:conv_halomass_sfr} that at low masses the SFR is always slightly over-predicted in lower resolution simulations, so the true convergence level is likely better than 0.5 dex. Note that the difference between the high and intermediate resolution simulations just above the convergence criterium for the intermediate resolution simulation (the vertical blue dotted line) is much smaller, so one expects that the difference between our high resolution simulations and even higher resolution simulations at our imposed particle limit will be smaller than 0.5 dex. Observe that the halo mass regime where the median SFR is zero is effectively removed when demanding a minimum number of 2000 dark matter particles per halo, because more than half of the haloes with a low number of particles do not have any gas particles with densities above the star formation threshold.

The build-up of stellar mass is influenced by the SFR at all epochs prior to the epoch at which it is measured. As all haloes were initially small and thus poorly resolved, the early build up of stellar mass is underestimated. We therefore expect the convergence of the (s)SFR as a function of stellar mass to be worse than that of the SFR as a function of total halo mass.

Fig.~\ref{fig:conv_stellarmass_ssfr} shows the sSFR as a function of stellar mass. The vertical cut-off at the low-mass end corresponds again to haloes for which the median SFR is zero. The mass range over which the sSFR appears approximately converged with respect to resolution corresponds to $\sim100$ star particles (indicated by the vertical dotted lines).  Convergence can be seen by comparing the solid black and blue lines rightwards of the the blue, dotted line and by comparing the solid blue and red curves rightwards of the red, dotted line. We note that, as expected, the same trends are found for SFR$/M_{\textrm{\scriptsize gas}}$ (not shown).

Finally, in Fig.~\ref{fig:conv_halomass_stellarfraction} we show the median stellar mass fraction as a function of halo mass. The vertical dotted lines indicate our adopted resolution limit of 2000 dark matter particles. The diagonal dotted lines indicate the stellar mass fraction for haloes consisting of 100 star particles, which is our resolution limit for plots with stellar mass on the horizontal axis. The fact that for a given resolution (i.e.\ line colour), the solid curve intersects the two dotted lines in about the same place, implies that the cuts of 100 star particles and 2000 dark matter particles are comparable for these simulations. Above the limit of 2000 dark matter particles, the stellar mass fractions are nearly converged. At lower resolution, the cut in dark matter particle number is more stringent than a the cut on the minimum number of star particles. Therefore, throughout the paper we use a minimum of 2000 dark matter particles when we plot quantities as a function of halo mass.

\begin{figure*}
\centering
\includegraphics[width=0.33\linewidth]{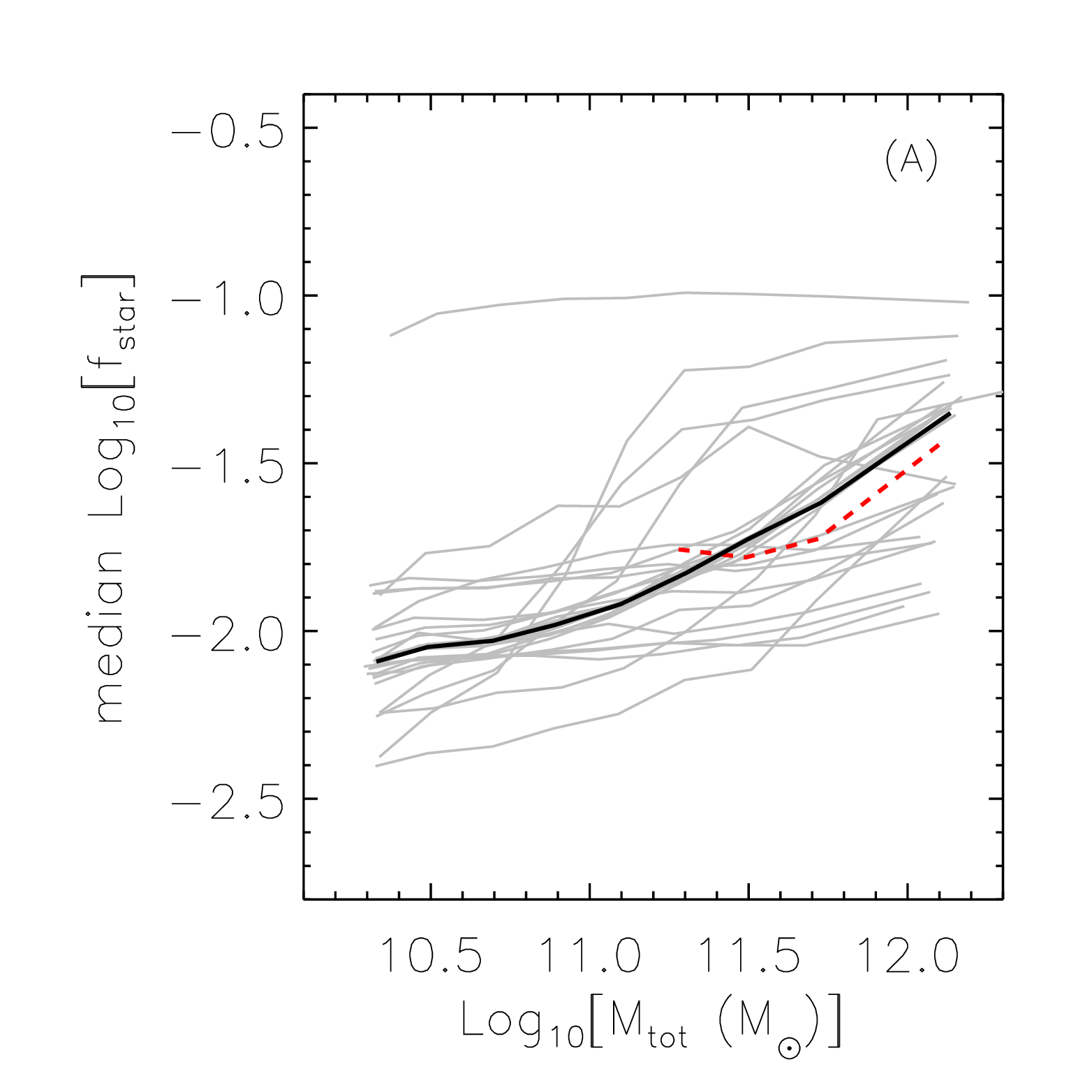}
\includegraphics[width=0.33\linewidth]{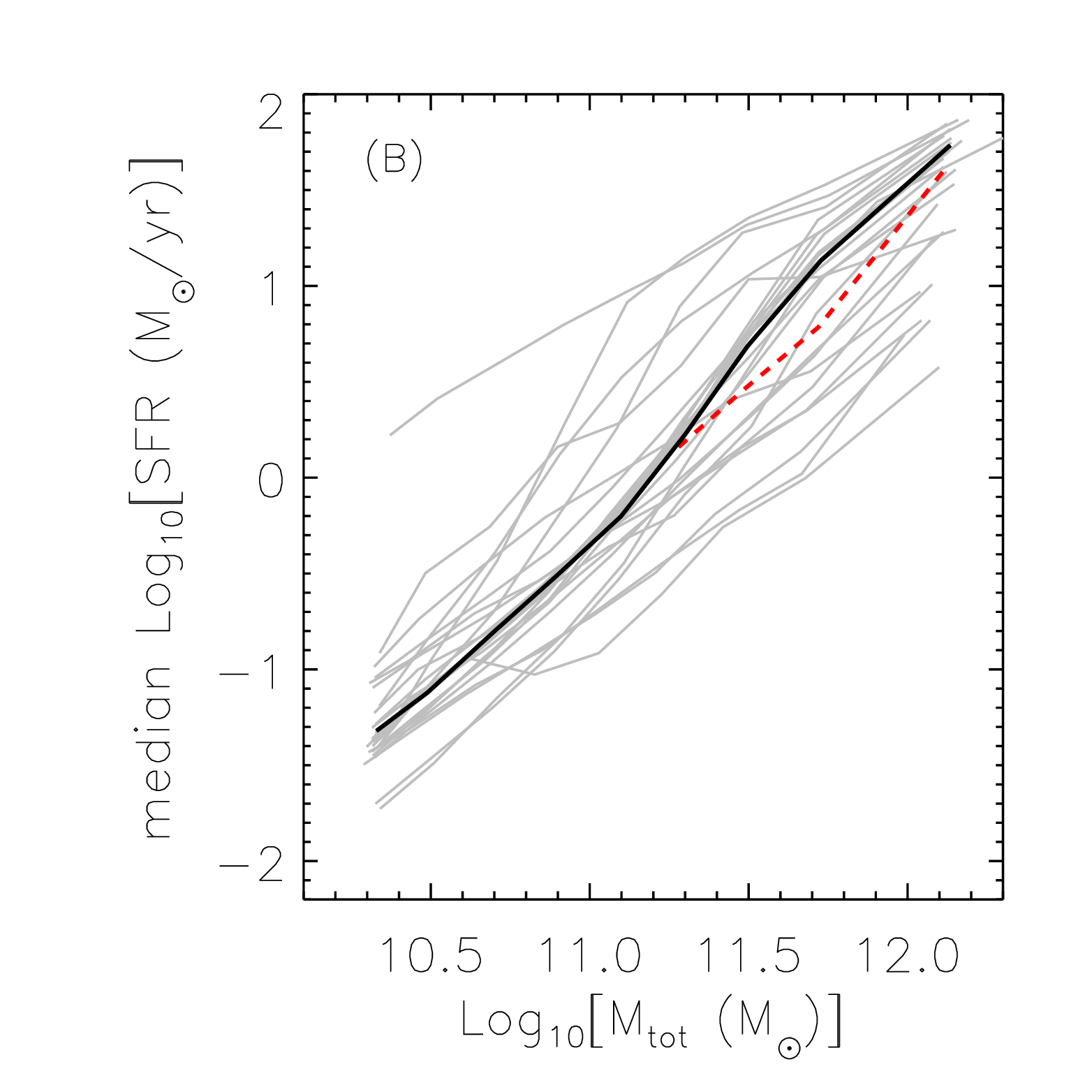}
\includegraphics[width=0.33\linewidth]{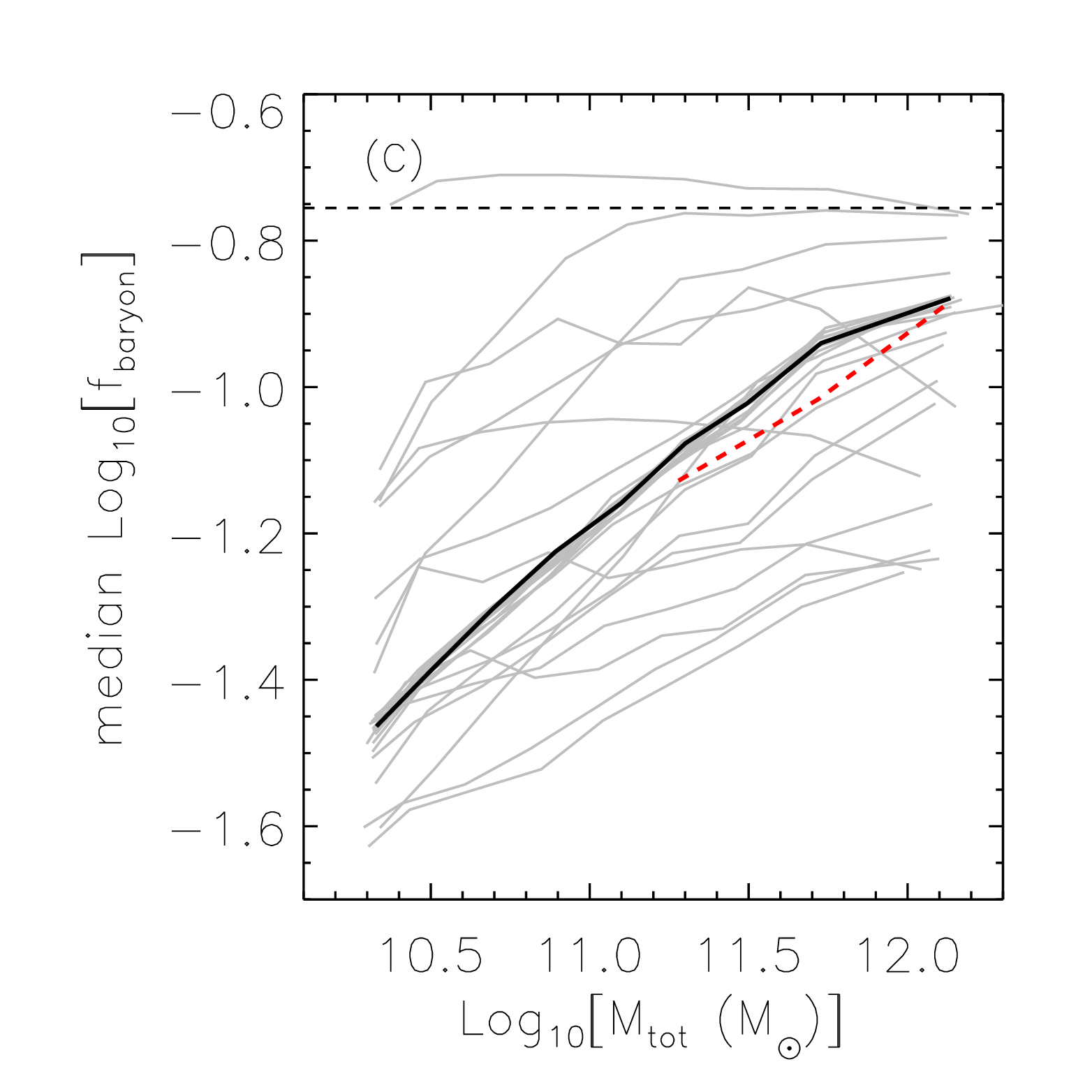} \\
\includegraphics[width=0.33\linewidth]{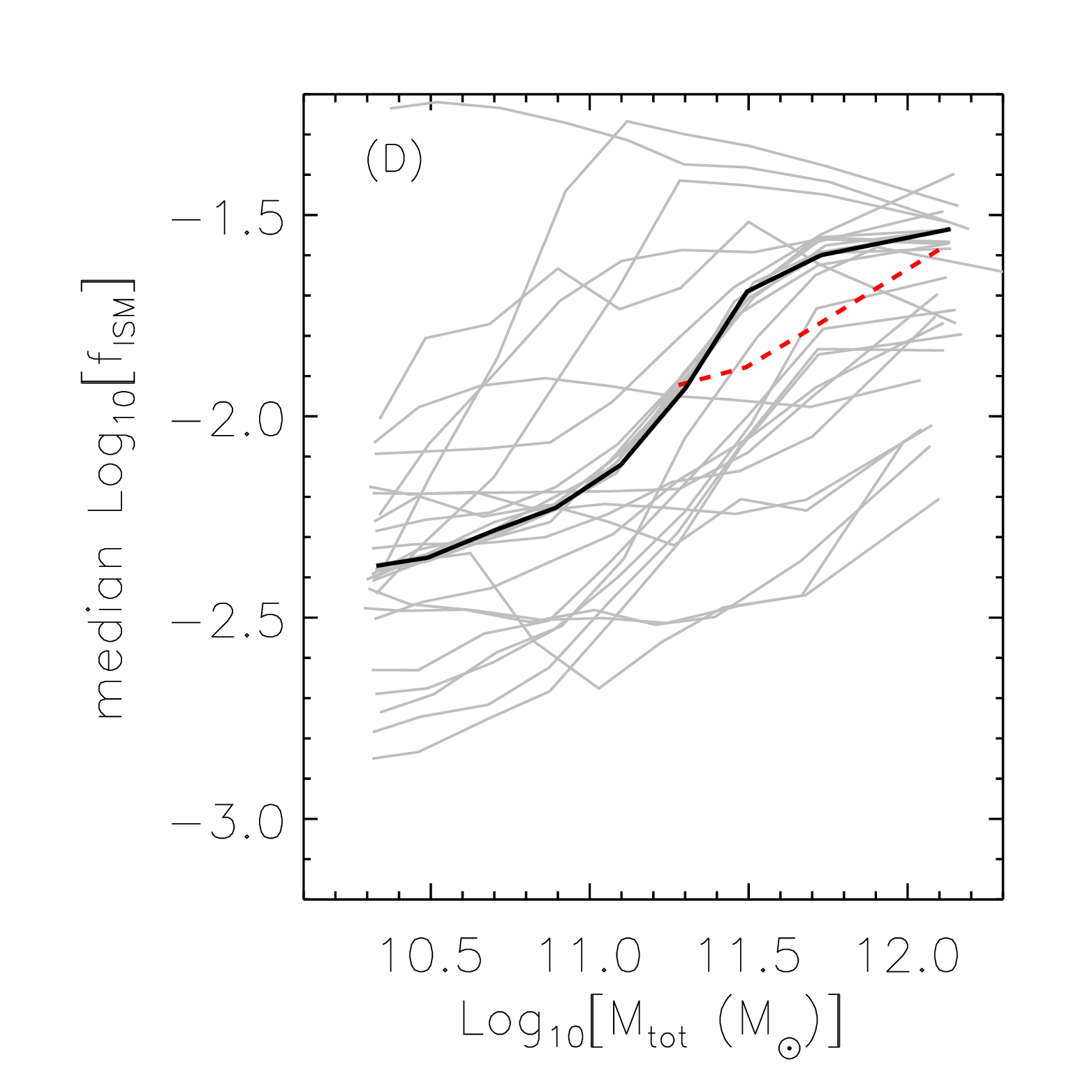}
\includegraphics[width=0.33\linewidth]{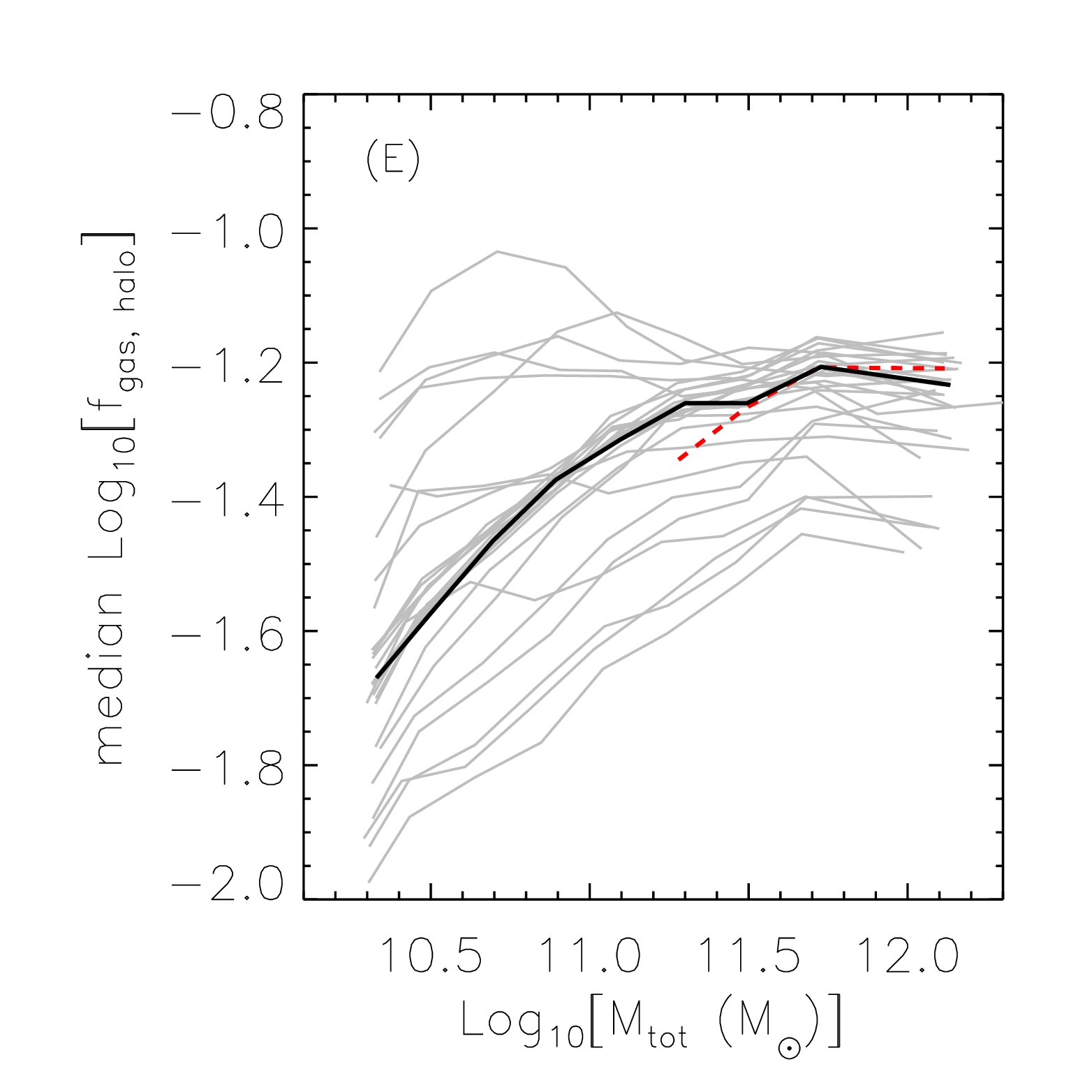}
\includegraphics[width=0.33\linewidth]{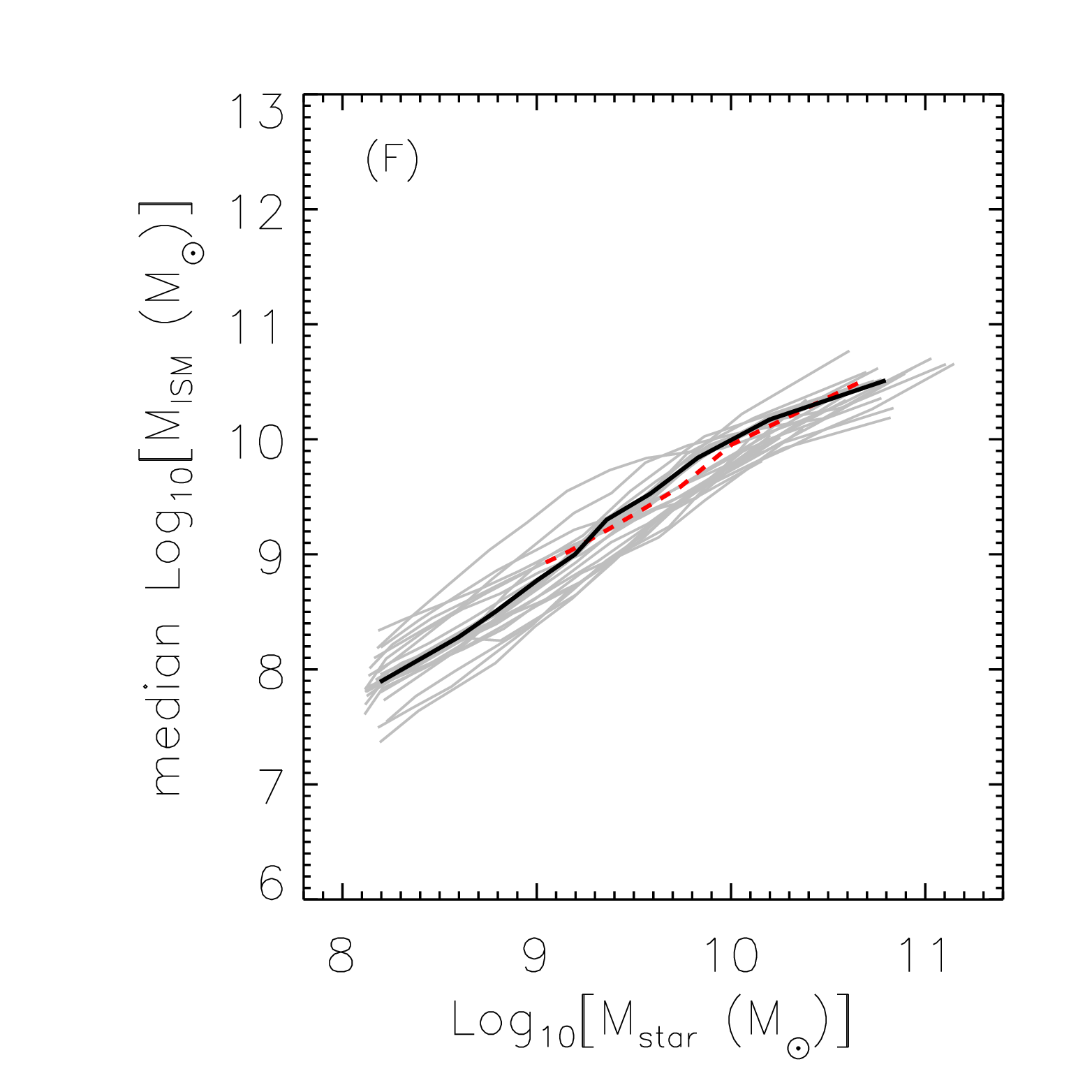} \\
\includegraphics[width=0.33\linewidth]{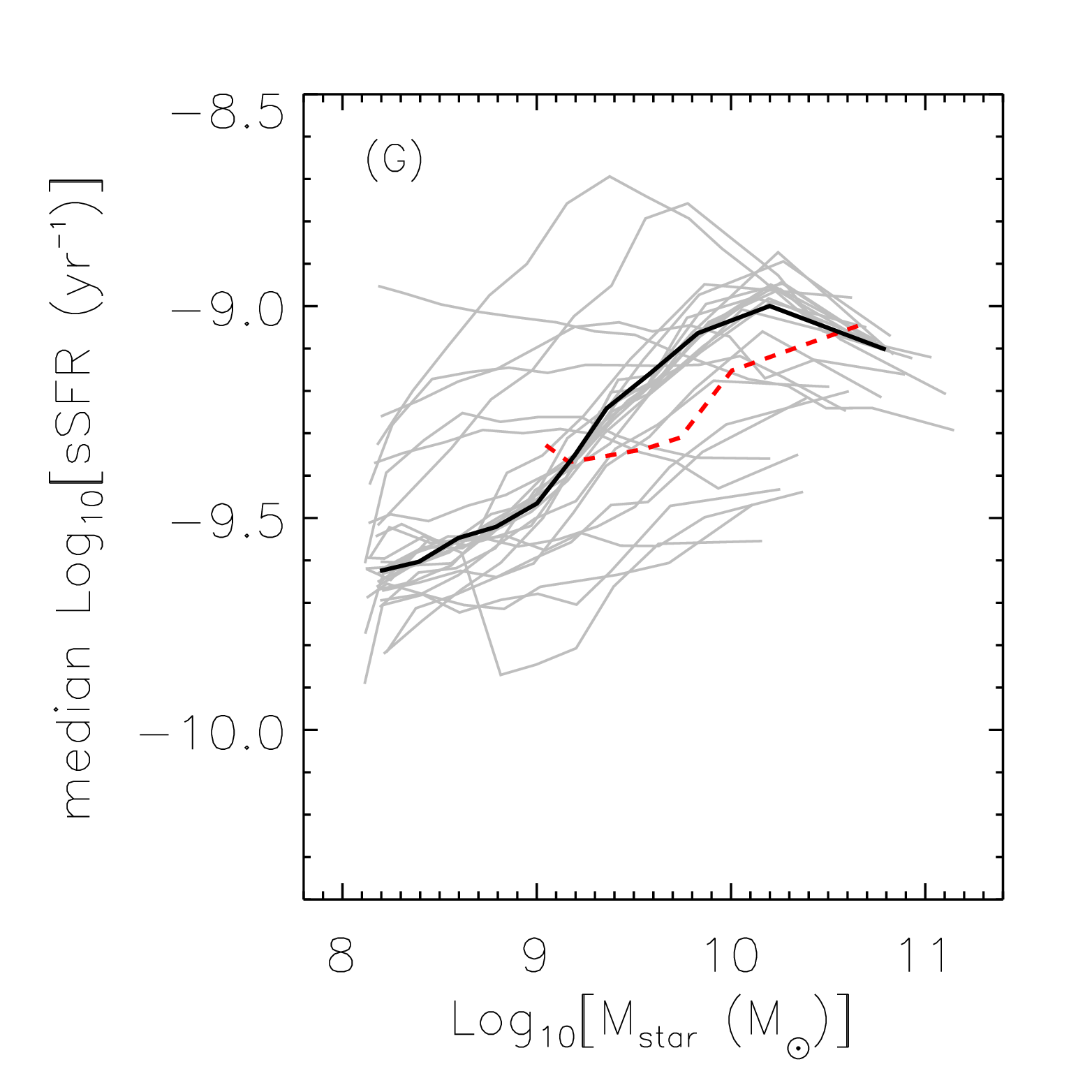}
\includegraphics[width=0.33\linewidth]{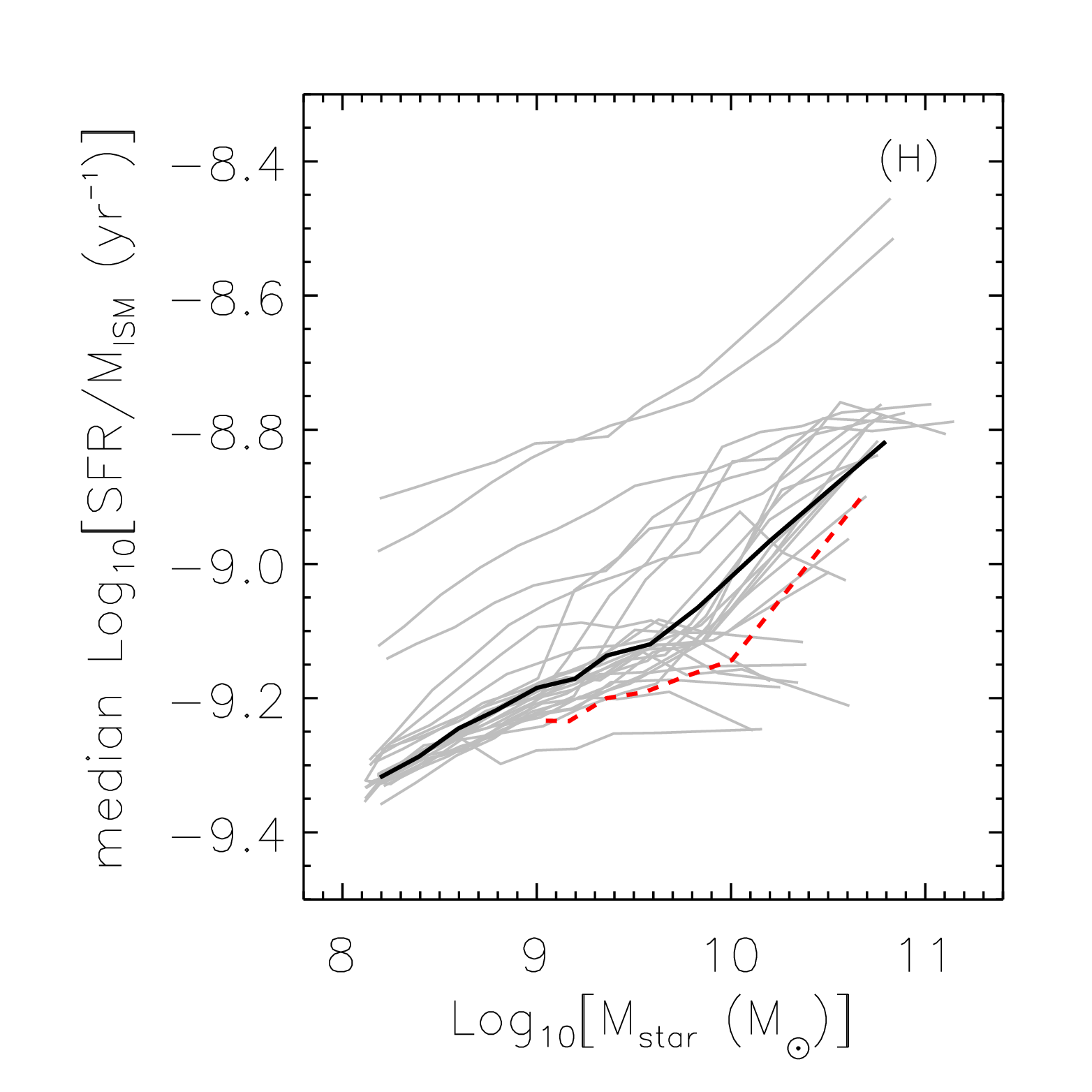}
\includegraphics[width=0.33\linewidth]{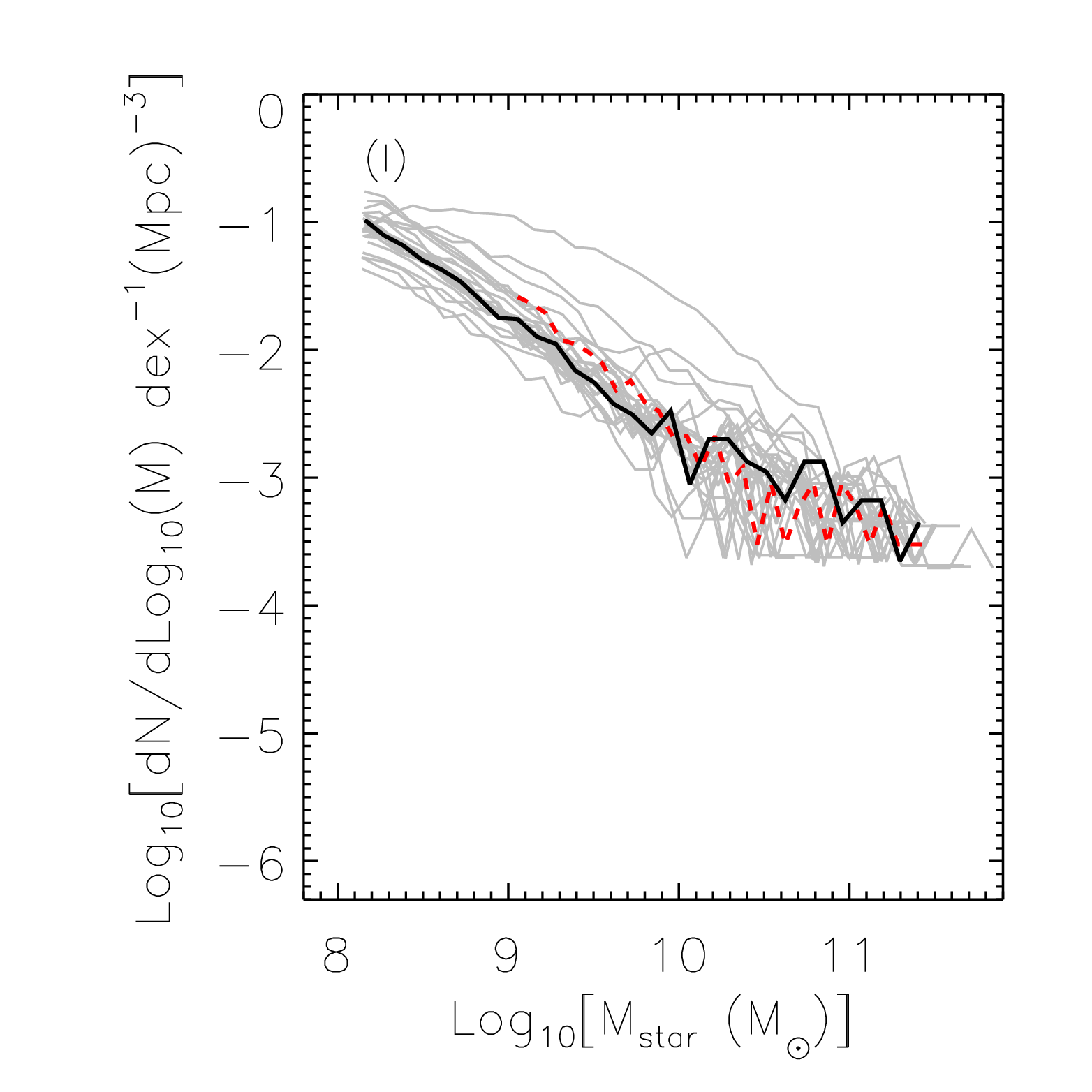} \\
\caption{As Fig.~\ref{fig:All_sims}, but for the reference input physics at two different resolutions: \textit{L025N512} (black solid lines) and \textit{L025N256} (red dashed lines), which has an 8 times lower mass resolution. For both sets of simulations the convergence criteria as used in the main body of the paper are used: a minimum of 2000 dark matter particles for panels (A) through (E) and a minimum of 100 star particles for panels (F) through (I). The thin grey lines are all physics variations at the same resolution as the black solid line. The spread due to physics variations, at fixed resolution, is comparable to or larger than the difference between simulation results at different resolution.} 
\label{fig:conv_all} 
\end{figure*}

In summary, our resolution requirements are as follows: To avoid biasing the results due to resolution effects, we impose a cut of 2000 dark matter particles when looking at relations with total halo mass and of 100 stars particles when investigating correlations with stellar mass. In Fig.~\ref{fig:conv_all} we show all simulations used in this paper and paper II in grey lines in the background, as well as the `\textit{REF}' simulation at the resolution used in the paper (\textit{L025N512}, black solid lines) and at lower resolution (\textit{L025N256}, red dashed lines). The spread in simulation results due to physics variations at fixed resolution is comparable to or larger than the difference between simulation results using the resolution that is used in the main body of the paper and a mass resolution that is 8 times lower than that. Even for results where the difference between two simulations with different physics at the same resolution is smaller than the difference between the simulations with the same physics but at different resolution, the difference between the simulations at the same resolution can still be significant. In such cases it becomes hard to tell whether the differences are due to a difference in resolution dependence of the two physics implementations or really only due to the physics. However, because they are run at the same resolution, it is unlikely that resolution affects the qualitative conclusions we draw from the comparison of the different models.

\section{The effect of halo definition}
\label{sec:physprop_halo}
Throughout this paper all of the halo masses that we quote are Friends-of-Friends \citep{davis85} masses, meaning that all particles associated with a given FoF halo contribute to its mass.  This particular definition of mass is somewhat arbitrary, so we check in this appendix whether using a different halo mass definition would significantly affect our results.  In particular, we compare FoF masses to those returned by {\sc SubFind} \citep{dolag09}, for both the main halos and the subhaloes.  {\sc SubFind} iteratively calculates, for each FoF halo, the mass of both baryonic and collisionless matter that is gravitationally bound to the same structure. 

\begin{figure}
   \begin{center}
   \resizebox{\hsize}{!}{\includegraphics{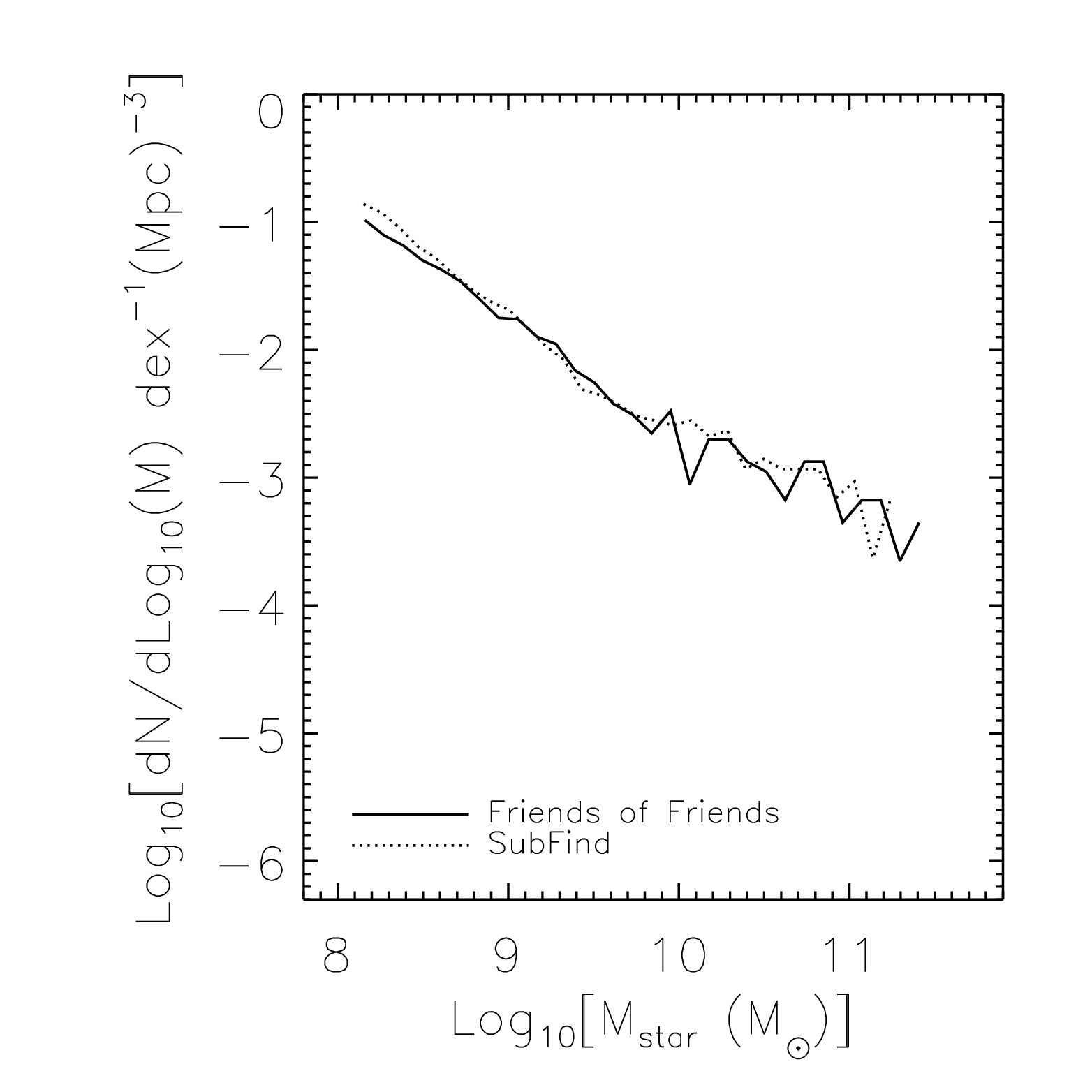}}\end{center}
    \caption{Galaxy stellar mass function for the `\textit{REF}' simulation with two different halo mass definitions.  The solid (dotted) curve shows the stellar mass function when FoF ({\sc SubFind}) haloes are used.  Results for mass functions are independent of the halo mass definition used, suggesting that any results plotted as a function of stellar mass are robust to changes in the halo mass definition.}{\label{fig:fof_vs_subfind_smf}
    }     
\end{figure}

In Fig.~\ref{fig:fof_vs_subfind_smf} we show the galaxy stellar mass function in FoF (solid curve) and {\sc SubFind} (dotted curve) haloes.  The two stellar mass functions are virtually identical.  This is largely because the stars live preferentially in the centres of haloes, and changing the halo mass definition affects mainly particles at the edges of the halo.  This plot suggests that all of our results that are plotted as a function of stellar mass are robust to the definition of halo used.

\begin{figure}
   \begin{center}
   \resizebox{\hsize}{!}{\includegraphics{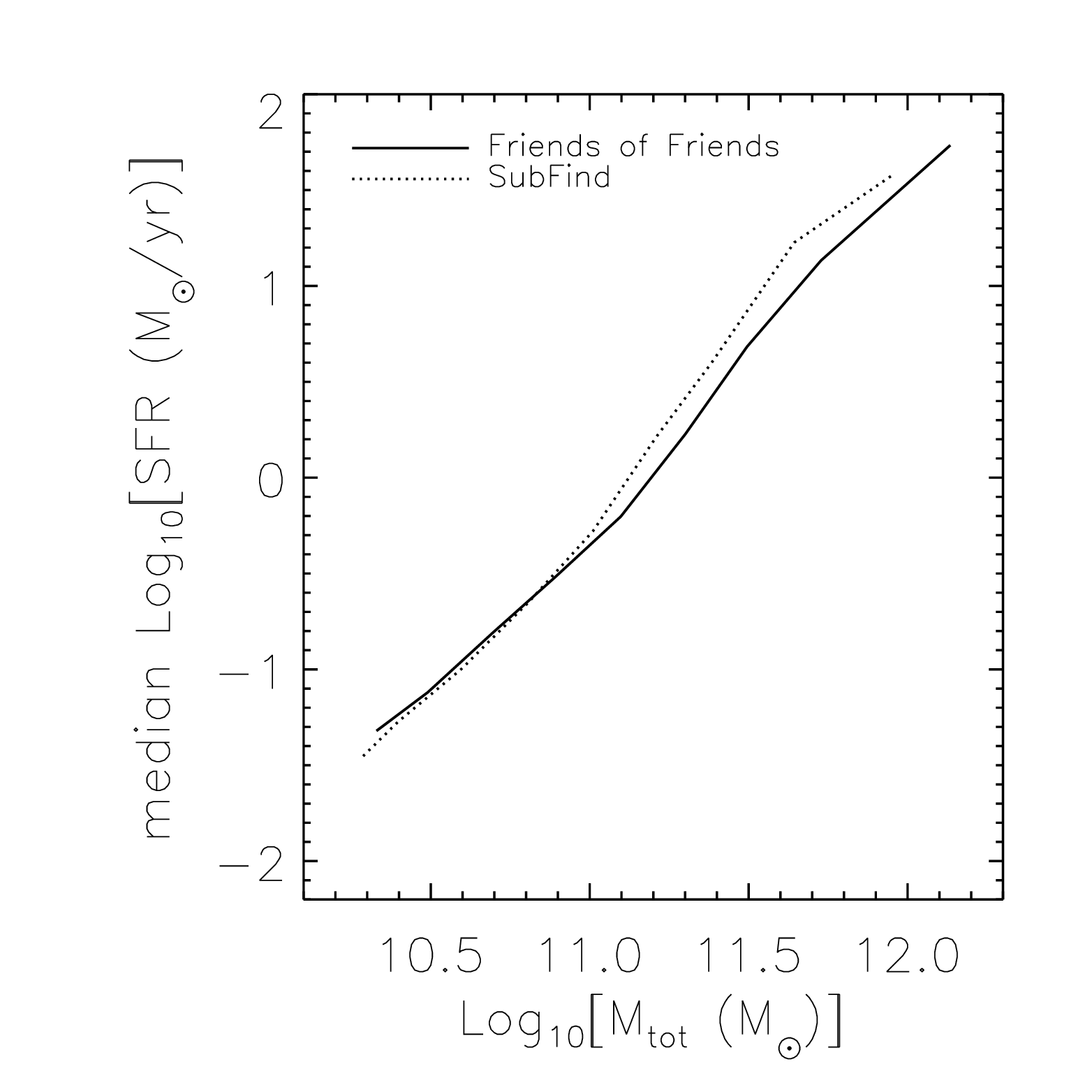}}\end{center}
    \caption{Relation between halo mass and galaxy star formation rate for the `\textit{REF}' simulation with two different halo mass definitions.  The solid (dotted) curve shows the relation for FoF ({\sc SubFind}) haloes.  Although at the highest masses, the different mass definitions give results that differ by up to 0.1 dex, this difference is far smaller than the magnitudes of the effects we are probing, and also affects all of the different simulations equally.}{\label{fig:fof_vs_subfind_sfr}
    }     
\end{figure}

The halo mass is somewhat more affected by the mass definition.  In Fig.~\ref{fig:fof_vs_subfind_sfr} we show the median star formation rate as a function of halo mass.  The solid curve shows the FoF haloes, and the dotted curve shows the {\sc SubFind} haloes.  Here we see almost perfect agreement at the low-mass end, but the overall effect of {\sc SubFind} at the high-mass end is to unbind haloes that are artificially linked together by the FoF algorithm, so at the highest masses the median halo mass can decrease by up to 0.1 dex.  We note, however, that this same effect will exist in all of the simulations analysed here, so although these curves may shift by a small amount, the main focus of this paper is on the relative differences between the simulations, and these remain totally unaffected by the halo mass definition. Note also that the {\sc Subfind} algorithm only removes mass from FoF haloes and never adds any. Many studies use the mass of a sphere containing an overdensity of $\sim 200$ as total mass for the main subhaloes, which results in halo masses closer to the FoF halo masses used in this paper.

\label{lastpage}

\end{document}